\pdfoutput=1
\documentclass[aps,prd,amsmath,floats,floatfix, twocolumn,
superscriptaddress,nofootinbib,showpacs,longbibliography]{revtex4-1}

\usepackage{comment}
\usepackage{xspace}
\usepackage{multirow}
\usepackage[caption=false]{subfig}
\usepackage[T1]{fontenc}
\usepackage[utf8]{inputenc}
\usepackage{lmodern}
\usepackage{verbatim}
\usepackage[dvipsnames, usenames]{xcolor}
\definecolor{linkcolor}{rgb}{0.0,0.3,0.5}
\usepackage[hypertexnames=false, unicode, colorlinks=true, linkcolor=linkcolor,
citecolor=linkcolor, filecolor=linkcolor,urlcolor=linkcolor,
pdfusetitle]{hyperref}
\usepackage[all]{hypcap}
\usepackage{graphicx}
\usepackage{xspace}
\usepackage{amssymb}
\usepackage[normalem]{ulem} 
\usepackage{bm} 
\usepackage{microtype}
\usepackage[english]{babel}
\usepackage{blindtext}
\usepackage{natbib}

\graphicspath{%
	{figs/}%
}

\DeclareMathAlphabet{\mathpzc}{OT1}{pzc}{m}{it}

\newcommand{\chieff}{\chi_{\mathrm{eff}}}

\newcommand{\bchif}{\bm{\chi}_{f}}
\newcommand{\bchiOne}{\bm{\chi}_{1}}
\newcommand{\bchiTwo}{\bm{\chi}_{2}}
\newcommand{\bvf}{\bm{v}_{f}}

\newcommand{\AEI}{\affiliation{Max Planck Institute for Gravitational Physics
		(Albert Einstein Institute), D-14476 Potsdam, Germany}}
\newcommand{\UMassDMath}{\affiliation{Department of Mathematics,
		University of Massachusetts, Dartmouth, MA 02747, USA}}
\newcommand{\CSCDR}{\affiliation{Center for Scientific Computing and Data Science Research,
		University of Massachusetts, Dartmouth, MA 02747, USA}}

\newcommand\MIT{\affiliation{LIGO Laboratory, Massachusetts Institute of
		Technology, Cambridge, Massachusetts 02139, USA}}
\newcommand{\MKI}{\affiliation{Department of Physics and Kavli Institute for
		Astrophysics and Space Research, Massachusetts Institute of Technology, 77
		Massachusetts Ave, Cambridge, MA 02139, USA}}
\newcommand{\monash}{\affiliation{School of Physics and Astronomy, Monash University, Vic 3800, Australia}}
\newcommand{\ozgrav}{\affiliation{OzGrav: The ARC Centre of Excellence for Gravitational Wave Discovery, Clayton VIC 3800, Australia}}
\newcommand{\RIT}{\affiliation{Center for Computational Relativity and Gravitation, Rochester Institute of Technology, Rochester, New York 14623, USA}}
\newcommand{\UTA}{\affiliation{University of Texas, Austin, TX 78712, USA}}
\newcommand{\UNLVphysics}{\affiliation{Department of Physics and Astronomy, University of Nevada, Las Vegas, 4505 South Maryland Parkway, Las Vegas, NV 89154, USA}}
\newcommand{\NCfA}{\affiliation{Nevada Center for Astrophysics, University of Nevada, Las Vegas, NV 89154, USA}}
\newcommand{\UOA}{\affiliation{Department of Statistics, The University of Auckland, New Zealand}}

\begin{document}

\title{Analysis of GWTC-3 with fully precessing numerical relativity surrogate models}

\author{Tousif Islam}
\email{tislam@umassd.edu}
\UMassDMath
\CSCDR

\author{Avi Vajpeyi}
\email{avi.vajpeyi@auckland.ac.nz}
\UOA
\ozgrav

\author{Feroz H. Shaik}
\UMassDMath
\CSCDR

\author{Carl-Johan Haster}
\UNLVphysics
\NCfA
\MIT
\MKI

\author{Vijay Varma}
\UMassDMath
\CSCDR
\AEI

\author{Scott E. Field}
\UMassDMath
\CSCDR

\author{Jacob Lange}
\RIT
\UTA

\author{Richard O’Shaughnessy}
\RIT

\author{Rory Smith}
\monash
\ozgrav


\hypersetup{pdfauthor={Islam et al.}}
\date{\today}

\begin{abstract}
The third Gravitational-Wave Transient Catalog (GWTC-3) contains 90 binary coalescence candidates detected by the LIGO-Virgo-KAGRA Collaboration (LVK). We provide a re-analysis of binary black hole (BBH) events using a numerical relativity (NR) waveform surrogate model, \texttt{NRSur7dq4}, that includes all $\ell \leq 4$ spin-weighted spherical harmonic modes as well as the complete physical effects of precession. Properties of the remnant black holes' (BH's) mass, spin vector, and kick vector are found using an associated remnant surrogate model \texttt{NRSur7dq4Remnant}. Both \texttt{NRSur7dq4} and \texttt{NRSur7dq4Remnant} models have errors comparable to numerical relativity simulations and allow for high-accuracy parameter estimates. We restrict our analysis to 47 BBH events that fall within the regime of validity of \texttt{NRSur7dq4} (mass ratios greater than 1/6 and total masses greater than $60 M_{\odot}$). While for most of these events our results match the LVK analyses that were obtained using the semi-analytical models such as \texttt{IMRPhenomXPHM} and \texttt{SEOBNRv4PHM}, we find that for more than 20\% of events the \texttt{NRSur7dq4} model recovers noticeably different measurements of black hole properties like the masses and spins, as well as extrinsic properties like the binary inclination and distance. For instance, GW150914\_095045 exhibits noticeable differences in spin precession and spin magnitude measurements. Other notable findings include one event (GW191109\_010717) that constrains the effective spin $\chieff$ to be negative at a 99.3\% credible level and two events (GW191109\_010717 and GW200129\_065458) with well-constrained kick velocities. Furthermore, compared to the models used in the LVK analyses, \texttt{NRSur7dq4} recovers a larger signal-to-noise ratio and/or Bayes factors for several events. While these findings have important astrophysical implications, some of the more interesting events are affected by transient detector noise, which can be challenging to remove from short-duration signals.
\end{abstract}

\maketitle
\section{Introduction}
\label{sec:introduction}
The discovery of GW150914\_095045 by the Advanced LIGO~\cite{Harry:2010zz} and
Virgo~\cite{VIRGO:2014yos} detectors marked the beginning of gravitational-wave
(GW) astronomy.  Since then, during the first three observing runs (referred to
as O1, O2, and
O3)~\cite{LIGOScientific:2018mvr,LIGOScientific:2020ibl,LIGOScientific:2021usb,LIGOScientific:2021djp},
the LVK Collaboration, consisting of LIGO and Virgo, together with the
KAGRA detector~\cite{KAGRA:2020tym}, has detected a total of $90$
gravitational wave events~\cite{LIGOScientific:2021djp}.
Several independent analyses of the public data have
further revealed another $\sim15$
events~\cite{Venumadhav:2019lyq,Nitz:2018imz,Nitz:2020oeq,Nitz:2021uxj}.
Analyses of the detected signals allow us to infer individual and population
properties of merging compact objects
~\cite{LIGOScientific:2018mvr,LIGOScientific:2020ibl,LIGOScientific:2021usb,Venumadhav:2019lyq,Nitz:2018imz,
Nitz:2020oeq,Nitz:2021uxj,2017ApJ_840L_24F,2019PhRvD_100d3012W,2019ApJ_882L_24A,Roulet:2020wyq,
LIGOScientific:2021psn,Roulet:2020wyq,vanSon:2022myr,Payne:2022xan,
Doctor:2019ruh,2021ApJ_912_98F}
including the existence of mass gaps and possible formation channels of binary
black hole (BBH) systems~\cite{Stevenson:2015bqa,LIGOScientific:2016vpg,Zevin:2017evb,LIGOScientific:2017apx}.
GW signals also encode information of the equation of state of the neutron-star
matter~\cite{LIGOScientific:2018cki,Most:2018hfd,Essick:2019ldf,LIGOScientific:2019eut},
the nature of
gravity~\cite{Yunes:2013dva,Carson:2019yxq,LIGOScientific:2016lio,LIGOScientific:2020tif,
LIGOScientific:2019fpa}, environments around compact objects~\cite{Barausse:2014pra},
and inform our understanding of cosmology by providing
an independent measurement of the Hubble
constant~\cite{LIGOScientific:2017adf,LIGOScientific:2019zcs,Cantiello:2018ffy}.

Inference of binary source properties from a detected signal relies on the
availability of expedient and accurate gravitational waveform models that are
capable of describing the entire coalescence from the inspiral through the
binary merger and ringdown of the remnant object. Current GW models can be
categorized into three sets: phenomenological
models ~\cite{husa2016frequency,khan2016frequency,london2018first,khan2019phenomenological,
hannam2014simple,khan2020including,Pratten:2020ceb,Estelles:2020osj,Estelles:2020twz,Estelles:2021gvs,Estelles:2021gvs,Hamilton:2021pkf, Ghosh:2024:PhRvD}, effective-one-body (EOB) models~\cite{bohe2017improved,cotesta2018enriching,cotesta2020frequency,pan2014inspiral,
babak2017validating,cotesta2020frequency,pan2014inspiral,babak2017validating,Ossokine:2020kjp,
Damour:2014sva,Nagar:2019wds,Nagar:2020pcj,Riemenschneider:2021ppj,Khalil:2023kep,Pompili:2023tna,Ramos-Buades:2023ehm,vandeMeent:2023ols},
and numerical relativity (NR) surrogate
models~\cite{Blackman:2015pia,Blackman:2017dfb,Blackman:2017pcm,varma2019surrogate,
varma2019surrogate2,Islam:2021mha}. The phenomenological and EOB waveform families are semi-analytical
models that rely on a physically motivated ansatz, with calibration parameters
that are fit to a set of NR simulations. NR surrogate models instead take a purely data-driven
approach by training the model directly on NR simulations without the need for
an ansatz. NR surrogates have been shown to be more accurate than
semi-analytical models within their common regime of validity
and capture the
underlying physics in the NR simulations (such as precession) at an accuracy
comparable to the simulations themselves~\cite{varma2019surrogate,
varma2019surrogate2, Blackman:2017pcm}. Consequently, NR surrogate
models can provide highly accurate and trustworthy information about BBH source
properties, especially as detector sensitivity improves~\cite{Kumar:2018hml,Islam:2020reh,
Varma:2021csh,LIGOScientific:2020ufj}.

Compared to phenomenological and EOB waveform models, however, surrogate models have a restricted regime of
validity as they rely on the availability of NR simulations for training.
The surrogate model we consider in this work,
\texttt{NRSur7dq4}~\cite{varma2019surrogate}, can only be evaluated for mass
ratios $q\geq1/6$, where $q:=m_{2,{\rm det}}/m_{1,{\rm det}}$; $m_{1,{\rm
det}}$ and $m_{2,{\rm det}}$ are the masses of the primary and secondary black
hole, respectively, in the detector frame. Furthermore, because of the
computational limitations of NR simulations, \texttt{NRSur7dq4} only supports
relatively short duration signals ($\sim 20$ orbits), which further restricts its
applicability to heavy binaries with a detector-frame total mass of
$M_{\rm det} := m_{1,{\rm det}} + m_{2,{\rm det}} \ge 60M_{\odot}$
These restrictions
leave 47 events from GWTC-3 that can be analyzed with the \texttt{NRSur7dq4}
waveform model (see Section \ref{sec:event_selection} for details).

In this paper, we present a thorough analysis of these 47 events using the
\texttt{NRSur7dq4} waveform model for binary source parameter inference and
an associated \texttt{NRSur7dq4Remnant} model to infer the mass, spin vector, and kick
vector of the final black hole remnant.
Our work builds upon previous studies that applied NR surrogate models to
special events such as the first observation~\cite{Kumar:2018hml},
the first observation of a high-mass ratio BBH merger~\cite{Islam:2020reh},
a highly precessing system~\cite{Varma:2022pld, Hannam:2021pit},
and an intermediate-mass black hole formed through a BBH merger~\cite{LIGOScientific:2020ufj}.
We also compare
(i) \texttt{NRSur7dq4} posteriors against publicly available LVK
posteriors~\cite{LIGOScientific:2021usb, LIGOScientific:2021djp, GWTC2.1_PE,
GWTC3_PE} obtained using the \texttt{IMRPhenomXPHM}~\cite{Pratten:2020ceb} and
\texttt{SEOBNRv4PHM}~\cite{Ossokine:2020kjp} models,
(ii) the recovered Bayes factors and differences in the signal-to-noise
ratios (SNRs) between \texttt{NRSur7dq4} and \texttt{IMRPhenomXPHM}, and
(iii) the remnant mass and spin magnitude estimates obtained for the three models.
We find that while, for most events, our results are consistent with the LVK analyses,
more than 20\% of events exhibit noticeably different measurements
and some differences, such as constraining GW191109\_010717's effective spin to be confidently negative, may have
important astrophysical implications.

The rest of the paper is organized as follows. Section~\ref{sec:methods}
presents our analysis framework, including strain data,
the \texttt{NRSur7dq4} model, and our
choice of priors, reference frame, and sampler settings.
In Sec.~\ref{sec:event_selection}, we discuss
how the set of 47 events that are analyzed have been selected.  We provide
an overview of the \texttt{NRSur7dq4} results in Sec.~\ref{sec:results} and
quantify the difference between \texttt{NRSur7dq4} and
\texttt{IMRPhenomXPHM}/\texttt{SEOBNRv4PHM} posteriors.
The events that show the largest discrepancy with the LVK results
are considered in more detail in Sec.~\ref{sec:special_events} and Sec.~\ref{sec:sky}.
In Sec.~\ref{sec:bayes_factors}, we use
Bayes factors and recovered SNRs to better understand whether the data has a
preference for a particular model, finding the data shows mild support
for the \texttt{NRSur7dq4} model. Constraints on the mass,
spin vector, and kick vector are considered in Sec.~\ref{sec:remnant}. Finally,
in Sec.~\ref{sec:discussion} we summarize our results and discuss
the implications of our findings.
Our posterior samples are publicly available at Ref.~\cite{NRSurCatalog}.

\section{Data analysis framework}
\label{sec:methods}
In this section, we summarize the Bayesian inference methods used in this study
(Sec. \ref{subsec:bayes}) and provide an overview of our data analysis framework.
Our setup is mostly consistent with settings
used to obtain the public LVK posteriors~\cite{LIGOScientific:2021usb,
LIGOScientific:2021djp, GWTC2.1_PE, GWTC3_PE}, with the relevant
differences detailed in Sec.~\ref{subsec:prior} and~\ref{sec:settings}

\subsection{Bayesian inference}
\label{subsec:bayes}
The GW signal from a coalescing quasi-circular compact BBH system in general
relativity can be completely characterized
by a set of 15 parameters that we denote $\Theta=\{\mathcal{I},\mathcal{E}\}$.
The parameter vector $\Theta$ consists
of eight intrinsic parameters $\mathcal{I}$ and seven extrinsic parameters
$\mathcal{E}$.  The vector
$\mathcal{I}:=\{m_1,m_2,\chi_1,\chi_2,\theta_1,\theta_2,\phi_{12},\phi_{\rm
JL}\}$ contains the intrinsic parameters that describe the binary: the
component masses $m_1$ and $m_2$ (with $m_1 \geq m_2$), dimensionless spin
magnitudes $\chi_1$ and $\chi_2$, spin tilt angles $\theta_1$ and $\theta_2$, and
two azimuthl spin angles $\phi_{12}$ and $\phi_{\rm JL}$. The definition of these
parameters is nicely summarized in Appendix E of Ref.~\cite{Romero-Shaw:2020owr}.  We further use
$\bm{\chi_1}$ and $\bm{\chi_2}$ to denote spin vectors for each component black
hole, $\bm{L}$ to denote the orbital angular momentum, and $\bm{J}$ to denote
the total angular momentum.  Moreover, we distinguish
between the source-frame and detector-frame masses by using the subscript `det'.
For example, $m_1$ is the source-frame mass and $m_{1,\rm det}$ the
detector-frame mass of the primary black hole. These masses are
related by $m_1 = m_{1,{\rm det}}/(1+z)$ where $z$ is the redshift of the
source.  The set of extrinsic parameters $\mathcal{E}=\{\theta_{\rm
JN},D_{\rm L},\alpha,\delta,\psi,\phi_c,t_c\}$ parameterizes the location and
orientation of the binary relative to the detectors as well as the coalescence time.
The angle between the total angular momentum of the binary and the
line-of-sight to the detector is denoted by $\theta_{\rm JN}$ (filling the role
of the `inclination angle') while the luminosity distance to the source is
denoted by $D_{\rm L}$.  Right ascension $\alpha$ and declination $\delta$
parameterize the location of the source in the sky, and $\psi$ is the polarization angle.  The
coalescence time is denoted by $t_c$ while $\phi_c$ indicates a reference
orbital phase.

We use Bayes' theorem to compute the \textit{posterior probability distribution} of
the binary parameters,
\begin{equation}
	p(\Theta | \{d_k\}, {\cal H}) = \frac{{\cal L} ( \{d_k\} | \Theta, {\cal H}) \pi(\Theta | {\cal H})}{\mathcal{Z}(\{d_k\} | {\cal H})} \,,
\end{equation}
where $\cal H$ is the signal hypothesis, $d_k(t)$ represents time-domain strain data
for the $k^{th}$ detector,
\begin{equation}
	d_k(t) = h_k(t; \Theta) + n_k(t) \,,
\end{equation}
which is assumed to be a sum of the true signal $h_k(t; \Theta)$ and noise $n_k(t)$
in each detector. The subscript ``k'' on our signal model denotes that the
observed signal will look different in different interferometers.
The posterior $p(\Theta | \{d_k\}, {\cal H})$ is the target for the parameter estimation analysis
while the model evidence,
\begin{equation}
	\mathcal{Z}(\{d_k\}|{\cal H}) = \int d\Theta \pi(\Theta | {\cal H}) \mathcal{L}(\{d_k\}|\Theta, {\cal H}) \,,
\end{equation}
is the target for hypothesis testing (sometimes referred to as model selection).
The prior probability, $\pi(\Theta | \cal H)$, is a prescribed probability distribution and represents our
initial assumptions about the parameters describing an individual binary. The likelihood function,
\begin{equation}
	\mathcal{L}(\{d_k\}|\Theta, {\cal H}) \propto \prod_{k} \mathrm{exp}\left( - \frac{1}{2}\langle d_k-h_k^{\cal H}(\Theta)|d_k-h_k^{\cal H}(\Theta) \rangle \right) \,,
\end{equation}
describes how well each set of $\Theta$ matches the data. Here, $h_k^{\cal H}(\Theta)$ is the signal
waveform generated from a specific waveform model as part of our hypothesis (we typically omit the superscript for brevity), and $\langle a | b\rangle$
is the noise-weighted overlap integral defined as
\begin{equation}\label{eq:inner_product}
	\langle a | b\rangle := 4 \Re\int_{f_{\rm low}}^{f_{\rm high}} \dfrac{\tilde{a}(f) \tilde{b}^*(f)}{S_n(f)} df,
\end{equation}
with $S_n(f)$ being the one-sided power spectral density (PSD) of the detector noise, a ``$\sim$'' indicates a
Fourier transform operation, and $*$ represents
the complex conjugate. The integration limits, $f_{\rm low}$ and $f_{\rm high}$, are chosen to reflect
the sensitivity bandwidth of the detectors; specific values are given in Sec.~\ref{sec:settings}.

To quantify how much more likely that the data is described by a signal and not a noise process, we compute the Bayes
factor,
\begin{equation}
\label{eq:bayes_factor}
	\mathcal{B}=\frac{\mathcal{Z_{\cal H}}}{\mathcal{Z}_n},
\end{equation}
where $\mathcal{Z_{\cal H}}=\mathcal{Z}(\{d_k\}|{\cal H})$ and $\mathcal{Z}_n$ denote, respectively, the evidence for a
signal model $\cal H$ and a noise-only model.

\subsection{Numerical relativity surrogate model}
\label{subsec:waveform_model}
Surrogate models use NR waveforms as training data and build a highly accurate interpolant
over the parameter space using a combination of reduced-order modeling~\cite{Field:2013cfa, Field:2011mf},
parametric fits, and non-linear transformations of the waveform data~\cite{Blackman:2017dfb}.
The \texttt{NRSur7dq4} model~\cite{varma2019surrogate} used in this paper is trained on 1528 NR simulations~\cite{varma2019surrogate} and
spans the 7-dimensional parameter space of spin-precessing binaries.
The model includes all $\ell \leq4$ spin-weighted spherical harmonic modes, as well as the
precession frame dynamics and spin evolution of the black holes.
While the model has been trained for mass ratio $1/4\le q\le1.0$ and spins $0.0 \le \chi_{1,2} \le 0.8$,
it can be extrapolated to $q\ge 1/6$ and $\chi_{1,2} \le1.0$~\cite{varma2019surrogate, Walker:2022zob}.
In its regime of validity, \texttt{NRSur7dq4} improves upon semi-analytical models by about an order of
magnitude in accuracy, in terms of mismatches against NR waveforms~\cite{varma2019surrogate, Walker:2022zob}.

We compare our results with public LVK posteriors~\cite{LIGOScientific:2021usb, LIGOScientific:2021djp}
obtained using the \texttt{IMRPhenomXPHM}~\cite{Pratten:2020ceb,Garcia-Quiros:2020qpx,Pratten:2020fqn}
and \texttt{SEOBNRv4PHM}~\cite{Ossokine:2020kjp,Cotesta:2018fcv} waveform models.
\texttt{IMRPhenomXPHM} is a phenomenological model that includes the effects of precession and the
$\{(\ell,m)\}=\{(2,\pm2), (2,\pm1), (3,\pm3), (3,\pm2),(4,\pm4)\}$ modes in the coprecessing frame.
\texttt{SEOBNRv4PHM} is an EOB model that also includes the effects of precession and the
$\{(\ell,m)\}=\{(2,\pm2), (2,\pm1), (3,\pm3),(4,\pm4), (5,\pm5)\}$ modes in the coprecessing frame modes.
Importantly, while both \texttt{IMRPhenomXPHM} and \texttt{SEOBNRv4PHM} are
calibrated against nonprecessing NR simulations, they are not calibrated
against precessing simulations.  Instead, they apply a frame-twisting
procedure to approximately transform an aligned-spin waveform in the
coprecessing frame to a precessing waveform in the inertial frame.

To infer the mass, spin vectors, and the kick velocity vector of the remnant black hole,
we use the remnant surrogate model, \texttt{NRSur7dq4Remnant}~\cite{Varma:2018aht,varma2019surrogate}.
The \texttt{NRSur7dq4Remnant} model is trained on the same set of
1528 NR simulations as \texttt{NRSur7dq4}, and is applicable in the same region of parameter space.
\texttt{NRSur7dq4Remnant} improves upon previous remnant models by an order of magnitude
in accuracy, in terms of errors in remnant properties with respect to NR simulations~\cite{varma2019surrogate}.

\subsection{Frame choice}
\label{subsec:frame_choice}
All time-dependent binary parameters, such as the BH spins and the system's orientation,
are defined in a frame such that at some reference time ($t_{\rm ref}$), the
$z$-axis is along the instantaneous angular momentum vector $\bm{L}$, the $x$-axis is along the line of
separation from the less massive BH to the more massive BH, and the $y$-axis completes
the right-handed triad.  The reference point is defined as the time ($t_{\rm
ref}$) during the binary evolution where the GW frequency in the coprecessing
frame (defined as twice the time derivative of Eq.~3 of
Ref.~\cite{varma2019surrogate}) passes $f_{\rm ref}=20$ Hz. Following
Ref.~\cite{Schmidt:2017btt}, we refer to this frame as the \emph{wave frame} at
$f_{\rm ref}=20$ Hz. We measure all binary parameters at this
reference frequency.

\subsection{GW strain data}
\label{subsec:strain}
The strain $d(t)$, PSD, and detector calibration uncertainty data are obtained from the LVK
public data releases~\cite{LIGOScientific:2019lzm, GWTC2.1_PE,
GWTC3_PE, GWTC2.1_Glitch, GWTC3_Glitch}. We use the de-glitched strain data for
certain events as summarized in Table \ref{Tab:special_events}.
The publicly available PSDs were generated using the on-source \texttt{BayesWave}
method~\cite{Littenberg:2014oda, Cornish:2014kda, Chatziioannou:2019zvs} while
the effect of frequency-dependent uncertainties in amplitude and phase of the interferometer calibration
on the parameter estimation of each event follows the methods of Refs.~\cite{calib,
Romero-Shaw:2020owr}. The signal's geocentric trigger time, duration, and sampling rate are
also taken from publicly available~\footnote{From each event's
\texttt{C01:IMRPhenomXPHM} and \texttt{C01:SEOBNRv4PHM} parameter estimation
results found within the respective \texttt{PEDataRelease\_mixed\_cosmo.h5} files.} data releases.
In particular, the event trigger times that can be separately obtained from
the Gravitational Wave Open Science Center (\href{https://www.gw-openscience.org/}{GWOSC})
were not used as they can be inconsistent~\footnote{The geocentric trigger times in GWOSC come from a search pipeline, which we found to be not sufficiently precise given our time-of-coalescence priors, especially for multi-modal distributions in the time-of-arrival parameter.}
with the times used for the LVK parameter estimation analyses.
The \texttt{GWTC-3} data
release~\cite{LIGOScientific:2021usb,LIGOScientific:2021djp, GWTC2.1_PE,
GWTC3_PE,GWTC2.1_Glitch,GWTC3_Glitch} further describe the methods used for
data conditioning and (where needed) deglitching.

\begin{table}
	\centering
	\begin{tabular}{c|c}
		\toprule
		Events with  \texttt{SEOBNRv4PHM} & Events with de-glitched\\
		posteriors missing in &strain data \\
		public LVK datasets &\\
		\hline
		& \\
		GW190413\_052954     &      GW190413\_134308   \\
		GW190413\_134308     &      GW190503\_185404   \\
		GW190421\_213856     &      GW190513\_205428   \\
		GW190426\_190642     &      GW190514\_065416   \\
		GW190521\_030229     &      GW190701\_203306   \\
		GW190602\_175927     &      GW191109\_010717    \\
		GW190803\_022701     &      GW200129\_065458   \\
		GW190828\_063405     &      \\
		GW190926\_050336     &      \\
		GW190929\_012149     & \\
		& \\
		\botrule
	\end{tabular}
\caption{
List of events for which \texttt{SEOBNRv4PHM} posteriors are absent
in the publicly released LVK posteriors~\cite{LIGOScientific:2021usb,
LIGOScientific:2021djp, GWTC2.1_PE, GWTC3_PE} as well as the events for which we use
publicly available de-glitched strain data~\cite{GWTC2.1_Glitch, GWTC3_Glitch,deglitch_ligodcc}.}
\label{Tab:special_events}
\end{table}

\subsection{Choice of prior}
\label{subsec:prior}
Our assumptions for the priors $\pi(\Theta | {\cal H})$ are identical to the
LVK analyses of the \texttt{GWTC-3} catalog~\cite{LIGOScientific:2018mvr,LIGOScientific:2020ibl,
LIGOScientific:2021usb,LIGOScientific:2021djp} with additional restrictions on the
mass ratio and the total mass of the binary:
\begin{itemize}
	\item We choose uniform priors in the detector-frame component masses subject to the following constraints on the BBH system:
	(i) the chirp mass satisfies $12M_{\odot} \leq \mathcal{M}_{c, {\rm det}}\leq 400M_{\odot} $,
	(ii) the mass ratio satisfies $0.167 \leq q \leq 1$, and
	(iii) the total mass satisfies $M_{\rm det} \geq 60 M_{\odot}$,
	where the chirp mass~\cite{PhysRevD.47.2198} is defined as $\mathcal{M}_{c, \rm det} =
	\frac{(m_{1, \rm det} m_{2, \rm det})^{3/5}}{(m_{1, \rm det}+m_{1, \rm det})^{1/5}}$.
	The second and third constraints are imposed to restrict the analysis to \texttt{NRSur7dq4}'s region of validity; see Sec.~\ref{sec:event_selection}.
	\item Uniform priors are used for the dimensionless spin magnitudes ($0.0 \leq \chi_{1,2} \leq 0.99$)
	of the binary, with spin orientations taken as uniform on the unit sphere.
    \item The prior on the luminosity distance assumes uniform source distribution in comoving volume
    and time as implemented in the \texttt{UniformSourceFrame} prior class in Ref~\cite{Romero-Shaw:2020owr}
    within $100 $ Mpc $ \leq D_{\rm L} \leq 10,000$ Mpc. For some events, we use a higher
    upper bound to ensure that the posterior is contained within the prior's range.
	\item For the orbital inclination angle $\theta_{\rm JN}$, we assume a uniform prior over $-1 \leq \cos(\theta_{\rm JN}) \leq 1$.
	\item Priors on the sky location parameters, right ascension $\alpha$ and declination $\delta$, are assumed
	to be uniform over the sky with periodic boundary conditions for $\alpha$.
	\item For the time of coalescence, we assume a uniform prior over $t_{\rm geo} -0.1 \le t_c \le t_{\rm geo} +0.1 $ where $t_{\rm geo}$ is the geocentric trigger time.
\end{itemize}

\subsection{Parameter estimation settings}
\label{sec:settings}

\begin{table}
	\centering
	\begin{tabular}{p{0.22\textwidth}|p{0.22\textwidth}}
		\toprule
		Sampler                  & Parameters          \\
		\hline
		\multirow{3}{*}{\texttt{Dynesty}~\cite{speagle2020dynesty}} & live-points $=2000$ \\
		& tolerance $=0.1$    \\
		& nact $=50$ \\
		\botrule
	\end{tabular}
	\caption{Dynesty static nested sampling configuration parameters. We use $2000$ live points and sample away from a current live point with random walks. We require that the number of random walk steps taken in each chain is at least 50 (=nact) times the autocorrelation length of the chain. Sampling continues until the estimated contribution of the remaining prior volume to the evidence is less than $0.1$ (=tolerance). }
	\label{Tab:sampler}
\end{table}

Performing a parameter estimation analysis requires specifying numerous configuration options whose specific settings can impact the final inferred source parameters. We have chosen our analysis settings to match the LVK analyses of \texttt{GWTC-3} as closely as possible based on
what is documented in Ref.~\cite{LIGOScientific:2019lzm} and in the public LVK data release~\cite{GWTC2.1_PE, GWTC3_PE,GWTC2.1_Glitch,GWTC3_Glitch}. We briefly describe the most important parameter estimation settings while pointing out differences when they arise. Our complete configuration and environment files are made publicly available~\cite{NRSurCatalog}.

To estimate source properties, we employ the publicly-available Bayesian inference libraries \texttt{parallel-bilby}~\cite{Smith:2019ucc} (version 1.0.1) and \texttt{bilby}~\cite{Ashton:2018jfp,Romero-Shaw:2020owr} (version 1.1.5) together with the \texttt{dynesty}~\cite{speagle2020dynesty} (version 1.0.1) nested sampling algorithm~\cite{Skilling:2006gxv}.
We report some of the most important sampler configuration settings in Table \ref{Tab:sampler}. We have varied these values for several events to check our posteriors are sufficiently converged. We have also varied the number of processes used in our \texttt{parallel-bilby} computation. As documented in App.~\ref{sec:pbilby}, we notice that an important quantity to consider is live-points-per-process.
When this number becomes too small, the computed posteriors are demonstrably
inaccurate.
Through extensive experimentation, we find that using 128 processes
(corresponding to $\approx 16$ live-points-per-process) provides robust
posterior computations. Finally, following the official LVK
analyses~\cite{GWTC2.1_PE, GWTC3_PE,GWTC2.1_Glitch,GWTC3_Glitch}, for each
event, we perform a total of four independent Bayesian inference runs with
different initial random seeds to account for statistical randomness.  We then
combine the posteriors from these four analyses weighted by their individual
evidences to compute the final posteriors.  As an additional consistency check,
for some events we have also compared posteriors
with different initial seeds, finding no significant differences between these runs.

When computing the overlap integral in Eq.(\ref{eq:inner_product}), we follow Ref.~\cite{GWTC2.1_PE, GWTC2.1_Glitch,GWTC3_PE, GWTC3_Glitch} and set the maximum frequency to be $f_{\rm high}=0.875 \times f_{\rm PSD}$, where $f_{\rm PSD}$ is the largest frequency in the publicly available PSD data~\cite{GWTC2.1_PE, GWTC3_PE}.
The factor of $0.875$ is used to minimize roll-off effects caused by the application of a tapering window to the time-domain data as implemented in \texttt{bilby}.
We set the minimum frequency to be $f_{\rm low}=20$Hz for most of the events, with notable exceptions being GW190521\_030229 ($f_{\rm low}=11$Hz for all detectors; following Ref.~\cite{LIGOScientific:2020iuh}) and GW190727\_060333 ($f_{\rm low}=50$Hz for the L1 detector; following Ref.~\cite{LIGOScientific:2021djp} to exclude data that could be corrupted by the presence of a nearby glitch).

We call the \texttt{NRSur7dq4} model through its \texttt{LALSimulation} interface~\cite{lalsim_bug}.
For generating \texttt{NRSur7dq4} waveforms, and as discussed in Sec.~\ref{subsec:frame_choice},
we always use a reference frequency of $f_{\rm ref}=20$Hz when
setting model parameters. Furthermore, we generate the full length of the
surrogate waveform (about 20 orbits) by passing $f_{\rm min}=0$ to the waveform generator function. The
important distinction between $f_{\rm low}$ and $f_{\rm min}$ is that $f_{\rm
low}$ sets the lower limit for the overlap integral~\eqref{eq:inner_product}, while $f_{\rm min}$ sets
the starting frequency of the waveform. By setting $f_{\rm min}=0$ we obtain the
longest possible signal from \texttt{NRSur7dq4}.

\begin{figure*}
    \includegraphics[width=0.97\textwidth]{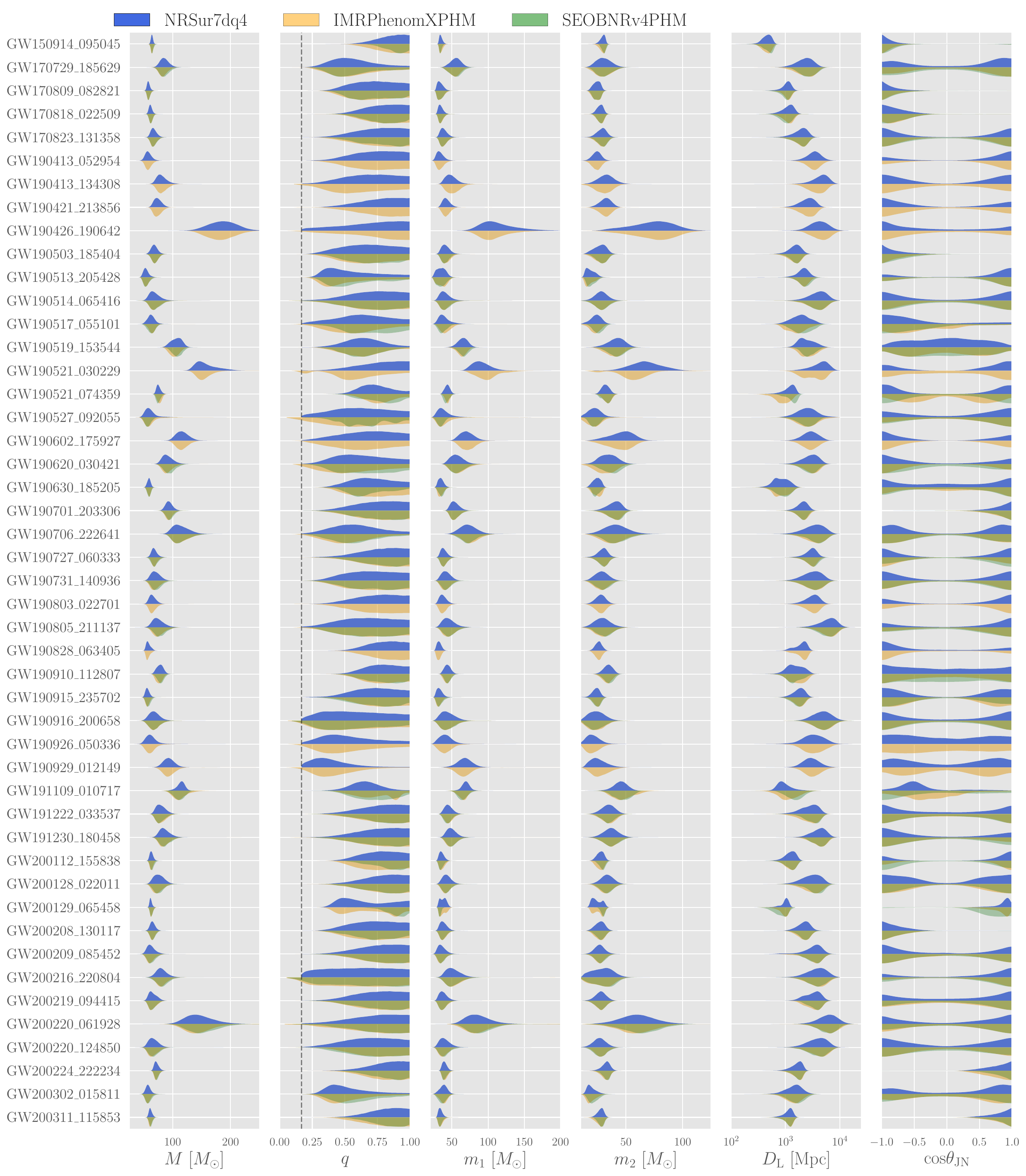}
\caption{\label{fig:violin_mass_dist}
Posteriors for the source-frame total mass $M$, mass ratio $q$, source-frame component
masses $m_1$, $m_2$, luminosity distance $D_{\rm L}$ and the cosine of the
inclination angle $\theta_{\rm JN}$ for all 47 events analyzed with
\texttt{NRSur7dq4} model (blue). For comparison, we also show the public LVK
posteriors~\cite{LIGOScientific:2021usb,LIGOScientific:2021djp, GWTC2.1_PE,
GWTC3_PE} obtained using \texttt{IMRPhenomXPHM} (orange) and
\texttt{SEOBNRv4PHM} (green) models. For some events, the \texttt{SEOBNRv4PHM} posteriors are
missing from the LVK release (see
Tab.~\ref{Tab:special_events}); the \texttt{SEOBNRv4PHM} results are, therefore, also absent in this and all
following figures for that subset of events. The grey dashed line represents the mass ratio cut of
$q=1/6$ used for \texttt{NRSur7dq4}.
Posteriors are reported in the wave frame (see Sec.~\ref{subsec:frame_choice}) at $f_{\rm ref}=20$ Hz. Further
details are given in Sec.~\ref{sec:results}.
}
\end{figure*}

\section{Selection of events}
\label{sec:event_selection}
One important limitation of \texttt{NRSur7dq4} is the restricted duration of
the waveforms provided by the model.  The model is only able to generate
relatively short duration waveforms corresponding to about $20$ orbits before
merger, making it difficult to analyze low-mass systems ($M_{\rm det} \lesssim 60 M_{\odot}$)
when using a lower cut-off frequency of $f_{\rm low}=20$ Hz, the current default value for
most LVK analyses~\cite{LIGOScientific:2021usb, LIGOScientific:2021djp, GWTC2.1_PE, GWTC3_PE}.
Furthermore, the model is valid only for binaries with a mass ratio in the range
$1/6 \le q \le 1$.  This restricts the number of events we can analyze with the
\texttt{NRSur7dq4} model.

We inspect the public LVK posteriors of the binary source
properties obtained using \texttt{IMRPhenomXPHM}~\cite{Pratten:2020ceb} and
\texttt{SEOBNRv4PHM}~\cite{Ossokine:2020kjp} waveform models for all events in
the \texttt{GWTC-3}~\cite{LIGOScientific:2021usb, LIGOScientific:2021djp, GWTC2.1_PE, GWTC3_PE} catalog.
We select the events for which at least 95\% of the samples in the posteriors
(inferred using the \texttt{IMRPhenomXPHM} model), have $M_{\rm det}$ larger than $60 M_{\odot}$
and mass ratio $q$ larger than $0.167$.  This ensures that these events fall within the domain of validity of the
\texttt{NRSur7dq4} model, which we also confirm after computing posteriors
with the \texttt{NRSur7dq4} model.
A total of 47 BBH events that satisfy these criteria will be used in our analysis; these are listed in Figure \ref{fig:violin_mass_dist}.

\section{Binary source parameter inference}
\label{sec:results}
We summarize our parameter inference results using \texttt{NRSur7dq4} and
\texttt{NRSur7dq4Remnant} for the 47 events selected from \texttt{GWTC-3}~\cite{LIGOScientific:2021usb, LIGOScientific:2021djp, GWTC2.1_PE, GWTC3_PE} according to the procedure laid out in Sec.~\ref{sec:event_selection}.
We then compare our results against the public LVK posteriors~\cite{LIGOScientific:2021usb, LIGOScientific:2021djp, GWTC2.1_PE, GWTC3_PE}.
While \texttt{GWTC-2}~\cite{LIGOScientific:2020ibl} also provides analyses of a subset of these events,
including some results obtained using \texttt{NRSur7dq4},  to compare against the largest possible sample with
the most consistent settings, we elide detailed comparisons to this earlier work for simplicity.
For similar reasons, we also omit detailed comparison against other	analyses of
these events~\cite{Venumadhav:2019lyq,Nitz:2018imz,Nitz:2020oeq,Nitz:2021uxj}.

\begin{figure*}
	\includegraphics[width=0.97\textwidth]{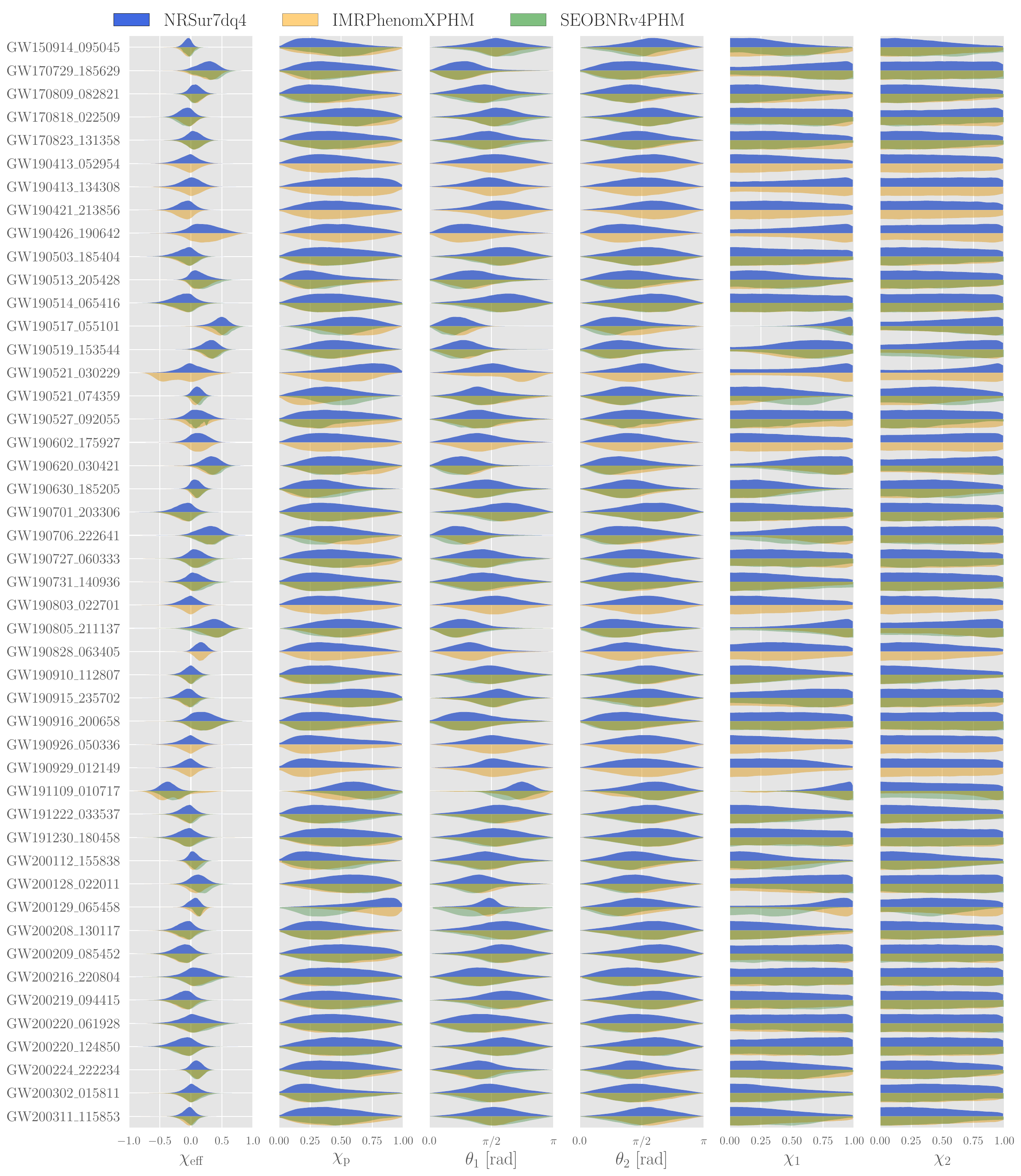}
\caption{\label{fig:violin_spin} 
Posteriors for the spin magnitudes $\chi_1$ and $\chi_2$, spin angles
$\theta_{1}$ and $\theta_{2}$, effective inspiral spin parameter $\chieff$, and
spin precession parameter $\chi_p$ for all 47 events analyzed with
\texttt{NRSur7dq4} model (blue). For comparison, we also show the public LVK
posteriors~\cite{LIGOScientific:2021usb,LIGOScientific:2021djp, GWTC2.1_PE,
GWTC3_PE} obtained using \texttt{IMRPhenomXPHM} in orange and
\texttt{SEOBNRv4PHM} (where available) in green.
Posteriors are reported in the wave frame (see Sec.~\ref{subsec:frame_choice}) at $f_{\rm ref}=20$ Hz.
We provide 3D visualizations of the full spin posteriors at Ref.~\cite{NRSurCatalog}.
Further details are given in Sec.~\ref{sec:results}.
}
\end{figure*}

\begin{figure*}
	\includegraphics[width=1.0\textwidth]{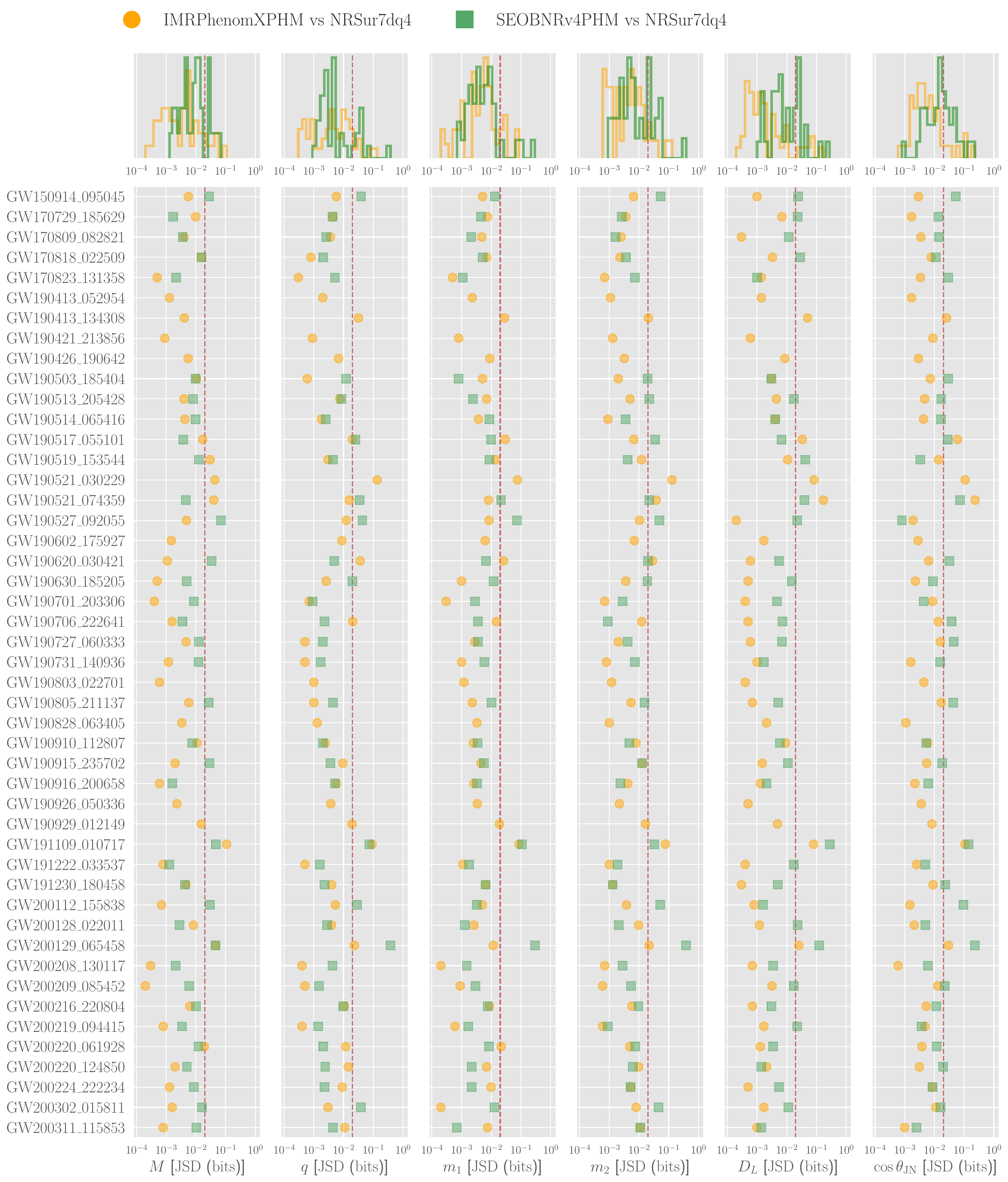}
\caption{\label{fig:jsdivs_intrinsic}
Jensen-Shannon divergence (JSD) values between the one-dimensional marginalized
posteriors of the source-frame total mass $M$, mass ratio $q$, source-frame component masses
$m_1$, $m_2$, luminosity distance $D_{\rm L}$ and the inclination angle
$\theta_{\rm JN}$ obtained using \texttt{NRSur7dq4} and the public LVK
posteriors~\cite{LIGOScientific:2021usb, LIGOScientific:2021djp,
GWTC2.1_PE, GWTC3_PE} obtained using \texttt{IMRPhenomXPHM} (blue circles) and
\texttt{SEOBNRv4PHM} (green squares, where available).  Dashed red lines
correspond to a JS divergence of 0.02, indicating significant differences
between these posteriors. Further details are discussed in Sec.~\ref{sec:results}.
}
\end{figure*}

\begin{figure*}
	\includegraphics[width=1.0\textwidth]{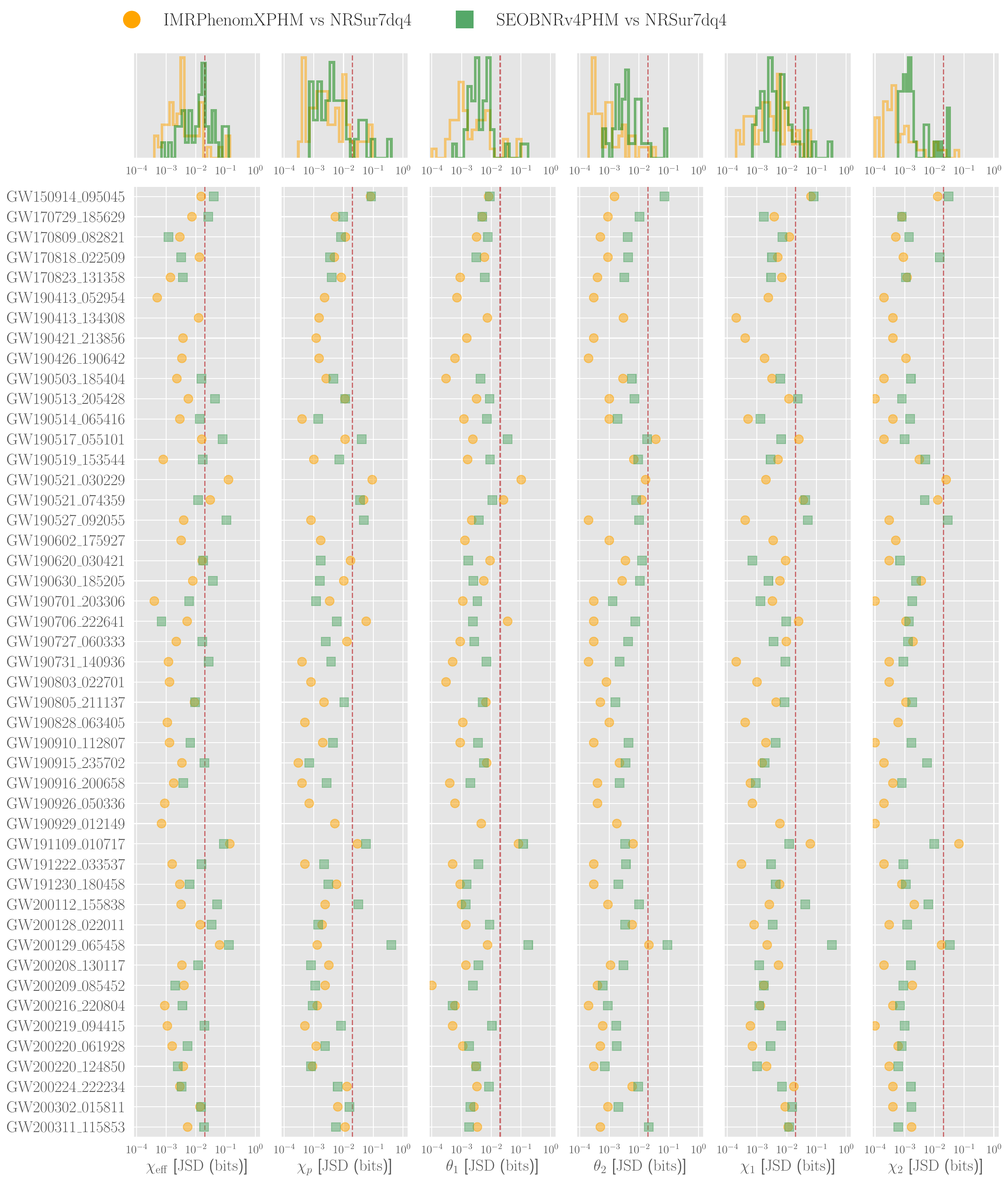}
\caption{\label{fig:jsdivs_spin}
Jensen-Shannon divergence (JSD) values between the one-dimensional marginalized
posteriors of the spin magnitudes $\chi_1$ and $\chi_2$, spin angles
$\theta_{1}$ and $\theta_{2}$, effective inspiral spin parameter $\chieff$ and
spin precession parameter $\chi_p$ obtained using \texttt{NRSur7dq4} and
the public LVK posterior samples~\cite{LIGOScientific:2021usb,
LIGOScientific:2021djp, GWTC2.1_PE, GWTC3_PE} obtained using
\texttt{IMRPhenomXPHM} (blue circles) and \texttt{SEOBNRv4PHM} (green squares,
where available). Dashed red lines correspond to a JS divergence of 0.02,
indicating significant differences between these posteriors. Further details
are discussed in Sec.~\ref{sec:results}.
	}
\end{figure*}

For each event we infer all 15 BBH source parameters discussed in Sec~\ref{subsec:bayes},
we present results~\footnote{A more comprehensive set of interactive figures is readily found on
the NRSurrogate Catalog website~\cite{NRSurCatalog}.}
for a smaller set of summary observables: the source-frame
total mass, mass ratio, dimensionless spin magnitudes $\chi_i$,  dimensionless
spin tilts $\theta_{i}$ between the spin vectors and the orbital angular
momentum, the effective inspiral spin parameter~\cite{Ajith:2009bn, Santamaria:2010yb, Vitale:2016avz},
\begin{equation}\label{eq:chi_eff}
	\chieff = \frac{m_1 \chi_1 \cos\theta_1 + m_2 \chi_2 \cos\theta_2}{m_1 + m_2} \,,
\end{equation}
and the transverse spin precession parameter~\cite{Hannam:2013oca, Schmidt:2014iyl},
\begin{equation}
	\chi_p = \max\left\{\chi_1 \sin \theta_1, \frac{q(4q+3)}{4+3q} \chi_2 \sin \theta_2\right\} \,.
\end{equation}

In Fig.~\ref{fig:violin_mass_dist}, we show the recovery of source-frame masses
as well as the distance and inclination angles; in the accompanying
Fig.~\ref{fig:violin_spin}, we show the corresponding results for the inferred
spin parameters. For comparison, in both figures we also show the results from the public
LVK posteriors~\cite{LIGOScientific:2021usb, LIGOScientific:2021djp, GWTC2.1_PE, GWTC3_PE} obtained using
the \texttt{IMRPhenomXPHM} and \texttt{SEOBNRv4PHM} models\footnote{Note that we
obtain \texttt{IMRPhenomXPHM} and \texttt{SEOBNRv4PHM} from the public data
release associated with the GWTC-3 catalog~\cite{LIGOScientific:2021usb, LIGOScientific:2021djp, GWTC2.1_PE, GWTC3_PE}.}. Note that \texttt{IMRPhenomXPHM}
posteriors are obtained with \texttt{bilby}~\cite{Ashton:2018jfp} while
\texttt{SEOBNRv4PHM} posteriors are obtained with
\texttt{RIFT}~\cite{Lange:2018pyp}. For the events listed in
Table~\ref{Tab:special_events}, \texttt{SEOBNRv4PHM} results are absent from the
public LVK data release~\cite{LIGOScientific:2021usb, LIGOScientific:2021djp, GWTC2.1_PE, GWTC3_PE}
and, therefore, are also absent in our comparisons.

In Figs.~\ref{fig:violin_mass_dist} and \ref{fig:violin_spin}, we find many
events for which noticeable differences exist between posteriors inferred with
\texttt{NRSur7dq4}, \texttt{IMRPhenomXPHM}, and \texttt{SEOBNRv4PHM}.
We use a standard diagnostic -- the Jensen-Shannon (JS)
divergence~\cite{Js61115} --  to quantify the difference between the
one-dimensional marginalized posteriors inferred with \texttt{NRSur7dq4} model
(this work) and the posteriors obtained using \texttt{IMRPhenomXPHM} and
\texttt{SEOBNRv4PHM} models (from the publicly available
LVK posteriors~\cite{LIGOScientific:2021usb, LIGOScientific:2021djp, GWTC2.1_PE, GWTC3_PE}).
Recall the JS divergence (JSD) between two
probability density functions $p(x)$ and $q(x)$,
\begin{eqnarray}
	{\rm JSD}(p,q) = \frac{{\rm KLD}(p||m) + {\rm KLD}(q||m)}{2},
\end{eqnarray}
is defined as a symmetrized extension of the Kullback-Leibler divergence~\cite{KLdivergence},
where $m(x) = 1/2 (p+q)$ is the point-wise mean of $p(x)$ and $q(x)$ and ${\rm KLD}$,
\begin{equation}
	{\rm KLD}(p||q) = \int p(x) \log_2\frac{p(x)}{q(x)} dx,
\end{equation}
is Kullback-Leibler divergence. When using
a base-2 logarithm JS divergence values are given in the units of bits.
A JS divergence value of $0$ bits signifies that the posteriors are identical, while a JS divergence
value of $1$ bit corresponds to posterior distributions that have no statistical overlap at all.
Values smaller than $2\times 10^{-3}$ bits can occur due to stochastic
sampling and indicate statistically indistinguishable posterior samples~\cite{Romero-Shaw:2020owr}.
The threshold for non-negligible bias varies by study, where JSD values of 0.007~\cite{LIGOScientific:2020ibl}
(which for a Gaussian corresponds to a 20\% shift in the mean), 0.02~\cite{Romero-Shaw:2020owr}, 0.05~\cite{LIGOScientific:2016wkq},
and 0.15~\cite{shaik2020impact} have all been used. In this paper, we will consider JSD values above 0.02 bits to
indicate important differences between posteriors recovered from two different BBH waveform models~\cite{Romero-Shaw:2020owr}.

In Figs.~\ref{fig:jsdivs_intrinsic}-\ref{fig:jsdivs_spin}, we show the JS divergence values between posterior distributions for the masses, spin, distance, and inclination angles (i.e., for the parameters shown in Figs.~\ref{fig:violin_mass_dist}-\ref{fig:violin_spin}), for each event.
We find that the JS divergence values between \texttt{NRSur7dq4} and the \texttt{IMRPhenomXPHM} and \texttt{SEOBNRv4PHM} models are mostly less than 0.02 bits suggesting good agreement.
However, for around $\sim 23$\% ($\sim 55$\%) of the analyzed events, JS divergence values between \texttt{NRSur7dq4} and \texttt{IMRPhenomXPHM} (\texttt{SEOBNRv4PHM}) are larger than 0.02 bits for at least one of the parameters shown in Figs.~\ref{fig:jsdivs_intrinsic}-\ref{fig:jsdivs_spin}.
Such differences can arise from waveform systematics, such as the missing physics in \texttt{IMRPhenomXPHM}/\texttt{SEOBNRv4PHM} from not being informed by precessing NR simulations.

Some of the observed differences may stem from the sampler techniques used: \texttt{bilby} was employed for \texttt{IMRPhenomXPHM}, \texttt{RIFT} for \texttt{SEOBNRv4PHM}, and \texttt{parallel-bilby} for \texttt{NRSur7dq4}. Differences between samplers are expected to be most significant when comparing results obtained with \texttt{RIFT} to those from \texttt{parallel-bilby}, as their underlying methods differ substantially. In contrast, \texttt{parallel-bilby} and \texttt{bilby} share the same nested sampling algorithm, leading to greater consistency between their results. In Appendix~\ref{sec:sampler_diagnostics}, we validate the agreement between our \texttt{parallel-bilby} parameter estimation setup and the LVK's \texttt{bilby} setup. Therefore, any discrepancies in inference results between \texttt{IMRPhenomXPHM} and \texttt{NRSur7dq4} are likely due to modeling systematics.

To ensure that sufficient sampling density exists in the publicly available \texttt{SEOBNRv4PHM} posteriors computed by \texttt{RIFT}, we confirmed that the Jensen-Shannon divergence (JSD) values reported in Figs.\ref{fig:jsdivs_intrinsic} and \ref{fig:jsdivs_spin} remain reliable even for events with a limited number of posterior samples. As a test, for each event we downsampled the \texttt{SEOBNRv4PHM} posterior by 80\% and recalculated all JSD values. The resulting JSD values changed by less than 0.01, with most differences being below 0.001. We further checked that the recomputed JSD values are visually indistinguishable from the results already presented in the figures. Thus, we conclude that the reported JSD values are robust and reliable, even for the apparently undersampled \texttt{SEOBNRv4PHM} posteriors. Further investigation, however, is needed to completely identify the specific sources of differences between posteriors computed with \texttt{SEOBNRv4PHM} and \texttt{NRSur7dq4}.

\subsection{Highlighted events}
\label{sec:special_events}

\begin{figure*}
	\includegraphics[width=1.0\textwidth]{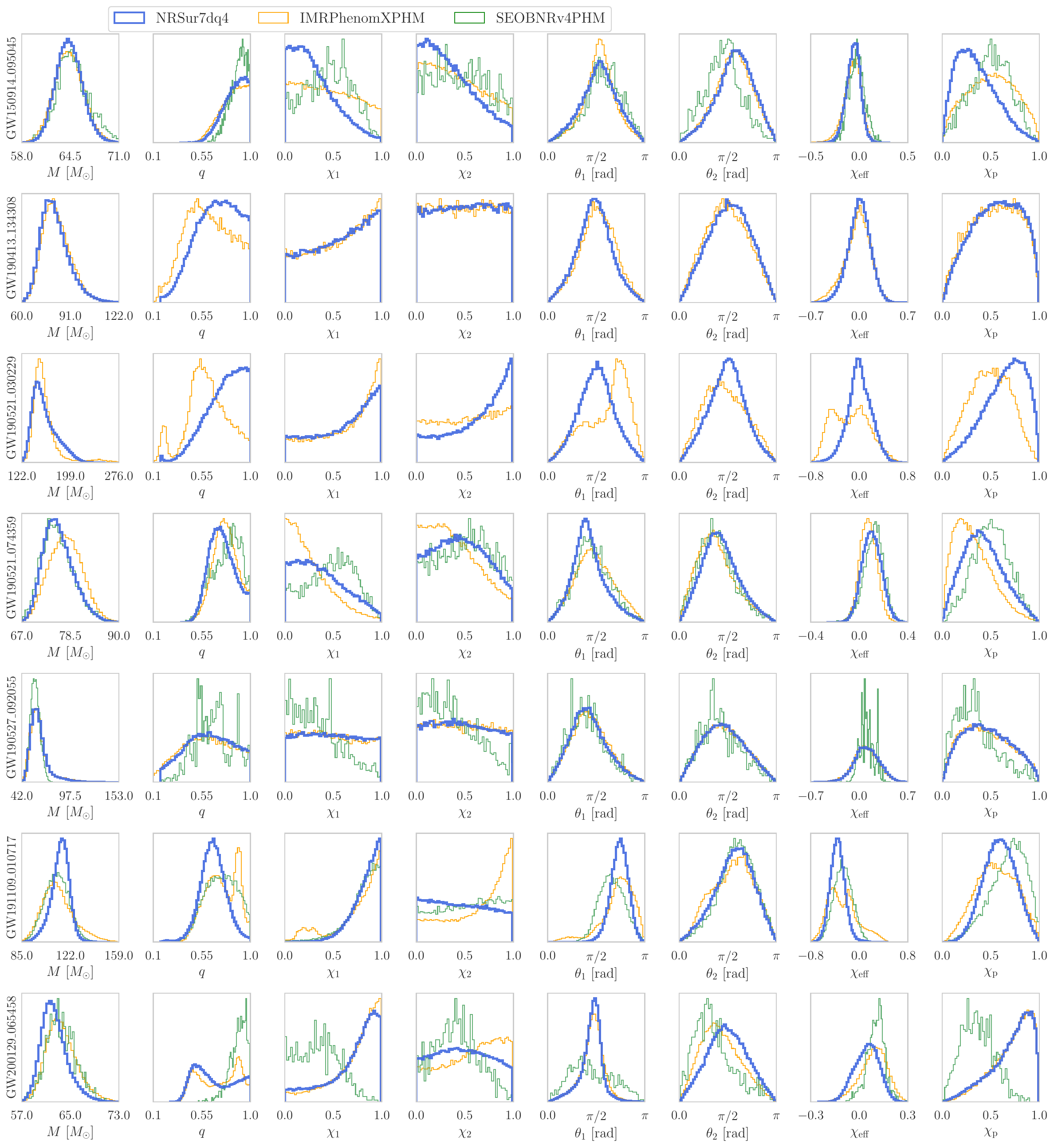}
\caption{\label{fig:special_events}
Posteriors for the source-frame total mass $M$, mass ratio $q$, spin magnitudes
$\chi_1$ and $\chi_2$, spin angles $\theta_{1}$ and $\theta_{2}$, effective
inspiral spin parameter $\chieff$, and spin precession parameter $\chi_p$ for a
list of seven events for which we infer the most significant differences
between results obtained using \texttt{NRSur7dq4} (blue histogram),
\texttt{IMRPhenomXPHM} (orange histogram) and \texttt{SEOBNRv4PHM} (green
histogram, where available). The under-sampled posteriors for
\texttt{SEOBNRv4PHM} is a consequence of an inefficient post-processing
procedure used in the \texttt{RIFT} code whenever calibration
uncertainties are accounted for~\cite{Payne:2019wmy, Payne:2020myg}.
Posteriors are reported in the wave frame (see Sec.~\ref{subsec:frame_choice}) at $f_{\rm ref}=20$
Hz. Further details are given in Sec.~\ref{sec:special_events}.
}
\end{figure*}

Having provided an overview of the comparison of \texttt{NRSur7dq4} posteriors
against \texttt{SEOBNRv4PHM} and \texttt{IMRPhenomXPHM} results in Sec.~\ref{sec:results}, we now
highlight certain events with noticeable differences in their inferred
posteriors.  For these events, Fig.~\ref{fig:special_events} shows the
posteriors for mass and spin parameters, while
Fig.~\ref{fig:special_events_extrinsic} focuses on the distance and inclination
posteriors (with the total mass shown for comparison). Median values of the
inferred source properties are given in Table~~\ref{tab:PEresults} in the Appendix.

We reiterate that, for some of the events, \texttt{SEOBNRv4PHM} posteriors
(obtained using the \texttt{RIFT} parameter estimation code) are missing in public
LVK posteriors, as listed in Table~\ref{Tab:special_events}.
Furthermore, for several events where \texttt{SEOBNRv4PHM} posteriors are
available due to an inefficient post-processing step incorporating the effects of calibration
uncertainties in the GW data, the number of
samples can be significantly reduced~\cite{Payne:2019wmy, Payne:2020myg}. This results
in the under-sampled posteriors for \texttt{SEOBNRv4PHM} visible for some events shown in
Fig.~\ref{fig:special_events} and Fig.~\ref{fig:special_events_extrinsic}.

\subsubsection{GW150914\_095045: first GW observation}
\label{subsubsec:GW15}
GW150914\_095045 was the first direct GW observation and, with an SNR of
$\sim26$~\cite{LIGOScientific:2016wkq}, remains one of the loudest events detected.  This event has
been extensively studied using a variety of waveform models over the past
several years
~\cite{LIGOScientific:2016wkq,Kumar:2018hml,Mateu-Lucena:2021siq,LIGOScientific:2016ebw}.  We find
that while most of the marginalized posterior distributions match across
waveform models, the distributions describing the spin parameters do not, as
can be seen in the top row of Fig.~\ref{fig:special_events}.

In Fig.~\ref{fig:special_events} (top row), we show that \texttt{NRSur7dq4} favors
smaller values for the spin magnitudes $\chi_1$ and $\chi_2$ compared to
\texttt{IMRPhenomXPHM} and \texttt{SEOBNRv4PHM}. We also observe significant
differences in the $\chi_p$ posteriors (Fig.~\ref{fig:GW15}), where \texttt{NRSur7dq4}
favors smaller $\chi_p$ values than \texttt{IMRPhenomXPHM} and \texttt{SEOBNRv4PHM}.
Fig.~\ref{fig:GW15_JSD} summarizes the JS divergence values for many of the most interesting
parameters.
Interestingly, as highlighted in Fig.~\ref{fig:GW15},
the \texttt{NRSur7dq4} posterior for $\chi_p$ matches more closely with earlier estimates from
\texttt{GWTC-1}~\cite{LIGOScientific:2016wkq} using the \texttt{IMRPhenomPv2}~\cite{Hannam:2013oca,
husa2016frequency, khan2016frequency} (which does not include higher order modes) and
\texttt{SEOBNRv3}~\cite{Pan:2013rra, Taracchini:2013rva, Babak:2016tgq} (which
do not include $\ell>2$ modes) models.
While we cannot offer a simple explanation for this tension between
models, especially as the different models also rely on different posterior
samplers\footnote{The parameter estimation analysis reported in \texttt{GWTC-1}~\cite{LIGOScientific:2016wkq}
primarily used the \texttt{LALInference}~\cite{Veitch:2014wba} package.},
this might point to systematic differences arising from the treatment of
precession or whenever subdominant harmonic modes play a role.

\subsubsection{GW190413\_134308}
For GW190413\_13430 we find noticeable differences in the mass ratio (second row of
Fig.~\ref{fig:special_events}), luminosity distance and inclination (first
column of Fig.~\ref{fig:special_events_extrinsic}).
The JS divergence values between the one-dimensional marginalized
posteriors recovered with \texttt{NRSur7dq4} and \texttt{IMRPhenomXPHM}
are 0.031 bits (for $q$), 0.05 bits (for $D_{\rm L}$), and 0.026 bits (for $\theta_{\rm JN}$).
We cannot compare with \texttt{SEOBNRv4PHM} because the
corresponding posteriors are missing from the most recent LVK release.

\subsubsection{GW190521\_030229}
\begin{figure*}
\includegraphics[width=\textwidth]{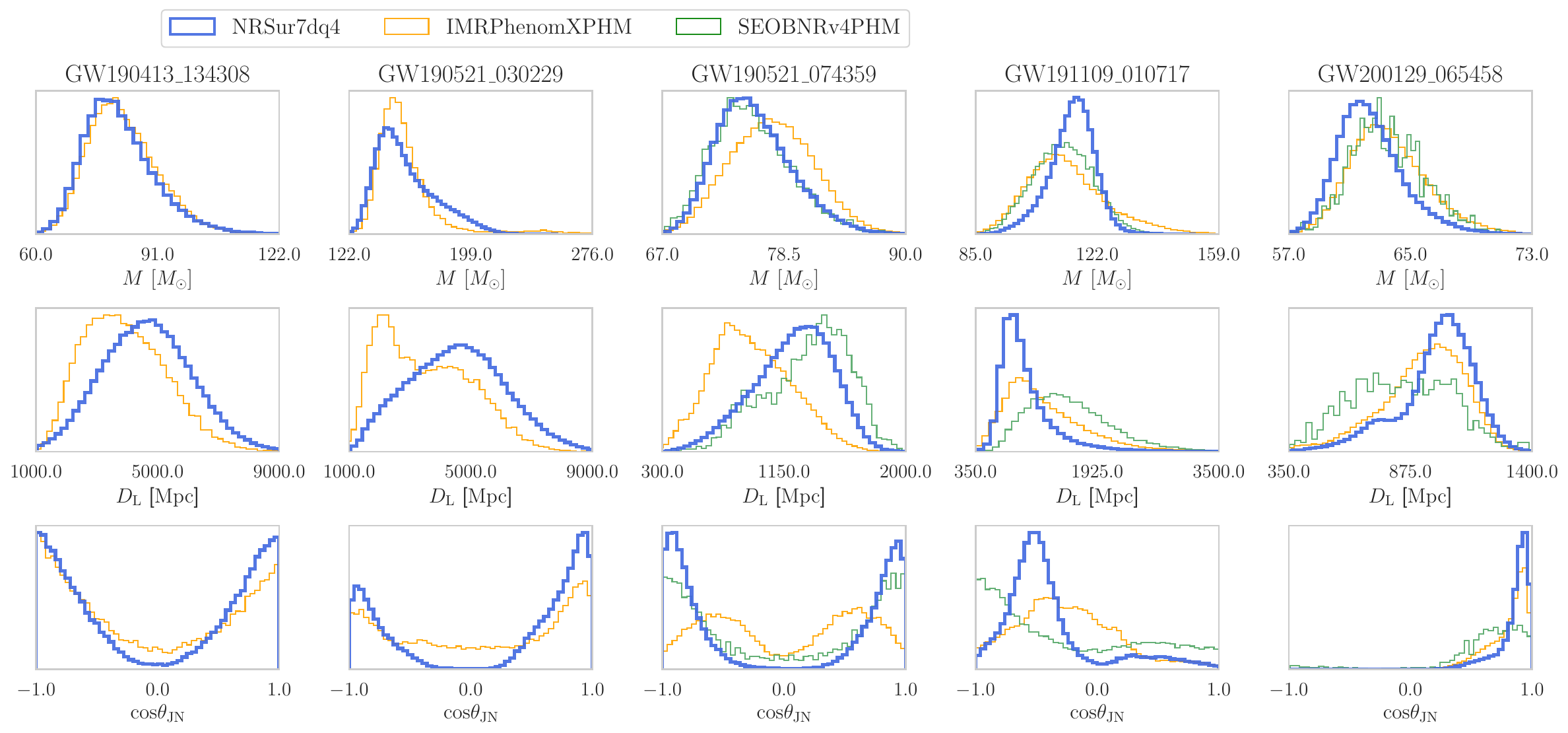}
\caption{\label{fig:special_events_extrinsic}
Posteriors for the source-frame total mass $M$, luminosity distance $D_{\rm L}$,
and cosine of the inclination angle $\theta_{\rm JN}$ for a list of five events
for which we see the largest differences between results obtained using
\texttt{NRSur7dq4} (blue histogram), \texttt{IMRPhenomXPHM} (orange histogram)
and \texttt{SEOBNRv4PHM} (green histogram, where available). Further details
are given in Sec.~\ref{sec:special_events}.
}
\end{figure*}

GW190521\_030229~\cite{LIGOScientific:2020iuh,2020ApJ_900L_13A} has unusually
high component masses, $85^{+21}_{-14}$ $M_\odot$ and $66^{+17}_{-18}$ $M_\odot$,
compared to other events.
Consequently, the observable signal contains only a few pre-merger cycles.
Furthermore, this
event shows tentative signs of precession~\cite{LIGOScientific:2020iuh,
2020ApJ_900L_13A, Biscoveanu:2021nvg, Estelles:2021jnz, Nitz:2020mga}. As a
result, systematic differences between waveform models in the
treatment of both precession and the merger-ringdown can play a more significant
role for signals like this one. For example, Refs.~\cite{LIGOScientific:2020ufj,
Estelles:2021jnz, Nitz:2020mga} already pointed out several systematic
differences between \texttt{NRSur7dq4} and other models for this event, and
the same can be seen in Figs.~\ref{fig:special_events} (third row) and
\ref{fig:special_events_extrinsic} (second column).
In particular, \texttt{IMRPhenomXPHM} prefers more unequal mass ratios with a
clear bimodal mass ratio distribution while the \texttt{NRSur7dq4} posterior shows visibly different features~\cite{Estelles:2021jnz, Nitz:2020mga} (third row of
Fig.~\ref{fig:special_events}).
Similarly, the posteriors for several spin
parameters, including $\chi_1$, $\chi_2$, $\theta_1$, $\theta_2$, $\chieff$ and
$\chi_p$ in Fig.~\ref{fig:special_events} indicate significant systematic
differences between \texttt{NRSur7dq4} and \texttt{IMRPhenomXPHM}.
The JS divergence values between the one-dimensional marginalized
posteriors recovered with \texttt{NRSur7dq4} and \texttt{IMRPhenomXPHM}
are 0.043 bits (for $M$), 0.136 bits (for $q$), 0.002 bits (for $\chi_1$),
0.025 bits (for $\chi_2$), 0.102 bits (for $\theta_1$), 0.017 bits (for $\theta_2$),
0.126 bits (for $\chieff$) and 0.092 bits (for $\chi_p$).
Among the extrinsic parameters, we find significant differences
between the posteriors for $D_{\rm L}$ and $\theta_{\rm JN}$ (second column of
Fig.~\ref{fig:special_events_extrinsic}), with JS divergence values of 0.083 bits
and 0.107 bits, respectively. While we cannot compare with \texttt{SEOBNRv4PHM} as the
corresponding posteriors are missing from the most recent LVK release
~\cite{LIGOScientific:2021usb, LIGOScientific:2021djp,GWTC2.1_PE, GWTC3_PE}, such
a comparison is shown in Ref.~\cite{LIGOScientific:2020ufj},
where significant differences were also noted between \texttt{SEOBNRv4PHM} and
\texttt{NRSur7dq4}.

\subsubsection{GW190521\_074359}
Figures \ref{fig:special_events} (fourth row) and
\ref{fig:special_events_extrinsic} (third column) show noticeable differences
between \texttt{NRSur7dq4} and the other models
for several parameters for this event. In particular, \texttt{NRSur7dq4} and
\texttt{SEOBNRv4PHM} yield similar posteriors for the source-frame total mass
$M$ whereas \texttt{IMRPhenomXPHM} favors slightly larger values for $M$.
\texttt{NRSur7dq4} also favors a slightly more asymmetric binary i.e., smaller values
for $q$ compared to the other two models. Furthermore, we find noticeable
differences in the $\chi_1$, $\chi_2$, and $\chi_p$ posteriors between
three models, with the \texttt{NRSur7dq4} falling broadly in between
\texttt{IMRPhenomXPHM} and \texttt{SEOBNRv4PHM}.  The JS divergence values
between \texttt{NRSur7dq4} and \texttt{IMRPhenomXPHM} (\texttt{SEOBNRv4PHM})
posteriors for $M$, $q$, $\chi_1$, $\chi_2$ and $\chi_p$ are 0.04 bits, 0.016 bits, 0.036 bits, 0.013 bits, and 0.047 bits (0.004 bits, 0.035 bits, 0.042 bits, 0.005 bits, and 0.036 bits), respectively.  We further find noticeable
differences between \texttt{NRSur7dq4} and \texttt{IMRPhenomXPHM}
(\texttt{SEOBNRv4PHM} [to a lesser extent]) posteriors for $D_{\rm L}$ and  $\theta_{\rm JN}$ (third
column of Fig.~\ref{fig:special_events_extrinsic}), with JS divergence values
of 0.172 bits and 0.234 bits (0.04 bits and 0.07 bits), respectively.

\subsubsection{GW190527\_092055}
GW190527\_092055 presents an interesting case as we find that, for almost all
parameters shown in Fig.\ref{fig:special_events} (fifth row),
\texttt{NRSur7dq4} and \texttt{IMRPhenomXPHM} yield consistent posteriors while
\texttt{SEOBNRv4PHM} posteriors show noticeable differences.  For example, both
\texttt{NRSur7dq4} and \texttt{IMRPhenomXPHM} posteriors for spin magnitudes
$\chi_1$ and $\chi_2$ are uninformative whereas \texttt{SEOBNRv4PHM} posteriors
show strong support for smaller values of $\chi_1$ and $\chi_2$ (fifth row of
Fig.~\ref{fig:special_events}).  The JS divergence values between
\texttt{NRSur7dq4} and \texttt{IMRPhenomXPHM} (\texttt{SEOBNRv4PHM}) posteriors
for $\chi_1$ and $\chi_2$ are 0.0004 bits and 0.0003 bits (0.05 bits and 0.03 bits) respectively.
Posteriors for the spin angles $\theta_{1}$ and $\theta_2$, however, match for
all models (fifth row of Fig.~\ref{fig:special_events}). However, the
\texttt{SEOBNRv4PHM} posteriors (obtained using the \text{RIFT} code) for this
event appear to be particularly under-sampled, making it difficult to
disentangle model systematics from sampler systematics. We note that Ref.~\cite{garron2023waveform} reanalyzed this event using \texttt{parallel-bilby} with \texttt{NRSur7dq4} and obtained results consistent with ours. However, Ref.~\cite{Dax:2022pxd} employed a machine-learning
based parameter estimation code~\cite{Dax:2021tsq} with importance sampling to reanalyze this event with \texttt{SEOBNRv4PHM}, finding better agreement between \texttt{SEOBNRv4PHM} and \texttt{IMRPhenomXPHM}.

\subsubsection{GW191109\_010717}
\label{sec:GW191109}
GW191109\_010717 is another event that shows interesting
and astrophysically important differences between
posteriors obtained using the \texttt{NRSur7dq4}, \texttt{IMRPhenomXPHM}, and
\texttt{SEOBNRv4PHM} models.  For mass ratio $q$ and spin
magnitude $\chi_1$, \texttt{IMRPhenomXPHM} posteriors show bimodalities whereas
\texttt{NRSur7dq4} and \texttt{SEOBNRv4PHM} posteriors do not (sixth row of Fig.~\ref{fig:special_events}).  Another noteworthy
observation is that \texttt{IMRPhenomXPHM} favors larger values for the
secondary spin magnitude $\chi_2$ (sixth row of Fig.~\ref{fig:special_events}),
whereas both \texttt{NRSur7dq4} and \texttt{SEOBNRv4PHM} present posteriors
for $\chi_2$ that are effectively uninformative.
Furthermore, \texttt{NRSur7dq4} shows a stronger preference for negative $\chieff$ at 99.3\%
credible level, compared to 95.9\% for
\texttt{SEOBNRv4PHM}~\footnote{Interestingly, Ref.~\cite{Ramos-Buades:2023ehm}
found that the newer \texttt{SEOBNRv5PHM} model shows a stronger preference for
negative $\chieff$ for GW191109\_010717, in agreement with \texttt{NRSur7dq4}.}
and 85.3\% for \texttt{IMRPhenomXPHM}. For \texttt{IMRPhenomXPHM}, we also see
a bimodality in $\chieff$, which likely results from the bimodality in $q$, due
to the correlation between these two parameters. Furthermore, we note that the
spin angle $\theta_{1}$ is well measured for this event but noticeably
different across waveform models. Finally, the $\chi_p$ posteriors are also
noticeably different across the models. The JS divergence values between
\texttt{NRSur7dq4} and \texttt{IMRPhenomXPHM} (\texttt{SEOBNRv4PHM}) posteriors
for $q$, $\chi_1$, $\chi_2$, $\chieff$, $\chi_p$ and $\theta_{1}$ are 0.091 bits,
0.062 bits, 0.067 bits, 0.139 bits, 0.029 bits and 0.082 bits (0.072 bits, 0.012 bits,
0.009 bits, 0.086 bits, 0.056 bits and 0.117 bits) respectively.

\begin{figure*}[htp]
	\subfloat[]{
		\includegraphics[width=\columnwidth]{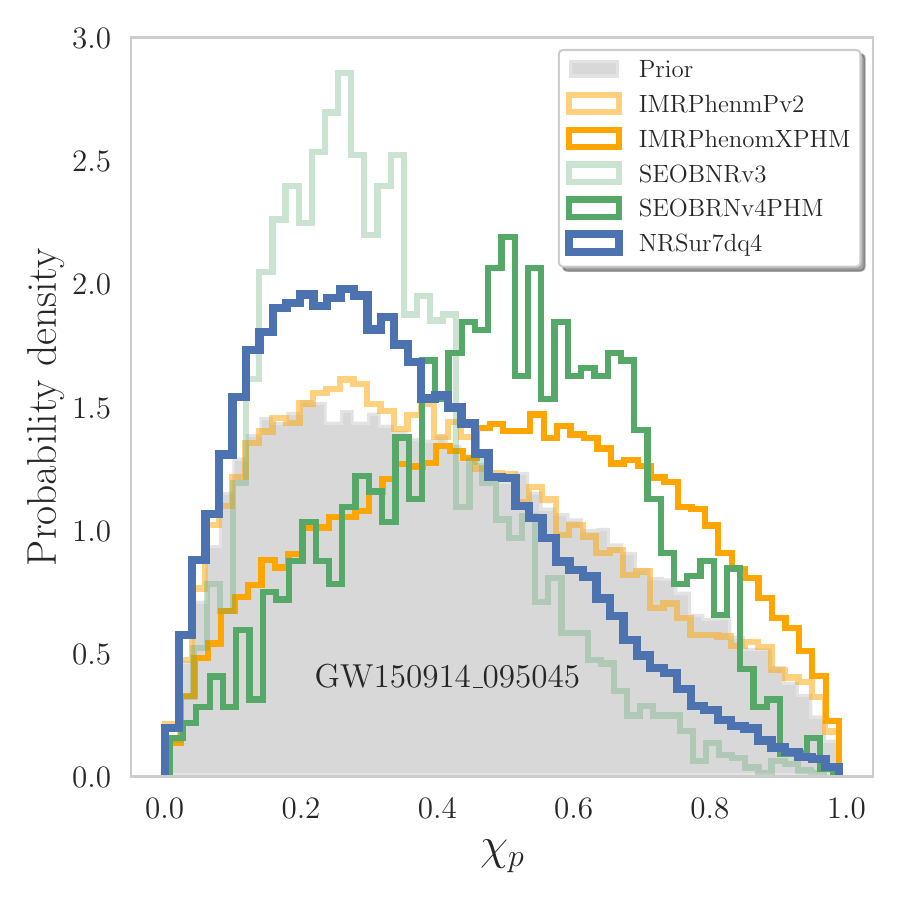}
		\label{fig:GW15}
	}
	\subfloat[]{
		\includegraphics[width=\columnwidth]{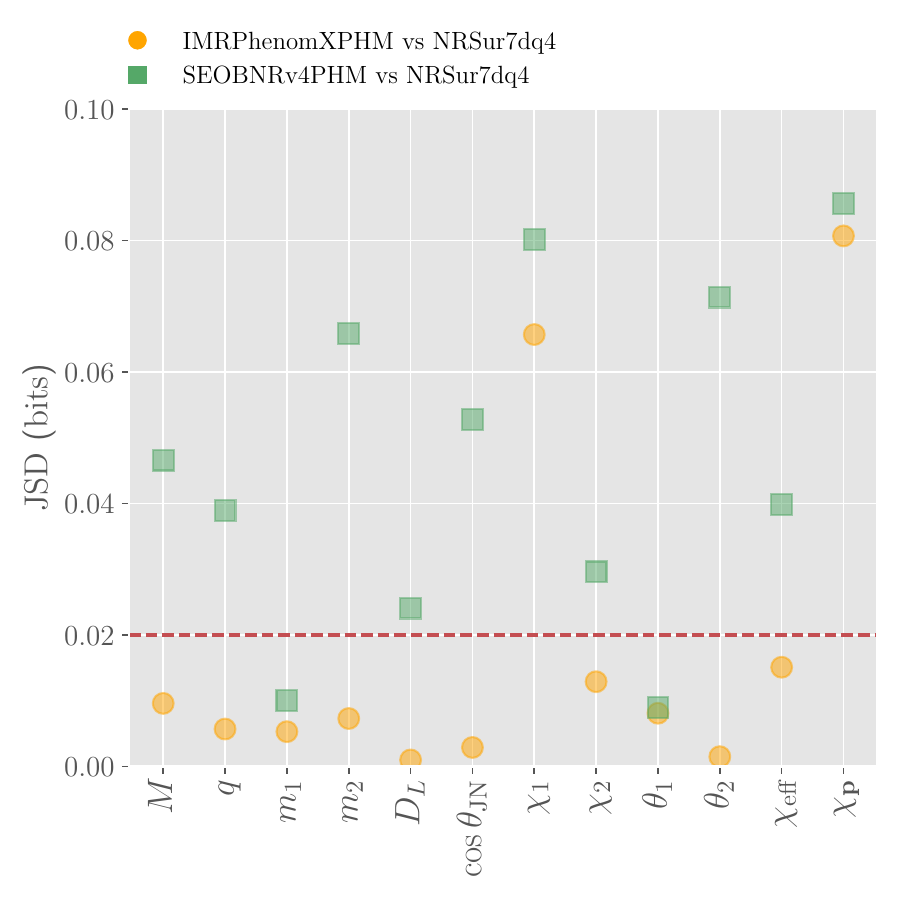}
		\label{fig:GW15_JSD}
	}
	\caption{\label{fig:GW15_All}
	(a) Posteriors and priors for the spin precession parameter $\chi_p$ (computed in the
		wave frame at $f_{\rm ref}=20$ Hz) of GW150914\_095045 obtained using
		\texttt{NRSur7dq4} (blue histogram), \texttt{IMRPhenomXPHM} (orange histogram)
		and \texttt{SEOBNRv4PHM} (green histogram). For comparison, we also show
		posteriors obtained in earlier LVK studies (\texttt{GWTC-1}) using
		\texttt{IMRPhenomPv2} (light orange) and  \texttt{SEOBNRv3} (light green).
		Somewhat surprisingly, the \texttt{NRSur7dq4} posterior agrees more closely with
		older \texttt{IMRPhenomPv2} and \texttt{SEOBNRv3} models that do not include $\ell >2$
		subdominant modes.
	(b) Jensen-Shannon divergence (JSD) values between the one-dimensional marginalized
	posteriors for a set of parameters (shown in Fig.\ref{fig:jsdivs_intrinsic} and Fig.\ref{fig:jsdivs_spin}) obtained using \texttt{NRSur7dq4} and the public LVK posterior samples~\cite{LIGOScientific:2021usb,
		LIGOScientific:2021djp, GWTC2.1_PE, GWTC3_PE} obtained using
	\texttt{IMRPhenomXPHM} (blue circles) and \texttt{SEOBNRv4PHM} (green squares). Dashed red lines correspond to a JS divergence of 0.02,
	indicating significant differences between these posteriors.
	Further details are given in Sec.~\ref{subsubsec:GW15}.
	}
\end{figure*}

Among the extrinsic parameters, \texttt{NRSur7dq4} provides more tightly constrained
posteriors for the luminosity distance $D_{\rm L}$ and inclination $\theta_{\rm
JN}$ (fourth column of Fig.~\ref{fig:special_events_extrinsic}).  The JS
divergence values between \texttt{NRSur7dq4} and \texttt{IMRPhenomXPHM}
(\texttt{SEOBNRv4PHM}) posteriors for $D_{\rm L}$ and $\theta_{\rm JN}$ are
0.081 bits and 0.107 bits (0.278 bits and 0.138 bits) respectively.

\begin{figure*}[htp]
	\subfloat[]{
		\includegraphics[width=0.95\columnwidth]{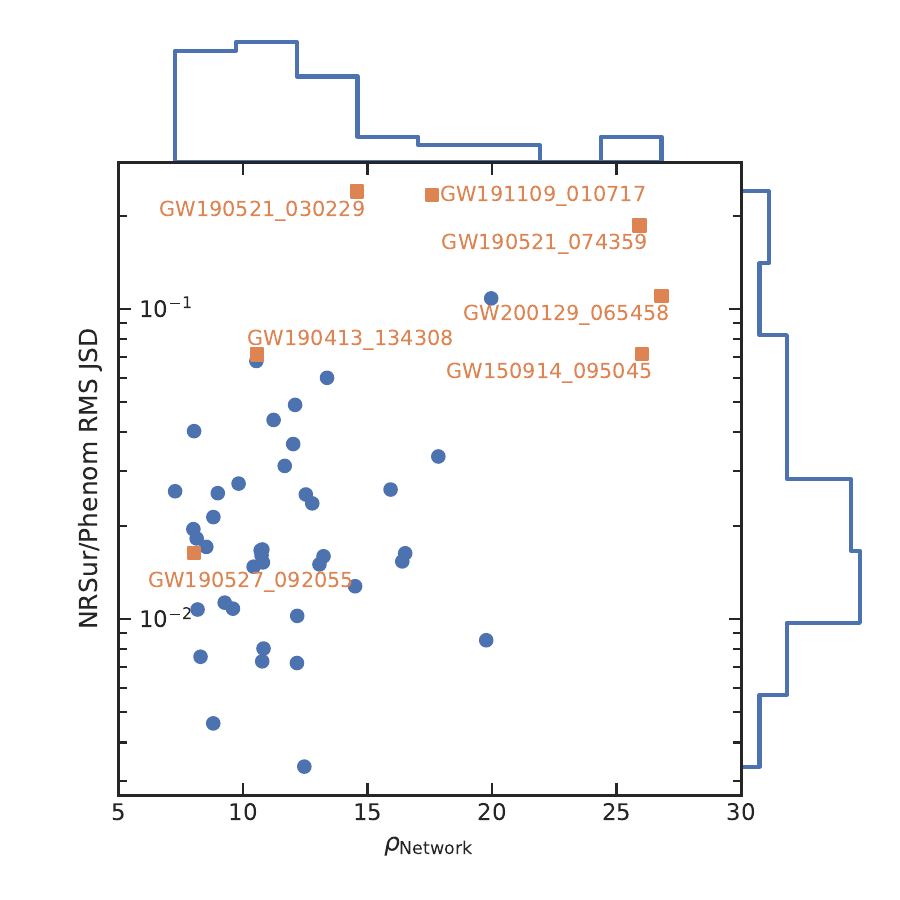}
		\label{fig:JSD_vs_SNR_phenom}
	}
	\subfloat[]{
		\includegraphics[width=0.95\columnwidth]{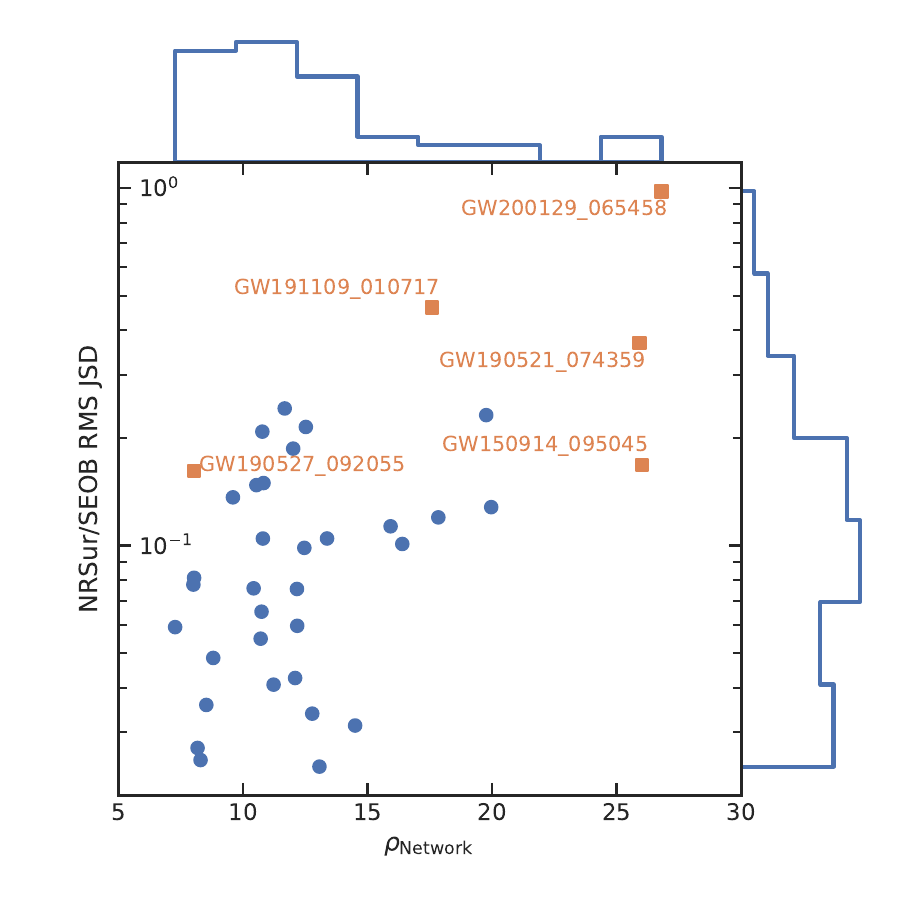}
		\label{fig:JSD_vs_SNR_eob}
	}
	\caption{\label{fig:JSD_vs_SNR} (a) Jensen-Shannon (JS) divergence (reported as the root mean square of the ten JSD values computed from the ten parameters shown in Figures~\ref{fig:violin_mass_dist} and \ref{fig:violin_spin}) between posteriors inferred using the \texttt{NRSur7dq4} and \texttt{IMRPhenomXPHM} models as a function of network matched-filter SNR. As expected, higher-SNR events generally exhibit larger JS divergence, indicating greater sensitivity to waveform model differences. Events highlighted in orange correspond to those discussed in Sec.~\ref{sec:special_events}. While most high-SNR events follow the expected trend, some deviate, suggesting regions of the binary black hole parameter space where these two models show less systematic disagreement.
	(b) Same as Fig.~\ref{fig:JSD_vs_SNR_phenom}, but comparing posteriors obtained with the \texttt{NRSur7dq4} and \texttt{SEOBNRv4PHM} models. The general trend remains, with larger JS divergence for higher-SNR events. Notably, however, GW190527\_092055 (SNR $\approx$ 8) stands out as an outlier, exhibiting a relatively large JS divergence. This event's posteriors, computed using the \texttt{RIFT} code, may be dominated by sampler effects.
	}
\end{figure*}

The strong preference for $\chieff<0$ can have important astrophysical implications,
as negative $\chieff$ is expected to be more common in dynamically formed
binaries than those formed through isolated evolution~\cite{Mandel:2009nx, Rodriguez:2016vmx}.
We note, however, that de-glitched strain data is used for this event.
Previous work on GW200129\_065458 has shown
potentially subtle issues can arise when using de-glitched strain data~\cite{Payne:2022spz}.
We refer to the dedicated glitch subtraction study presented for this and several other events
in Ref.~\cite{Davis:2022ird} and Ref.~\cite{LIGOScientific:2021sio}.
Importantly, Ref.~\cite{LIGOScientific:2021sio} found (see their App.~A) that transient non-gaussian noise or glitches affecting the data
around the time of this event led to false violations of GR in the tests conducted in that work. Such effects could also impact the inference of $\chieff$ (see e.g. App. B of Ref.~\cite{Davis:2022ird}).

\subsubsection{GW200129\_065458}
Next, we look at GW200129\_065458 - another event where glitch subtraction was
necessary~\cite{GWTC3_Glitch, Payne:2022spz}, thereby complicating a straightforward interpretation of the
signal. This event has many interesting properties: it has
a network matched-filter SNR of 26.8~\cite{LIGOScientific:2021djp} making it the
loudest detected BBH signal, it has observable spin-induced orbital precession~\cite{Hannam:2021pit},
and the post-merger remnant BH has a large recoil velocity~\cite{Varma:2022pld}.

Comparing the posteriors for
\texttt{NRSur7dq4}, \texttt{SEOBNRv4PHM}, and \texttt{IMRPhenomXPHM}, we find
noticeable differences for several parameters in Fig.~\ref{fig:special_events}
(seventh row) and Fig.~\ref{fig:special_events_extrinsic} (fifth column).
For example, \texttt{NRSur7dq4} and \texttt{IMRPhenomXPHM} posteriors exhibit
varying degrees of bimodality in mass ratio $q$ while \texttt{SEOBNRv4PHM}
posteriors are uni-modal.  Furthermore, \texttt{NRSur7dq4} and
\texttt{IMRPhenomXPHM} favor smaller (more unequal) values
of $q$ than \texttt{SEOBNRv4PHM}.  We also find significant differences in
$\chi_p$ estimates between \texttt{NRSur7dq4} and \texttt{SEOBNRv4PHM} (seventh
row of Fig.~\ref{fig:special_events}) while \texttt{IMRPhenomXPHM} posteriors
are consistent with \texttt{NRSur7dq4}.  We note that our \texttt{NRSur7dq4}
posteriors for $\chi_p$ match the results obtained in
Ref.~\cite{Hannam:2021pit}.  Similarly, for $\theta_1$ and $\theta_{2}$,
\texttt{NRSur7dq4} and \texttt{SEOBNRv4PHM} posteriors show noticeable
differences whereas \texttt{IMRPhenomXPHM} broadly agrees with
\texttt{NRSur7dq4}. The JS divergence values between \texttt{NRSur7dq4} and
\texttt{IMRPhenomXPHM} (\texttt{SEOBNRv4PHM}) posteriors for $q$, $\chi_1$,
$\chi_2$, $\chieff$, $\chi_p$, $\theta_1$ and $\theta_2$ are 0.023 bits, 0.002 bits,
0.017 bits, 0.063 bits, 0.001 bits, 0.008 bits and 0.008 bits (0.378 bits,
0.331 bits, 0.033 bits, 0.129 bits, 0.405 bits, 0.174 bits and 0.09 bits)
respectively.

Finally, we also find significant differences between \texttt{NRSur7dq4} and
\texttt{IMRPhenomXPHM} (\texttt{SEOBNRv4PHM}) posteriors for $D_{\rm L}$ and
$\theta_{\rm JN}$ (fifth column of Fig.~\ref{fig:special_events_extrinsic}),
with JS divergence values of 0.025 bits and 0.029 bits (0.124 bits and 0.228 bits),
respectively. We note that \texttt{NRSur7dq4} favors larger values for $D_{\rm L}$ than both the
\texttt{IMRPhenomXPHM} and \texttt{SEOBNRv4PHM} models.

\begin{figure}[htp]
	\subfloat[GW190727\_060333]{
		\includegraphics[width=\columnwidth]{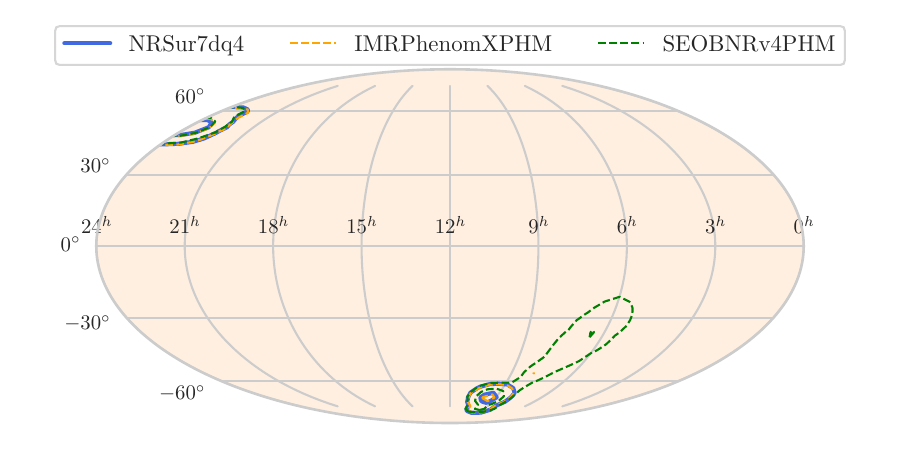}
		\label{fig:skymap_GW190727_060333}
	}\\
	\subfloat[GW200220\_061928]{
		\includegraphics[width=\columnwidth]{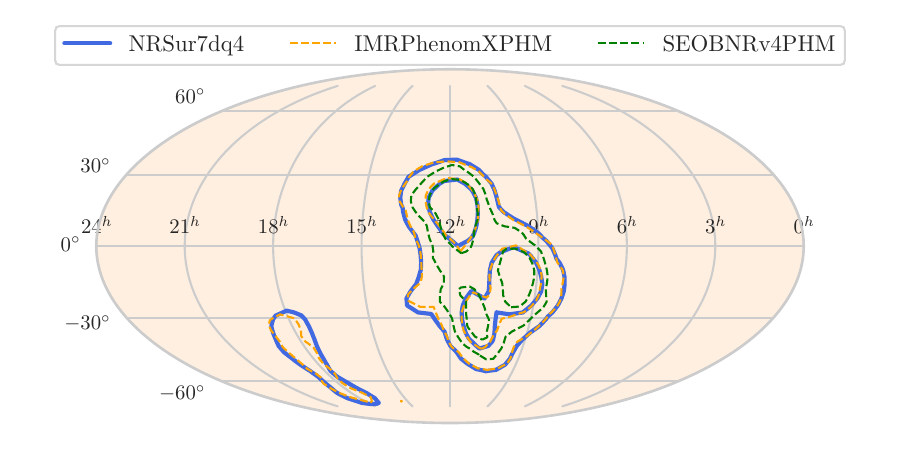}
		\label{fig:skymap_GW200220_061928}
	}\\
	\subfloat[GW200129\_065458]{
		\includegraphics[width=\columnwidth]{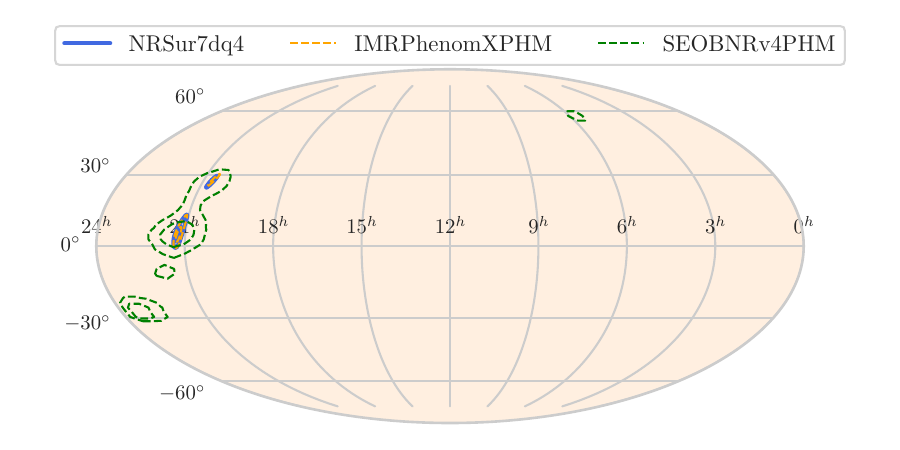}
		\label{fig:skymap_GW200129_065458}
	}
\caption{
\label{fig:skymap}
Skymaps for three events where noticeable differences are observed between
\texttt{NRSur7dq4} (blue), \texttt{IMRPhenomXPHM} (orange), and
\texttt{SEOBNRv4PHM} (green). We note that the discrepancies between \texttt{SEOBNRv4PHM} and the other models may arise from a different sampler used for the analysis with \texttt{SEOBNRv4PHM}. Further details are given in Sec~\ref{sec:sky}.
}
\end{figure}

\begin{figure*}
	\includegraphics[width=1.0\textwidth]{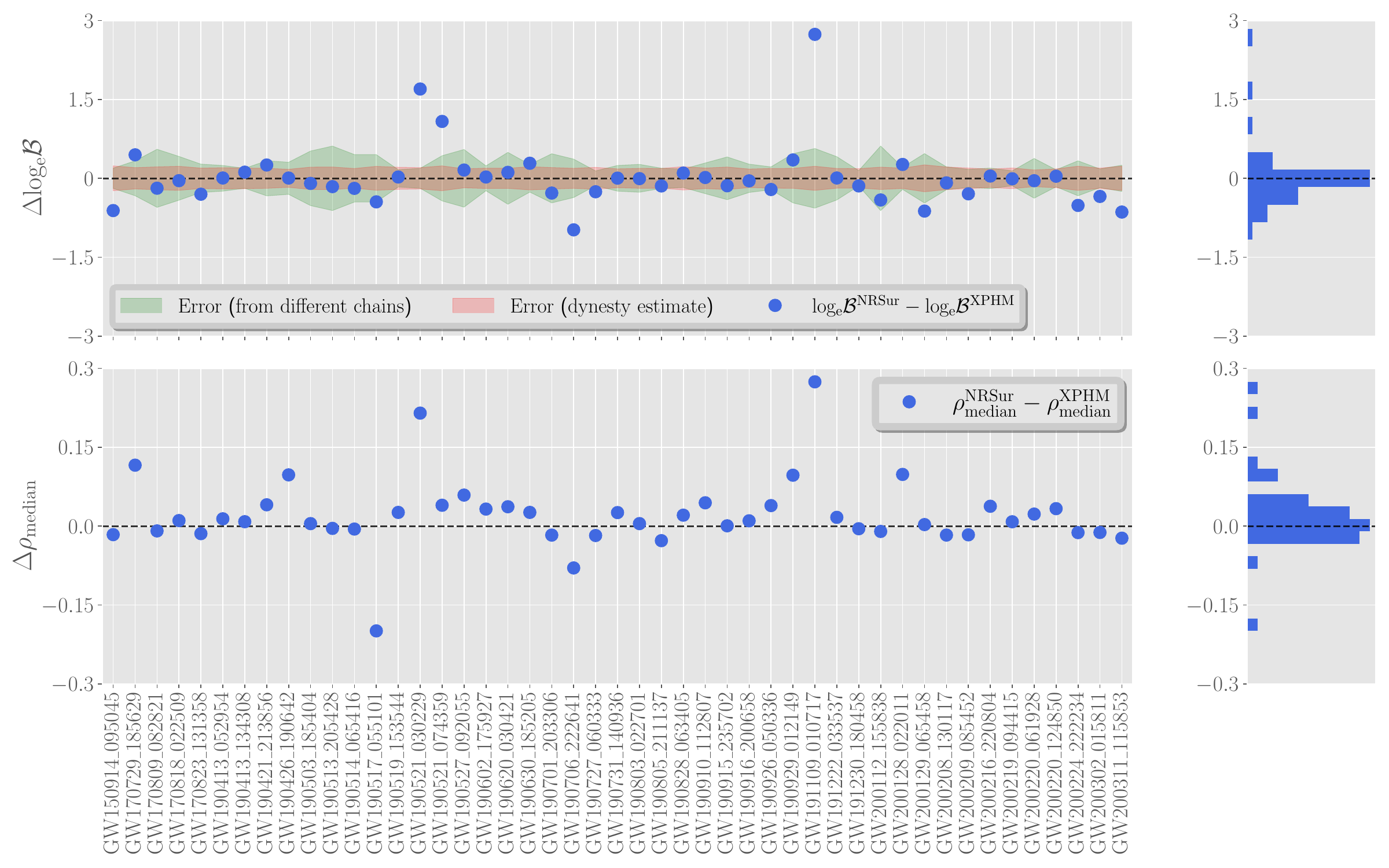}
\caption{\label{fig:SNR}
Differences in the recovered log Bayes factors $\Delta {\rm log_e} \mathcal{B}$
(\textit{upper panel}) and the median SNR $\Delta \rho_{\rm median}$
(\textit{lower panel}) between \texttt{NRSur7dq4} (referred to as `NRSur') and
\texttt{IMRPhenomXPHM} models (referred to as `XPHM') for all 47 events.
These events were analyzed using the prior described in Sec.~\ref{subsec:prior}, which is restricted to a smaller portion of the parameter space as compared to the official LVK prior but matches the one used for our
\texttt{NRSur7dq4} analysis. We also show
the estimated uncertainty in the calculation of the log evidence
from \texttt{parallel-bilby} as a shaded red patch (\textit{upper panel}).
On the other hand, the shaded green patch (\textit{upper panel}) shows the
uncertainties estimated from four independent Bayesian inference runs with
different random seeds. Further details are given in Sec.~\ref{sec:bayes_factors}.
	}
\end{figure*}

\subsection{Dependence of posterior discrepancies on signal strength}
\label{sec:JSD_vs_SNR}

One generally expects the JS divergence between posteriors inferred using different waveform models to be larger for higher-SNR signals and smaller for lower-SNR signals. This follows from the fact that lower-SNR events provide less informative data, leading to posteriors that remain closer to the prior distribution. Additionally, certain parameters -- such as the secondary spin, $\chi_2$ -- are known to be more difficult to measure, resulting in typically lower JS divergence values compared to parameters like the total mass or effective spin, which tend to be better constrained.

We examine the dependence of the JS divergence on the network SNR for posteriors obtained using the \texttt{NRSur7dq4} model. Figure~\ref{fig:JSD_vs_SNR_phenom} confirms the expected trend: higher-SNR events generally exhibit greater differences between posteriors inferred with the \texttt{NRSur7dq4} and \texttt{IMRPhenomXPHM} models. Notably, five of the seven highlighted events from Sec.~\ref{sec:special_events} have SNRs exceeding 14, placing them in the upper tail of the SNR distribution. However, some high-SNR events deviate from this pattern, suggesting that in certain regions of the binary black hole parameter space, the \texttt{NRSur7dq4} and \texttt{IMRPhenomXPHM} models exhibit relatively low systematic differences.

A similar trend is observed when comparing \texttt{NRSur7dq4} and \texttt{SEOBNRv4PHM}, as shown in Fig.~\ref{fig:JSD_vs_SNR_eob}. One particularly notable outlier is GW190527\_092055, which has a relatively low SNR of about 8 yet exhibits a large JS divergence between the \texttt{NRSur7dq4} and \texttt{SEOBNRv4PHM} posteriors. As discussed in Sec.~\ref{sec:special_events}, the posteriors for this event, obtained using the \texttt{RIFT} code, may be dominated by sampler effects.

\subsection{Sky localization}
\label{sec:sky}
Next, we investigate whether \texttt{NRSur7dq4} posteriors yield better
constrained sky localization when compared to the public
LVK posteriors obtained using the \texttt{IMRPhenomXPHM} and
\texttt{SEOBNRv4PHM} models~\cite{LIGOScientific:2021usb,
LIGOScientific:2021djp, GWTC2.1_PE, GWTC3_PE}.
For most events, \texttt{NRSur7dq4}, \texttt{IMRPhenomXPHM}, and
\texttt{SEOBNRv4PHM} offer largely consistent skymaps.

In Fig.~\ref{fig:skymap}, we
show the recovered skymaps for three events for which we notice the largest
differences; the contours show regions containing the central 50\% and 90\% of
the two-dimensional posterior distribution over sky angles - right ascension
$\alpha$ and declination $\delta$. For each of the three events considered in Fig.~\ref{fig:skymap}, we find
that the skymaps for \texttt{NRSur7dq4} and \texttt{IMRPhenomXPHM} are
consistent, while \texttt{SEOBNRv4PHM} shows a significant difference.
It is important to recall that
\texttt{NRSur7dq4} and \texttt{IMRPhenomXPHM} posteriors are computed using
\texttt{parallel-bilby} and \texttt{bilby}, respectively, and both employ the same
\texttt{dynesty} sampler. The \texttt{SEOBNRv4PHM} posteriors are obtained
with \texttt{RIFT}, which employs different sampling techniques over the
extrinsic parameters. This is a potential reason for the differences in the
skymap posteriors of \texttt{SEOBNRv4PHM} compared to the other models.

\section{Model selection}
\label{sec:bayes_factors}
To understand whether the data prefers a particular waveform model,
one can compare the Bayes factors, given in Eq.~\eqref{eq:bayes_factor}, for different
models. For simplicity, we assume all models are equally likely, thereby
sidestepping the issue of setting prior model odds. Even with this
simplification, meaningfully comparing Bayes factors is
complicated by the fact that
the prior used for \texttt{NRSur7dq4} in our study
is restricted to a smaller portion of the parameter space~\footnote{This restricted prior,
which is described in Sec.~\ref{subsec:prior}, is sufficiently large to contain the full
extent of the posterior. Yet the integral appearing in Eq.~\eqref{eq:bayes_factor} is
carried out over the prior's domain.}
as compared to the ones used
for \texttt{IMRPhenomXPHM} and \texttt{SEOBNRv4PHM} in the LVK analyses~\cite{LIGOScientific:2021usb,
LIGOScientific:2021djp, GWTC2.1_PE, GWTC3_PE}.
Therefore, we re-analyze all 47 events considered in this paper using the
\texttt{IMRPhenomXPHM} model with the same restricted priors and sampler
settings (see Sec.\ref{sec:settings}) used for the \texttt{NRSur7dq4} runs.
We do not perform any new parameter
estimation runs with \texttt{SEOBNRv4PHM} due to the model's
high computational cost. Note that these
\texttt{IMRPhenomXPHM} results obtained with redistricted priors are used only for meaningfully
comparing the Bayes factors and SNRs in Fig.~\ref{fig:SNR} and Fig.~\ref{fig:extrapolation_fraction}.
For all other comparisons made throughout this paper, we use  \texttt{IMRPhenomXPHM} results from the
public LVK posteriors~\cite{LIGOScientific:2021usb, LIGOScientific:2021djp,
GWTC2.1_PE, GWTC3_PE}.

\subsection{Bayes factors}

We compute the differences,
\begin{equation} \label{eq:bayes_diff}
	\Delta \log_e\mathcal{B}=\log_e\mathcal{B}^{\rm NRSur} - \log_e\mathcal{B}^{\rm XPHM} \,,
\end{equation}
where Eq.~\eqref{eq:bayes_diff} is motivated by the identity
\begin{equation} \label{eq:evidence_diff}
\frac{\mathcal{Z}(d|{\cal H}^{\rm NRSur})}{\mathcal{Z}(d|{\cal H}^{\rm XPHM})} = \frac{\mathcal{B}^{\rm NRSur}}{\mathcal{B}^{\rm XPHM}} \,.
\end{equation}
Here  $\mathcal{B}^{\rm NRSur}$ is the Bayes factor of \texttt{NRSur7dq4} over
noise hypothesis and $\mathcal{B}^{\rm XPHM}$ is the Bayes factor
of \texttt{IMRPhenomXPHM} over noise hypothesis using the same priors used for the
\texttt{NRSur7dq4} analyses; see
Eq.~(\ref{eq:bayes_factor}).

The log Bayes factor differences between \texttt{NRSur7dq4}
and \texttt{IMRPhenomXPHM} is shown in the upper panel of
Fig.~\ref{fig:SNR}.  We further provide associated error estimates for $\Delta
\log_e\mathcal{B}$ (shaded red region), obtained using the error estimates for
the logarithm of the waveform-model evidence provided by \texttt{dynesty}, and adding them in
quadrature. We note, however, that this error band provides only a rough guide
to indicate the accuracy achieved when computing the
Bayes factor~\cite{skilling_2006, skiling_2004, speagle2020dynesty,dynesty_2022}.
Indeed, we should view \texttt{dynesty}'s method for computing Bayes factors probabilistically,
and the error estimation as a probabilistic statement (say, 1-sigma interval) instead of an error bound.
For instance, in Fig.~\ref{fig:SNR}, we compare \texttt{dynesty}'s error estimate with the
difference between the maximum and minimum log-Bayes factors (shaded green
region) from the four independent Bayesian inference runs with different random
seeds performed for each event (see Sec.~\ref{sec:settings}).  This provides an
an alternative estimate of the errors in the log-Bayes factors~\footnote{We obtain
this estimate from the \texttt{IMRPhenomXPHM} runs and add in
quadrature to itself to reflect the error estimate for $\Delta
\log_e\mathcal{B}$.}.  We find that, for at least some events, the difference in
log Bayes factor between the runs with different seeds can be larger than the
error estimate in log Bayes factor provided by
\texttt{dynesty}/\texttt{parallel-bilby}. This is not surprising in light of
\texttt{dynesty}'s probabilistic error estimation~\cite{skilling_2006, skiling_2004, speagle2020dynesty,dynesty_2022},
but it does mean that for any particular event a log Bayes factor value can
fluctuate outside of the red shaded region (\texttt{dynesty} error estimate) due to stochastic sampling.
The histogram's bin size has been set to roughly a 1-sigma interval.

Combining all 47 events in Fig.~\ref{fig:SNR}, the
cumulative $\Delta \log_e\mathcal{B}$ value for \texttt{NRSur7dq4} over
\texttt{IMRPhenomXPHM} is 0.54, suggesting an overall mild
preference for \texttt{NRSur7dq4}. We find three events,
GW191109\_010717, GW190521\_030229 and GW190521\_074359, where there is a clear
preference for \texttt{NRSur7dq4}, with $\Delta \log_e\mathcal{B}$ values of
2.73, 1.69, and 1.08, respectively. All three of these events were highlighted in
Sec.~\ref{sec:special_events} as showing noticeable differences in posteriors.
Excluding these three events, the histogram in the top-right panel of Fig.~\ref{fig:SNR} indicates
that (i) most events show no preference for either model, and (ii) many events show a very mild
preference for \texttt{IMRPhenomXPHM}, although, as noted above,
the true errors in our Bayes factor computation may be larger than those indicated in Fig.~\ref{fig:SNR}
thereby spoiling a clear interpretation of small values for any particular event. GW190706\_222641 shows the
largest preference for \texttt{IMRPhenomXPHM}, with a $\Delta
\log_e\mathcal{B}$ value of $-0.98$.

In summary, while the data shows a very mild preference
for \texttt{NRSur7dq4} over \texttt{IMRPhenomXPHM},
there are clear outlier events and secular trends in the distribution
of Bayes factors. Outlier events, trends,
and the robustness of our results are considered in App.~\ref{sec:outliers}.

\subsection{Network SNR}
Next, we compute posteriors for the network-matched filter SNR, $\rho$, recovered
by the \texttt{NRSur7dq4} model for each event. We then compare the median
values of the posteriors for $\rho$ against the ones obtained using
\texttt{IMRPhenomXPHM}. In the lower panel of Fig.~\ref{fig:SNR}, we
report the difference,
\begin{equation}
	\Delta \rho_{\rm median} = \rho^{\rm NRSur}_{\rm median} - \rho^{\rm XPHM}_{\rm median} \,,
\end{equation}
between the median network matched filter SNR recovered
by \texttt{NRSur7dq4} and \texttt{IMRPhenomXPHM}.
In this case, as indicated in
the histogram in the bottom-right panel of Fig.~\ref{fig:SNR},
we find \texttt{NRSur7dq4} typically recovers larger SNRs.
GW191109\_010717
and GW190521\_030229 show the largest preference for \texttt{NRSur7dq4}, with
differences in median SNR of 0.28 and 0.22, respectively. On the other hand,
GW190517\_055101 shows the largest preference for
\texttt{IMRPhenomXPHM}, with a difference in median SNR of -0.20.
Outlier events, trends, and the robustness of our results
are considered in App.~\ref{sec:outliers}.

\begin{figure*}
	\includegraphics[width=\textwidth]{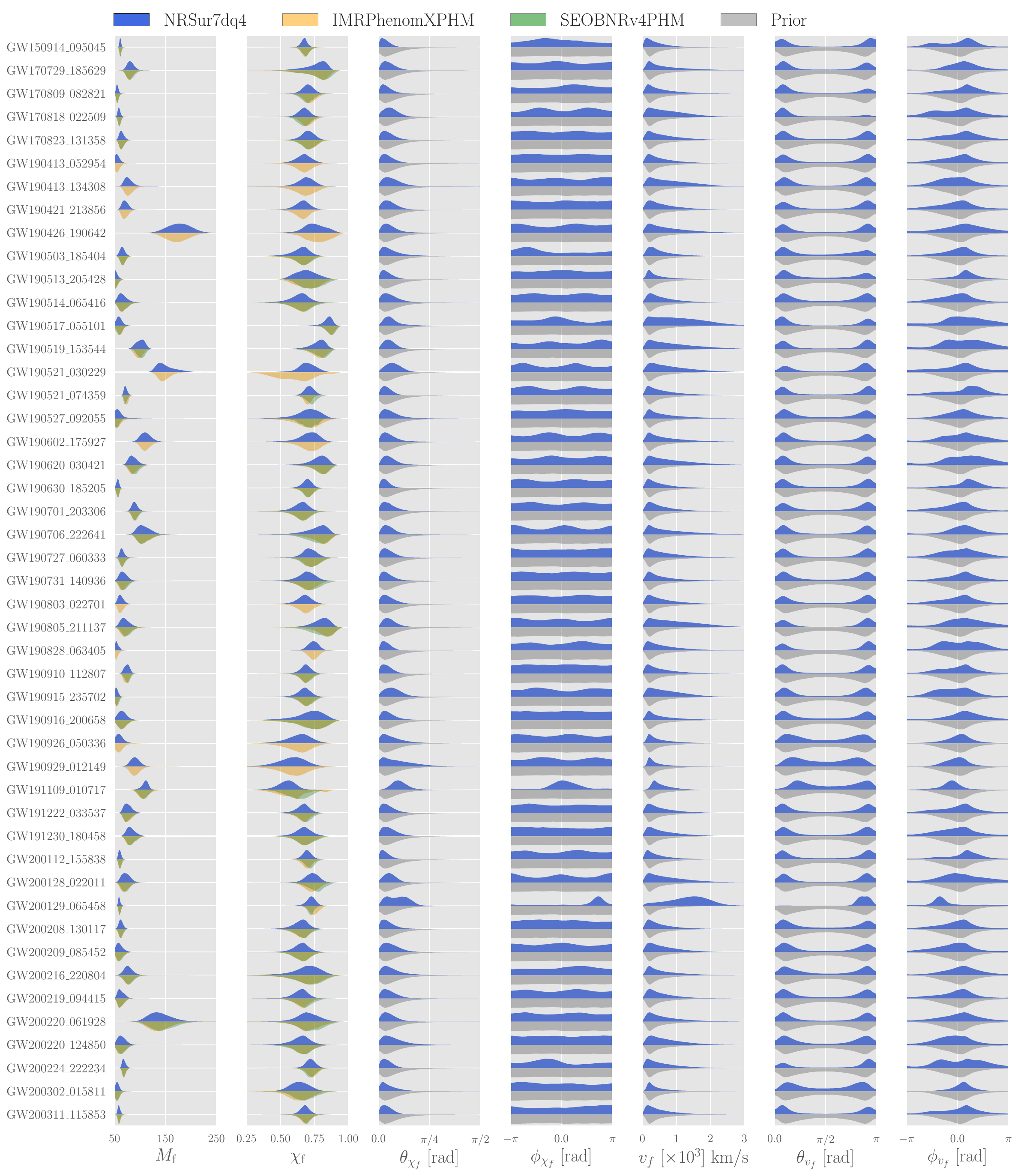}
\caption{\label{fig:remnant}
Posteriors for the source-frame remnant mass $M_f$, the remnant spin magnitude
$\chi_f$, the remnant kick magnitude $v_f$, and the spherical polar
$\{\theta_{\chi_f},\theta_{v_f}\}$ and azimuthal angles
$\{\phi_{\chi_f},\phi_{v_f}\}$ of $\bchif$ and $\bvf$ for all 47 events
analyzed with the \texttt{NRSur7dq4} model (blue). For comparison, we also show
the $M_f$ and $\chi_f$ posteriors from the public LVK data
release~\cite{LIGOScientific:2021usb,LIGOScientific:2021djp, GWTC2.1_PE,
GWTC3_PE} obtained using \texttt{IMRPhenomXPHM} (orange) and
\texttt{SEOBNRv4PHM} (green, where available).  For the rest of the parameters,
we show the effective priors (in gray); the difference between the prior and
posterior can be used to assess how informative the data are about these
parameters. The spin and kick angles are shown in the wave frame at $t_{\rm
ref}=-100 \, M_{\rm det}$.
We provide 3D visualizations of the full remnant spin and kick posteriors at Ref.~~\cite{NRSurCatalog}.
Further details are given in Sec.~\ref{sec:remnant}.
}
\end{figure*}

\begin{figure*}
	\includegraphics[width=0.94\textwidth]{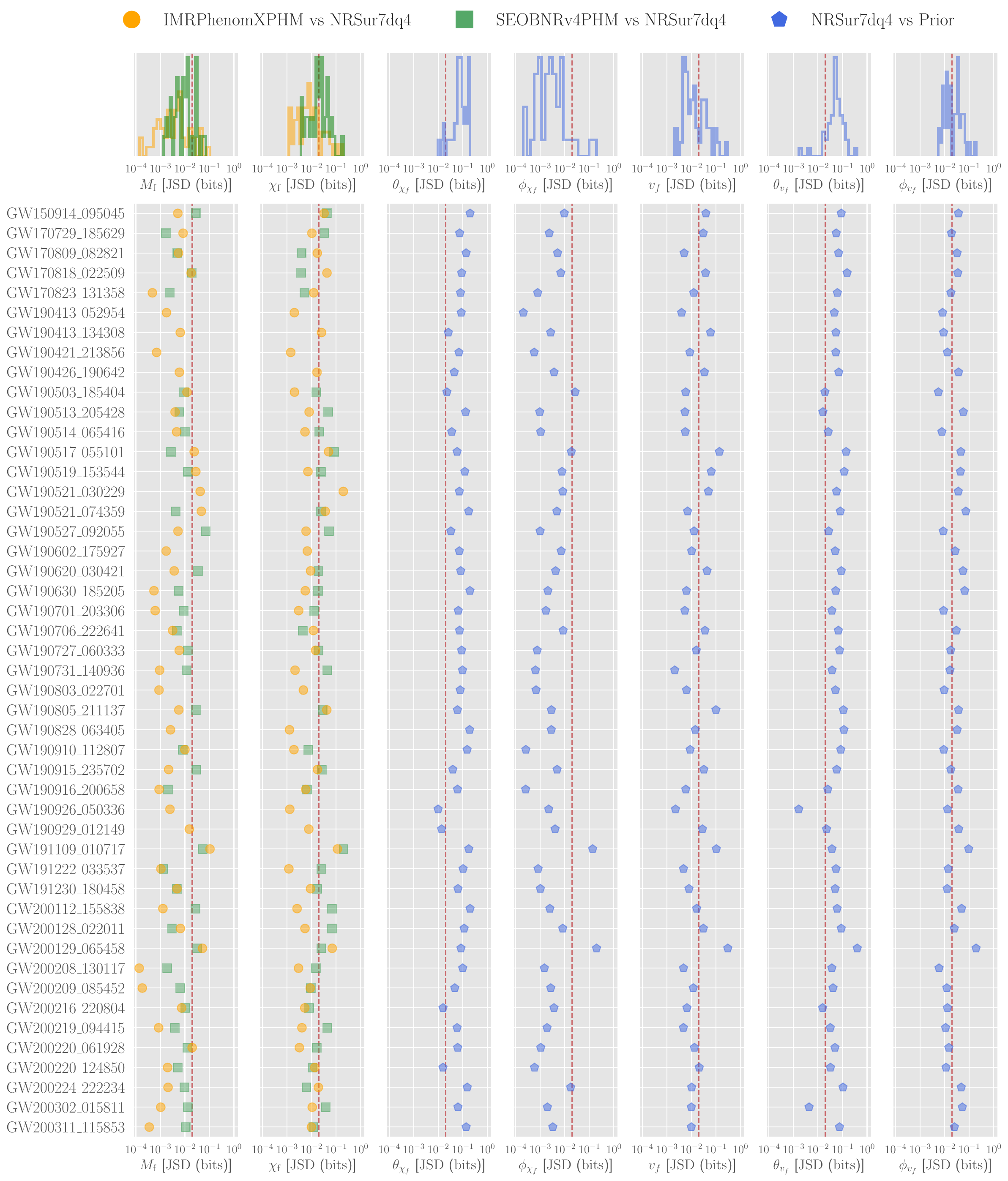}
\caption{\label{fig:jsdivs_remnant}
The first two columns show the JS divergence (JSD) for the source-frame remnant
mass $M_f$ and spin magnitude $\chi_f$ between the \texttt{NRSur7dq4}
posteriors and the \texttt{IMRPhenomXPHM} (orange circles) and
\texttt{SEOBNRv4PHM} (green squares, where available) posteriors from the LVK
release~\cite{LIGOScientific:2021usb, LIGOScientific:2021djp, GWTC2.1_PE,
GWTC3_PE}.  The remaining columns show the JS divergence for the remaining
remnant parameters (same as Fig.~\ref{fig:remnant}) between the posterior and
prior for \texttt{NRSur7dq4} (blue pentagons).  Dashed red lines indicate a JS
divergence of 0.02, indicating significant differences between
\texttt{NRSur7dq4} and \texttt{IMRPhenomXPHM}/\texttt{SEOBNRv4PHM} posteriors
for the first two columns, and informative \texttt{NRSur7dq4} posteriors for
the remaining columns. Further details are given in Sec.~\ref{sec:remnant}.
}
\end{figure*}

\section{Inference of the remnant properties}
\label{sec:remnant}

In addition to constraining the properties of the component BHs in the binary,
we infer the properties of the final BH left behind after the merger, in
particular, its source-frame mass $M_f$, spin vector $\bchif$, and recoil velocity vector
$\bvf$. During its evolution, the binary radiates energy, angular momentum, and
linear momentum. The radiated energy and angular momentum are reflected in
$M_f$ and $\bchif$, respectively. The radiated linear momentum causes a shift
in the binary's center of mass in the opposite direction, imparting a recoil
velocity or a ``kick'' $\bvf$ to the remnant BH.  While $\bchif$ is restricted to
be either parallel or anti-parallel to the orbital angular momentum $\bm{L}$ for
nonprecessing binaries, it can be arbitrarily oriented for precessing binaries.
On the other hand, while the kicks for nonprecessing binaries are typically
restricted to $\lesssim 300$ km/s and along the orbital plane, kicks for
precessing binaries can reach magnitudes up to $\sim 5000$
km/s~\cite{Campanelli:2007cga,Gonzalez:2007hi,Lousto:2011kp,Gonzalez:2006md}
with arbitrary orientations. However, as we will discuss below, for precessing
binaries, the direction of $\bchif$ is preferentially along, while the
direction of $\bvf$ is preferentially along or opposite, the direction of
$\bm{L}$ near merger.

Remnant BH properties have important applications for astrophysics and
fundamental physics. The remnant mass and spin magnitude are important for
tests of general relativity using GWs~\cite{LIGOScientific:2021sio}, as the
the remnant mass entirely determines frequencies in the ringdown and
spin. The kick magnitude is important for placing observational
constraints~\cite{Varma:2022pld, Varma:2020nbm, Varma:2021xbh,
LIGOScientific:2020ufj, Doctor:2019ruh, Mahapatra:2021hme} on the rate of
hierarchical mergers in dense environments: repeated mergers are a means to
form heavy BHs in nature, but if the kick exceeds the escape velocity of the
host environment, the remnant BH after the first merger would simply get
ejected and not participate in another merger. Finally, the remnant spin
direction and kick direction~\cite{Varma:2022pld, CalderonBustillo:2022ldv} can
be useful to study binaries that may be formed in active galactic nuclei disks,
as the final BH's spin orientation with respect to the disk as well as its
motion can impact potential electromagnetic counterparts~\cite{Graham:2020gwr}.
The kick direction also shows up when computing the
Doppler-shifted remnant mass, which
may play a role in future high-accuracy ringdown tests of GR~\cite{Varma:2020nbm, Varma:2022pld, moore2016black, Mahapatra:2023htd}.

Therefore, while the public LVK results~\cite{LIGOScientific:2021usb,
LIGOScientific:2021djp, GWTC2.1_PE, GWTC3_PE} only include the remnant mass and
spin magnitudes, we provide posterior samples for the full spin and kick
vectors. We report the source-frame mass $M_f$, spin vector $\bm{\chi}_f$ and
kick velocity vector $\bm{v}_f$ of the remnant black hole for each of the 47
events we analyze in Fig.~\ref{fig:remnant}.

\subsection{Analysis framework}
\label{sec:remnant_analysis_framework}

We follow the prescription outlined in Refs.~\cite{Varma:2020nbm,
Varma:2022pld, varma2019surrogate, Varma:2018aht}: starting with the posterior
samples for the spins and detector-frame component masses
obtained using the \texttt{NRSur7dq4} model, we evaluate the associated remnant
surrogate model \texttt{NRSur7dq4Remnant} which provides estimates for
$M_{f,{\rm det}}$, $\bchif$ and $\bvf$, from which we compute
$M_f = M_{f,{\rm det}}/(1+z)$.  These samples are included in our public release at Ref.~\cite{NRSurCatalog}.
Note that the remnant spin and
kick vectors in our public release are defined in the same frame as the
component BHs, i.e. the wave frame at $f_{\rm ref}=20$ Hz, as described in
Sec.~\ref{subsec:frame_choice}.

However, when visualizing the remnant spin and kick directions in this section,
we adopt a different frame that is more naturally suited for discussing
remnant properties: the wave frame at $t_{\rm ref}=-100 \, M_{\rm det}$ as
proposed in Ref.~\cite{Varma:2021csh}. This frame is similar to the wave frame
at $f_{\rm ref}=20$ Hz (see Sec.~\ref{subsec:frame_choice}), except that the
reference point is chosen to be the dimensionless time of $t_{\rm ref}=-100 \,
M_{\rm det}$ before the peak waveform amplitude (defined in Eq.5 of
Ref.~\cite{varma2019surrogate}). Because this reference point is always very
close to the merger (typically within 2-4 GW cycles~\cite{Varma:2021csh}), it
provides a more natural frame to define remnant spin and kick vectors than the
wave frame at $f_{\rm ref}=20$ Hz (which can occur up to $\sim40$ GW cycles
before the mergers for the events considered in this work).

\begin{figure*}
	\includegraphics[width=1.0\textwidth]{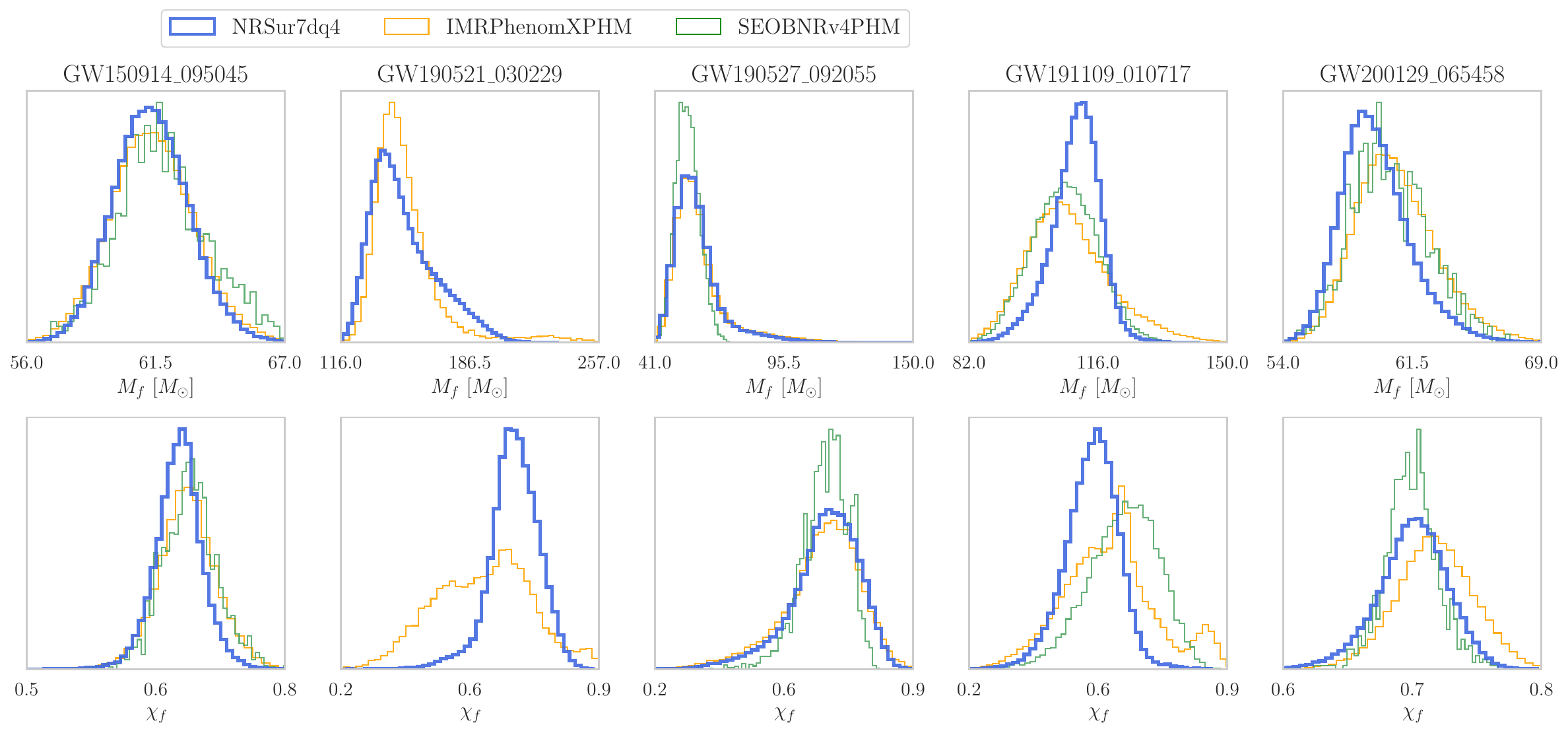}
\caption{\label{fig:special_events_remnants}
Posteriors of source-frame mass and spin magnitude of the remnant BH for five
events where we see the most prominent differences between results obtained
using \texttt{NRSur7dq4} (blue histogram), \texttt{IMRPhenomXPHM} (orange
histogram) and \texttt{SEOBNRv4PHM} (green histogram, where available). Further
details are given in Sec.~\ref{subsec:mf_chif}.
}
\end{figure*}

\begin{figure}
	\includegraphics[width=\columnwidth]{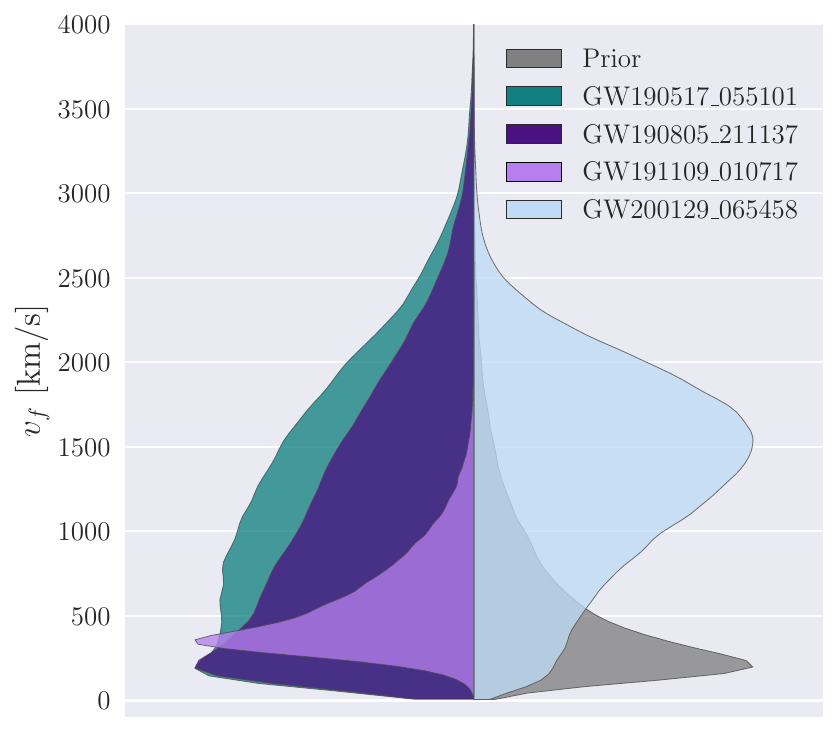}
\caption{\label{fig:superkick}
Kick magnitude constraints for four events with the largest JS divergence
between the posterior and prior. We show the posteriors and the effective prior
for comparison. GW200129\_065458 and GW191109\_010717 show a preference away
from $v_f=0$. Further details are given in Sec.~\ref{subsec:kick}.
}
\end{figure}

\begin{figure*}[htp]
    \subfloat[$\theta_{\chi_f}$]{
            \includegraphics[width=\columnwidth]{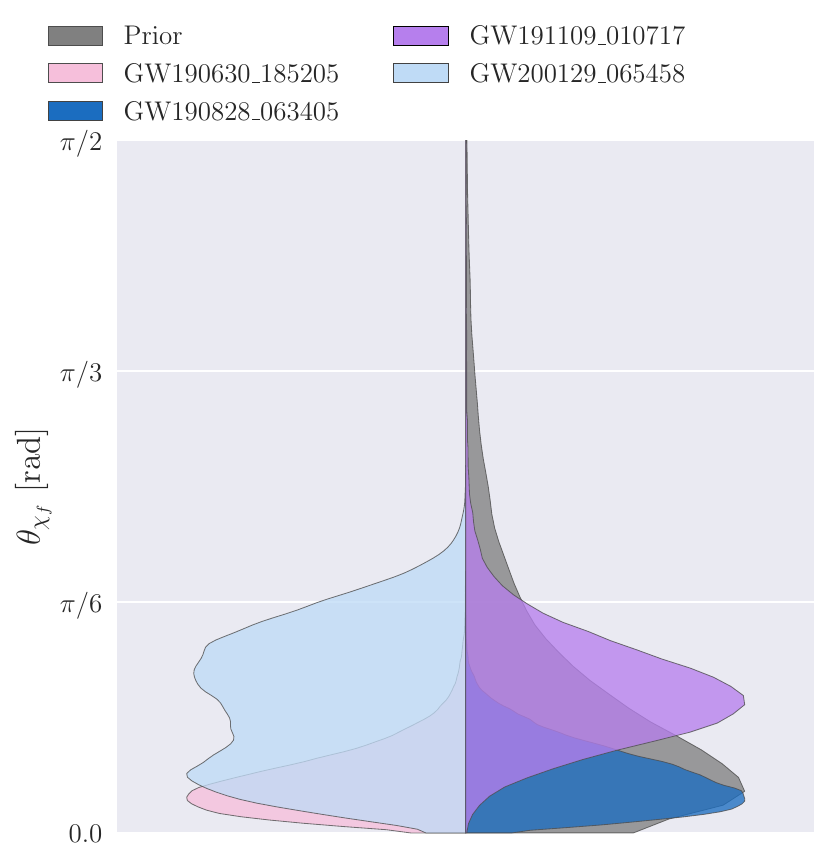}
            \label{fig:theta_chif}
    }
    \subfloat[$\theta_{v_f}$]{
            \includegraphics[width=\columnwidth]{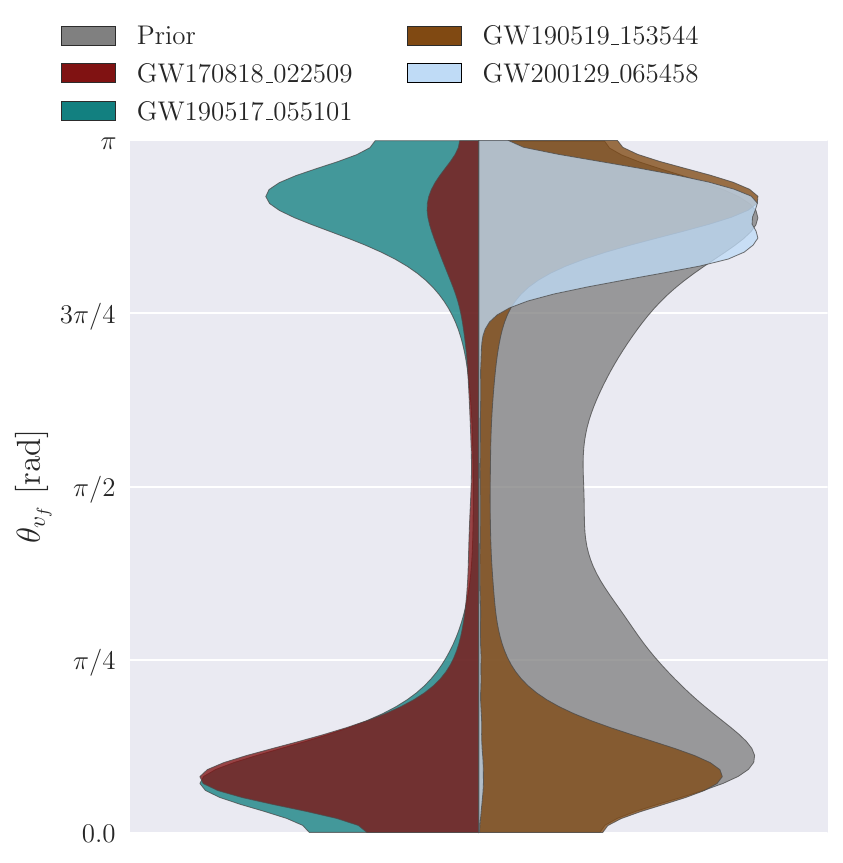}
            \label{fig:theta_vf}
    }\\
    \subfloat[$\phi_{\chi_f}$]{
            \includegraphics[width=\columnwidth]{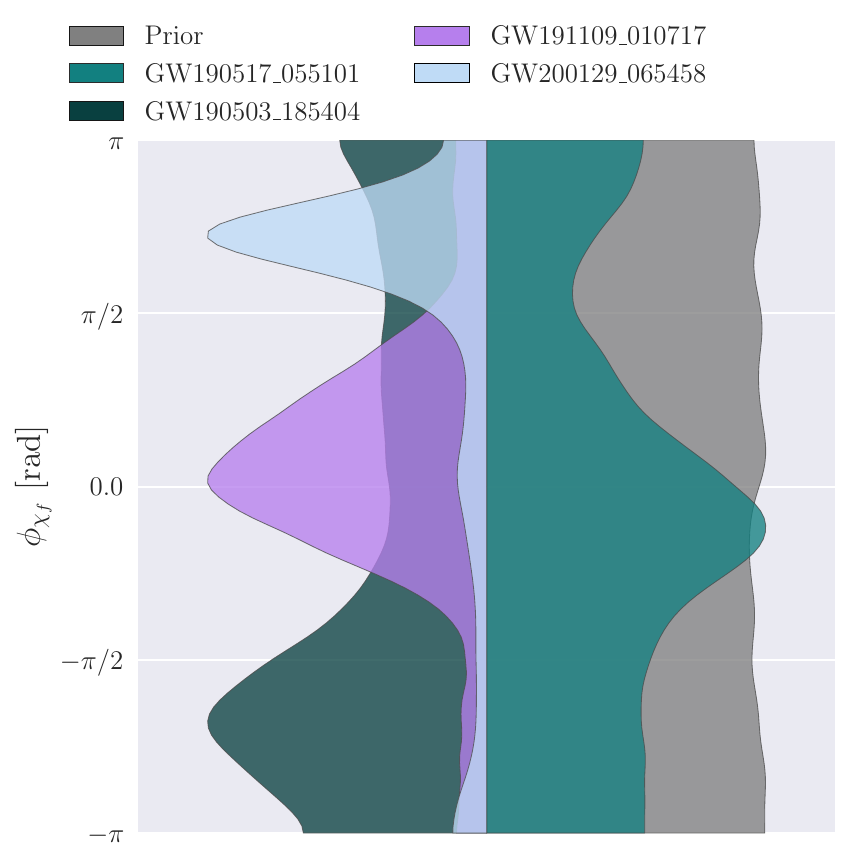}
            \label{fig:phi_chif}
    }
    \subfloat[$\phi_{v_f}$]{
            \includegraphics[width=\columnwidth]{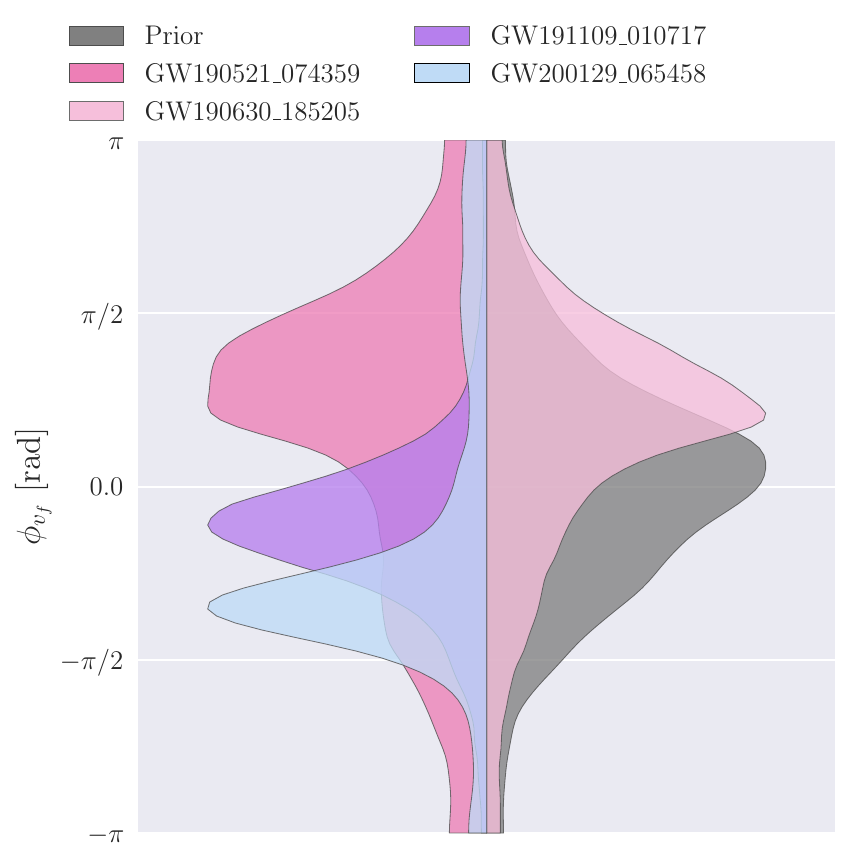}
            \label{fig:phi_vf}
    }
\caption{\label{fig:remnant_angles}
Posteriors for the polar angles $\{\theta_{\chi_f},\theta_{v_f}\}$ and
azimuthal angles $\{\phi_{\chi_f},\phi_{v_f}\}$ of the remnant black holes'
spin and kick directions for four events; these events are picked either
because they show the highest JS divergence between posterior and prior, or
because the posterior peaks noticeably away from the peak of the prior. We also
show the effective priors in gray. The spin and kick angles are shown in the
wave frame at $t_{\rm ref}=-100 \, M_{\rm det}$.  Further details are in the
text (Sec.~\ref{subsec:remnant_angles}).
}
\end{figure*}

For example, in the wave frame at $t_{\rm ref}=-100 \, M_{\rm det}$, the
direction of $\bchif$ is preferentially oriented close to the $z$-axis. This
can be explained as follows: the direction of $\bchif$ can be approximated by the
direction~\cite{Barausse:2009uz, Hofmann:2016yih} of the total angular momentum
$\bm{J}=\bm{L} + m_1^2 \bchiOne + m_2^2 \bchiTwo$, but $\bm{J}$ is typically
dominated (excluding the special case of transitional precession) by the
contribution from $\bm{L}$ rather than the contribution from the spins. As a
result, the direction of $\bchif$ is preferentially oriented close to $\bm{L}$
near merger, which is along the $z$-axis of the wave frame at $t_{\rm ref}=-100
\, M_{\rm det}$ (see Sec.~\ref{subsec:frame_choice}). This is reflected in the
prior for the remnant spin direction, as we will see in Sec.~\ref{subsec:remnant_angles}.  Similarly, the wave frame at $t_{\rm ref}=-100
\, M_{\rm det}$ is well-suited to discuss the kick direction as well, as the
kick is known to be preferentially orientated close to or opposite to $\bm{L}$
near merger~\cite{Campanelli:2007cga, Gonzalez:2007hi, Lousto:2011kp,
Gonzalez:2006md, Varma:2018aht}.  For this reason, while our public release
~\cite{NRSurCatalog} will contain remnant spin and kick vectors defined in the
wave frame at $f_{\rm ref}=20$ Hz for consistency with the frame used for the
component BH spins, in Figs.~\ref{fig:remnant}, \ref{fig:jsdivs_remnant},
\ref{fig:special_events_remnants}, \ref{fig:superkick},
\ref{fig:remnant_angles}, we will adopt the wave frame at $t_{\rm ref}=-100 \,
M_{\rm det}$.  As described in Ref.~\cite{Varma:2021csh}, the two frames are
related by a transformation described by the dynamics of \texttt{NRSur7dq4},
which is provided by the model~\cite{varma2019surrogate}.

\subsection{Remnant mass and spin magnitude}
\label{subsec:mf_chif}
Figure~\ref{fig:remnant} summarizes our constraints on the remnant properties
for all 47 events considered in our analysis (cf. Sec.~\ref{sec:event_selection}).
The first two columns show the source-frame
remnant mass $M_f$ and spin magnitude $\chi_f$ for \texttt{NRSur7dq4} with
\texttt{NRSur7dq4Remnant}, along with the corresponding constraints from the
LVK public release for \texttt{IMRPhenomXPHM} and \texttt{SEOBNRv4PHM}
~\cite{LIGOScientific:2021usb, LIGOScientific:2021djp, GWTC2.1_PE, GWTC3_PE}.
In the LVK results, the remnant mass and spin magnitude are computed following
Ref.~\cite{mcdaniel2016}, using the remnant models of
Refs.~\cite{Healy:2016lce, Jimenez-Forteza:2016oae, Hofmann:2016yih}.  While
these models include some precession corrections, they are not informed by
precessing NR simulations. Therefore, differences with respect to
our estimates of $M_f$ and $\chi_f$ can arise from
differences in the remnant models as well as the waveform
model used to infer binary source parameters.

The first two columns of Fig.~\ref{fig:jsdivs_remnant} show the JS divergence
between our posteriors for $M_f$ and $\chi_f$, and the public LVK samples
for \texttt{IMRPhenomXPHM} and \texttt{SEOBNRv4PHM}.  For about 13 \% (17\% ) of
events, the JS divergence between \texttt{NRSur7dq4} and \texttt{IMRPhenomXPHM}
(\texttt{SEOBNRv4PHM}) for $M_f$ rises above 0.02 bits, indicating noticeable
differences.  Similarly, for about 19\% (38\%) of events, the JS divergence
between \texttt{NRSur7dq4} and \texttt{IMRPhenomXPHM} (\texttt{SEOBNRv4PHM})
for $\chi_f$ rises above 0.02 bits. The events with the most prominent differences
are highlighted in Fig.~\ref{fig:special_events_remnants}.

Constraints on the remaining remnant parameters (the direction of $\bchif$, and
the vector $\bvf$) are not provided in the public LVK
samples~\cite{LIGOScientific:2021usb, LIGOScientific:2021djp, GWTC2.1_PE,
GWTC3_PE}. Therefore, in Figs.~\ref{fig:remnant} and \ref{fig:jsdivs_remnant},
we compare our posteriors for these parameters with their \emph{effective
priors}, to judge how informative the data are about these quantities.
Following Refs.~\cite{Varma:2020nbm, Varma:2022pld}, effective prior samples
for $M_f$ (not shown), $\bchif$ and $\bvf$ are obtained by evaluating the
\texttt{NRSur7dq4Remnant} model on samples drawn from the prior on
$\{m_1,m_2,\bm{\chi}_1,\bm{\chi}_2\}$ (see Sec.~\ref{subsec:prior}). In the
following figures, the prior samples for $\bchif$ and $\bvf$ are also
transformed to the wave frame at $t_{\rm ref}=-100 \, M_{\rm det}$.  To
parameterize the remnant spin and kick directions, we adopt the standard
spherical polar angles $\theta_{{\chi}_f}$ and $\theta_{v_f}$, and azimuthal
angles $\phi_{{\chi}_f}$ and $\phi_{v_f}$, computed in the wave frame at
$t_{\rm ref}=-100 \, M_{\rm det}$.

\subsection{Recoil velocity magnitude}
\label{subsec:kick}
The fifth column of Fig.~\ref{fig:remnant} shows our constraints on the kick
magnitude $v_f$ along with the corresponding prior. As expected from
Ref.~\cite{Varma:2020nbm}, for most events the posterior is largely
indistinguishable from the prior, meaning that the data are not informative
about the kick magnitude. In Fig.~\ref{fig:jsdivs_remnant}, we find that the JS
divergence between the posterior and prior of $v_f$ rises above 0.02 bits for
about 36\% of the events considered. The events with the four highest JS
divergence values are highlighted in Fig.~\ref{fig:superkick}. Notably, two of these
events show a clear preference away from $v_f=0$ compared to the prior:
GW200129\_065458 with $v_f \sim 1392^{+848}_{-1085}$ km/s and a JS divergence
of 0.305 bits, and GW191109\_010717 with $v_f \sim 485^{+668}_{-252}$ km/s and a
JS divergence of 0.103 bits. Here, we report the median and 90\% symmetric
credible interval. While this finding can have important astrophysical
implications, especially for hierarchical mergers, we again point out that both
GW200129\_065458 and GW191109\_010717 suffered from being coincident with a
detector glitch, which can be challenging to remove from short signals reliably
such as these, potentially impacting inferences about precession and
kicks~\cite{Payne:2022spz}.

\subsection{Remnant spin and kick velocity directions}
\label{subsec:remnant_angles}
Finally, the remaining columns in Figs.~\ref{fig:remnant} and
\ref{fig:jsdivs_remnant} report the \texttt{NRSur7dq4} posteriors and priors
for the remnant spin and kick direction parameters. For each parameter
($\theta_{{\chi}_f}$, $\theta_{v_f}$, $\phi_{{\chi}_f}$ and $\phi_{v_f}$), we
highlight four interesting events in Fig.~\ref{fig:remnant_angles}. These
events are picked either because they show the highest JS divergence between
posterior and prior, or because the posterior peaks noticeably away from the
peak of the prior.

Note that $\theta_{{\chi}_f}=0$ means that the spin is directed along $\bm{L}$
at $t_{\rm ref}=-100 \, M_{\rm det}$. As mentioned above, binaries have a
preference for $\theta_{{\chi}_f}=0$ in this frame, and this is reflected in
the prior for $\theta_{{\chi}_f}$ having a strong preference for zero
(Fig.~\ref{fig:theta_chif}). Similarly, Fig.~\ref{fig:theta_vf} shows that the
prior for $\theta_{v_f}$ has a strong preference for zero or $\pi$, which was
also motivated above. Finally, to orient the reader, we remind that
$\phi_{{\chi}_f}=0$ ($\phi_{v_f}$=0) indicates that the in-plane component of
the remnant spin (kick) is coaligned with a vector from the less massive BH to the
more massive BH at $t_{\rm ref}=-100 \, M_{\rm det}$.
In Fig.~\ref{fig:remnant_angles}, we note that while the prior for
$\phi_{{\chi}_f}$ is broad, the prior for $\phi_{v_f}$ shows a preference for
zero.  These features highlight why the wave frame at $t_{\rm ref}=-100 \,
M_{\rm det}$ is a suitable frame for discussing remnant spin and kick
directions.

In Fig.~\ref{fig:jsdivs_remnant}, we find that the JS divergence between the
posterior and prior for $\theta_{{\chi}_f}$, rises above 0.02 bits for about 91.4\%
of the events considered, indicating that the data are informative about this
parameter for most events. Among the highlighted events in
Fig.~\ref{fig:theta_chif}, GW190630\_185205, GW190828\_063405 and GW191109\_010717
show a stronger preference for
$\theta_{{\chi}_f}\sim0$ than the prior, GW200129\_065458
has a slightly bimodal posterior extending to $\theta_{{\chi}_f} \sim \pi/6$,
and GW191109\_010717 peaks at $\theta_{{\chi}_f} \sim \pi/12$, clearly
away from the peak of the prior.

Next, for $\theta_{v_f}$, we find that JS divergence between the posterior and
prior rises above 0.02 bits for about 89.3\% of the events considered, indicating
that this parameter may be more informative than the kick magnitude $v_f$
itself. Among the highlighted events in Fig.~\ref{fig:theta_vf},
GW170818\_022509 prefers a kick directed roughly along $\bm{L}$ at $t_{\rm
ref}=-100 \, M_{\rm det}$, while GW200129\_065458 prefers the opposite. Here,
we note that while GW200129\_065458 shows a measurable kick magnitude,
GW170818\_022509 does not (see Fig.~\ref{fig:remnant}). Our findings are
broadly consistent with the discussion of the measurability of the kick direction
in Refs.~\cite{Varma:2020nbm, Varma:2022pld, CalderonBustillo:2022ldv}, but
more work may be needed to understand how to interpret $\theta_{v_f}$
constraints for events where the kick magnitude is unmeasured.

Finally, the JS divergence between the posterior and prior for
$\phi_{{\chi}_f}$ and $\phi_{v_f}$ rises above 0.02 bits for about 6\%, and 51\% of
the events considered, respectively. Therefore, the data are mostly
uninformative about $\phi_{{\chi}_f}$, but can be used to constrain
$\phi_{v_f}$.  Among the
highlighted events in Fig.~\ref{fig:phi_chif} and \ref{fig:phi_vf}, GW191109\_010717
and GW200129\_065458 show the strongest constraints; clearly the kick direction is better
measured than the kick magnitude.

\section{Conclusion}
\label{sec:discussion}

The third Gravitational-Wave Transient Catalog
contains 90 binary coalescence candidates detected by the LIGO-Virgo-KAGRA Collaboration.
In this paper, we identify a set of 47 events from the \texttt{GWTC-3}
catalog~\cite{LIGOScientific:2021usb, LIGOScientific:2021djp, GWTC2.1_PE,
GWTC3_PE} that falls within the domain of validity ($q \geq 1/6$ and $M_{\rm
det} \geq 60 M_{\odot}$) of the \texttt{NRSur7dq4} waveform model
(Sec.~\ref{sec:event_selection}). Within this domain, the \texttt{NRSur7dq4} model about
an order-of-magnitude more accurate than the models used in the the official LVK
analysis and includes the full physical effects of precession.
We use the Bayesian inference
code \texttt{parallel-bilby} to
estimate the source properties for these BBH events with the \texttt{NRSur7dq4} model.
We compare source properties inferred using the \texttt{NRSur7dq4} model and public LVK
posteriors obtained using the \texttt{IMRPhenomXPHM} and \texttt{SEOBNRv4PHM}
models~\cite{LIGOScientific:2021usb, LIGOScientific:2021djp, GWTC2.1_PE,
GWTC3_PE} (Sec.~\ref{sec:results}).  The difference between the resulting posterior samples have
been quantified using JS divergence, a common measure of the
statistical distance in information content between probability distributions.
We find many events for which noticeable differences exist between posteriors
obtained with \texttt{NRSur7dq4}, \texttt{IMRPhenomXPHM}, and
\texttt{SEOBNRv4PHM} models. Below are the key
take-aways from our results:
\begin{itemize}
\item While the posteriors are consistent for the majority of events, we find a number of
    events for which posteriors for \texttt{NRSur7dq4} are noticeably different
    from the posteriors for \texttt{IMRPhenomXPHM}/\texttt{SEOBNRv4PHM}. In
    particular, for $\sim 23$\% ($\sim 55$\%) of the analyzed events, JS
    divergence between \texttt{NRSur7dq4} and \texttt{IMRPhenomXPHM}
    (\texttt{SEOBNRv4PHM}) exceeds a commonly used threshold of 0.02 bits, indicating
    non-negligible differences, for at
    least one of the following parameters: total mass $M$, mass ratio $q$,
    component masses $m_1, m_2$, spin magnitudes $\chi_1, \chi_2$, spin tilts
    $\theta_{1}, \theta_2$, the effective inspiral spin $\chieff$, the spin
    precession parameter $\chi_p$, luminosity distance $D_{\rm L}$, and
    inclination angle $\theta_{\rm JN}$ (Sec.~\ref{sec:results}).
    For many of these cases, multiple parameters exceed this threshold,
    and for a handful of events, the JS divergence values are above 0.1 (and in a few cases above 0.2)
    indicating substantial differences. The most interesting GW events are summarized
    in Sec.~\ref{sec:special_events}.
\item Even for the first GW signal, GW150914\_095045, we find noticeable
    differences in $\chi_p$ and $\chi_1$ measurements between \texttt{NRSur7dq4},
    \texttt{IMRPhenomXPHM} and \texttt{SEOBNRv4PHM}.
    Interestingly, our \texttt{NRSur7dq4} estimates show better agreement with
    earlier LVK results obtained with \texttt{SEOBNRv3} and \texttt{IMRPhenomPv2}.
    Fig.~\ref{fig:GW15_All} summarizes some of these observations.
\item For GW191109\_010717, \texttt{NRSur7dq4} shows a stronger preference for
    negative $\chieff$ at 99.3\% credible level, compared to 95.9\% for
    \texttt{SEOBNRv4PHM} and 85.3\% for \texttt{IMRPhenomXPHM}
    (Sec.~\ref{sec:GW191109}). This is consistent with
    Ref.~\cite{Ramos-Buades:2023ehm}, where a similar preference was found for
    the newer \texttt{SEOBNRv5PHM} model. The preference for $\chieff<0$ can
    have important astrophysical implications, as negative $\chieff$ is
    expected to be more common in dynamically formed binaries than those formed
    through isolated evolution. However, some caution is warranted as this
    event suffered from a detector glitch.
\item The events showing the most notable differences in the posteriors for the
    component BH properties are highlighted in Sec.~\ref{sec:special_events}.
    Furthermore, by comparing the Bayes factors and recovered SNRs between
    \texttt{NRSur7dq4} and \texttt{IMRPhenomXPHM} with the same prior settings
    for all 47 events, we find that there is a mild preference for
    \texttt{NRSur7dq4} over \texttt{IMRPhenomXPHM}
    (Sec.~\ref{sec:bayes_factors}).
\item We find several events where the remnant mass and spin magnitude
    posteriors are noticeably different between \texttt{NRSur7dq4} and
    \texttt{IMRPhenomXPHM}/\texttt{SEOBNRv4PHM}, which can have implications
    for tests of general relativity (Sec.~\ref{subsec:mf_chif}).
\item We provide kick magnitude posteriors for all 47 events, which can be
    useful for constraining the formation rate of heavy BHs through repeated
    mergers (Sec.~\ref{subsec:kick}). We find that the kick magnitude is
    informative for GW191109\_010717 and GW200129\_065458, with
    GW200129\_065458 showing a preference for a large kick, as noted by
    Ref.~\cite{Varma:2022pld}. However, once again, some caution is warranted
    as both of these events suffered from detector glitches.
\item Finally, we also provide posteriors for the remnant spin and kick
    directions for all 47 events and discuss possible astrophysical
    applications of these measurements in Sec.~\ref{subsec:remnant_angles}.
\end{itemize}

The differences in the posteriors for \texttt{NRSur7dq4},
\texttt{IMRPhenomXPHM} and \texttt{SEOBNRv4PHM} suggest that waveform
systematics are already important for GW data analysis. These differences
can become compounded when the posterior samples
are used in hierarchical analyses like
constraining astrophysical populations or tests of general relativity.
Systematic differences can arise from the differences in the modeling approach
as well as the physics included; for example, while \texttt{NRSur7dq4} is
trained directly on precessing NR simulations, \texttt{IMRPhenomXPHM} and
\texttt{SEOBNRv4PHM} are only informed by nonprecessing simulations.

However, we note that our comparisons are based on posteriors obtained
using different Bayesian Inference codes: \texttt{bilby} for \texttt{IMRPhenomXPHM}, \texttt{RIFT}
for \texttt{SEOBNRv4PHM}, and \texttt{parallel-bilby} for \texttt{NRSur7dq4}.
This may introduce additional systematics in our attempts to make meaningful
comparisons. For example, some of the \texttt{SEOBNRv4PHM} posteriors appear
undersampled due to an inefficient post-processing step to include calibration
uncertainties in \texttt{RIFT} (Fig.~\ref{fig:special_events}). Furthermore, we
find that \texttt{SEOBNRv4PHM} posteriors for the skymaps are significantly
different for a number of events while \texttt{NRSur7dq4} and
\texttt{IMRPhenomXPHM} match closely (Fig.~\ref{fig:skymap}). A more careful
study will be necessary to fully disentangle waveform from sampler
systematics.

Nonetheless, our results provide further motivation to improve all waveform
models. In particular, it is important to
extend the region of validity of NR surrogate models to include more unequal mass
ratios as well as longer inspirals. As the detector sensitivities
improve~\cite{Purrer:2019jcp}, systematic biases in estimating binary source
parameters could limit important applications like BH astrophysics,
dark siren cosmology~\cite{LIGOScientific:2021aug}, and fundamental tests of
general relativity.

Our results are publicly accessible~\cite{NRSurCatalog}. We additionally provide an application programming interface (API) to access our data programmatically. The API is documented in the GitHub
repository for this work~\cite{NRSurCatalog}. The catalog's website utilized software from the \texttt{TESS-Atlas} project~\cite{tess_atlas}.

\begin{acknowledgments}
We thank Michael Puerrer, David Keitel, Angel Garron, Gregorio Carullo, Christopher Berry, Collin Capano, Connor Kenyon and Gaurav Khanna for helpful discussions throughout the project.
T.I., F.H.S, and S.E.F acknowledge support from NSF Grants Nos. PHY-2110496, DMS-2309609, DMS-1912716, and by UMass Dartmouth's Marine and Undersea Technology (MUST) Research Program funded by the Office of Naval Research (ONR) under Grant No. N00014-23-1–2141.
Part of this work is additionally supported by the Heising-Simons Foundation, the Simons Foundation, and NSF Grants Nos. PHY-1748958.
A.V. is supported by  Marsden Fund Grant MFP-UOA2131, administered by the Royal Society Te Aparangi.
C.-~J.~H. acknowledges the support from the Nevada Center for Astrophysics, from NASA Grant No. 80NSSC23M0104, and the NSF through the Award No.~PHY-2409727.
V.V.~acknowledges support from NSF Grant No. PHY-2309301, and the European
Union’s Horizon 2020 research and innovation program under the Marie
Skłodowska-Curie grant agreement No.~896869.
ROS acknowledges support from NSF Grants No. PHY-2012057, PHY-2309172, and the Simons Foundation.
Simulations were performed on CARNiE at the Center for Scientific Computing and Data science Research (CSCDR) of UMassD, which is supported by the Office of Naval Research (ONR)/Defense University Research Instrumentation Program (DURIP) Grant No.\ N00014181255, the UMass-URI UNITY supercomputer supported by the Massachusetts Green High Performance Computing Center (MGHPCC), and
on the XSEDE/ACCESS resource Anvil at the Rosen Center For Advanced Computing through Allocation No.~PHY990002.

This material is based upon work supported by NSF's LIGO Laboratory which is a major facility fully funded by the National Science Foundation.
This research has made use of data or software obtained from the Gravitational Wave Open Science Center (gw-openscience.org), a service of LIGO Laboratory, the LIGO Scientific Collaboration, the Virgo Collaboration, and KAGRA. LIGO Laboratory and Advanced LIGO are funded by the United States National Science Foundation (NSF) as well as the Science and Technology Facilities Council (STFC) of the United Kingdom, the Max-Planck-Society (MPS), and the State of Niedersachsen/Germany for support of the construction of Advanced LIGO and construction and operation of the GEO600 detector. Additional support for Advanced LIGO was provided by the Australian Research Council. Virgo is funded, through the European Gravitational Observatory (EGO), by the French Centre National de Recherche Scientifique (CNRS), the Italian Istituto Nazionale di Fisica Nucleare (INFN) and the Dutch Nikhef, with contributions by institutions from Belgium, Germany, Greece, Hungary, Ireland, Japan, Monaco, Poland, Portugal, Spain. The construction and operation of KAGRA are funded by Ministry of Education, Culture, Sports, Science and Technology (MEXT), and Japan Society for the Promotion of Science (JSPS), National Research Foundation (NRF) and Ministry of Science and ICT (MSIT) in Korea, Academia Sinica (AS) and the Ministry of Science and Technology (MoST) in Taiwan.

A portion of this work was carried out while a subset of the authors (T.I., F.S., S.F., V.V., C.J.H. and R.S.) were in residence at the Institute for Computational and Experimental Research in Mathematics (ICERM) in Providence, RI, during the Advances in Computational Relativity program. ICERM is supported by the National Science Foundation under Grant No. DMS-1439786.
\end{acknowledgments}

\appendix
\section{Diagnostic checks for sampler}
\label{sec:sampler_diagnostics}

\subsection{Posterior's dependence on processes}
\label{sec:pbilby}

\begin{table*}
	\begin{tabular}{r | c | c | c | c | c | c | c | c | c | c }
		\hline\hline
		&$M (M_\odot)$ &$q$ &$m_1 (M_\odot)$ &$m_2 (M_\odot)$ &$\chi_1$ &$\chi_2$ &$\chi_{\rm eff}$ &$\chi_p$ &$D_{\rm L}$ (Gpc) &$\theta_{\rm JN}$ (rad) \\
		\hline \hline
		{GW150914\_095045}&&&&&&&&&&\\
		\texttt{NRSur7dq4}& $64.3 ^{+3.0}_{-2.8}$& $0.86 ^{+0.12}_{-0.21}$& $34.61 ^{+4.47}_{-2.69}$& $29.73 ^{+2.74}_{-4.50}$& $0.26 ^{+0.47}_{-0.24}$& $0.32 ^{+0.52}_{-0.29}$& $-0.05 ^{+0.10}_{-0.12}$& $0.33 ^{+0.41}_{-0.25}$& $0.5 ^{+0.1}_{-0.1}$& $2.72 ^{+0.31}_{-0.46}$\\
		\texttt{IMRPhenomXPHM}& $64.3 ^{+3.4}_{-3.2}$& $0.85 ^{+0.14}_{-0.22}$& $34.90 ^{+4.92}_{-2.98}$& $29.38 ^{+3.15}_{-4.99}$& $0.43 ^{+0.49}_{-0.39}$& $0.40 ^{+0.51}_{-0.36}$& $-0.06 ^{+0.11}_{-0.15}$& $0.51 ^{+0.38}_{-0.40}$& $0.5 ^{+0.1}_{-0.1}$& $2.69 ^{+0.32}_{-0.47}$\\
		\texttt{SEOBNRv4PHM}& $64.7 ^{+4.2}_{-3.2}$& $0.90 ^{+0.09}_{-0.15}$& $34.28 ^{+3.58}_{-2.23}$& $30.52 ^{+2.82}_{-3.29}$& $0.46 ^{+0.37}_{-0.40}$& $0.46 ^{+0.46}_{-0.41}$& $-0.02 ^{+0.12}_{-0.11}$& $0.50 ^{+0.31}_{-0.35}$& $0.5 ^{+0.1}_{-0.2}$& $2.71 ^{+0.32}_{-2.12}$\\
		\hline
		{GW190413\_134308}&&&&&&&&&&\\
		\texttt{NRSur7dq4}& $80.1 ^{+18.0}_{-11.4}$& $0.70 ^{+0.27}_{-0.34}$& $47.99 ^{+14.65}_{-10.26}$& $32.54 ^{+11.71}_{-11.64}$& $0.59 ^{+0.37}_{-0.52}$& $0.50 ^{+0.44}_{-0.45}$& $0.00 ^{+0.25}_{-0.29}$& $0.57 ^{+0.35}_{-0.42}$& $4.6 ^{+2.5}_{-2.3}$& $1.58 ^{+1.28}_{-1.29}$\\
		\texttt{IMRPhenomXPHM}& $81.3 ^{+16.8}_{-11.8}$& $0.60 ^{+0.35}_{-0.32}$& $51.28 ^{+16.59}_{-12.59}$& $30.43 ^{+11.71}_{-12.74}$& $0.60 ^{+0.36}_{-0.54}$& $0.49 ^{+0.44}_{-0.44}$& $-0.01 ^{+0.28}_{-0.38}$& $0.55 ^{+0.36}_{-0.41}$& $3.8 ^{+2.5}_{-1.8}$& $1.85 ^{+1.01}_{-1.52}$\\
		\texttt{SEOBNRv4PHM}& -& -& -& -& -& -& -& -& -& -\\
		\hline
		{GW190521\_030229}&&&&&&&&&&\\
		\texttt{NRSur7dq4}& $153.7 ^{+41.5}_{-19.8}$& $0.78 ^{+0.20}_{-0.34}$& $89.27 ^{+23.62}_{-16.52}$& $67.05 ^{+23.73}_{-22.02}$& $0.66 ^{+0.30}_{-0.59}$& $0.69 ^{+0.28}_{-0.61}$& $0.00 ^{+0.32}_{-0.32}$& $0.67 ^{+0.26}_{-0.42}$& $4.5 ^{+2.7}_{-2.5}$& $0.88 ^{+1.94}_{-0.62}$\\
		\texttt{IMRPhenomXPHM}& $153.1 ^{+42.2}_{-16.2}$& $0.59 ^{+0.33}_{-0.38}$& $98.40 ^{+33.58}_{-21.71}$& $57.25 ^{+27.14}_{-30.09}$& $0.71 ^{+0.26}_{-0.63}$& $0.53 ^{+0.42}_{-0.48}$& $-0.14 ^{+0.50}_{-0.45}$& $0.49 ^{+0.33}_{-0.35}$& $3.3 ^{+2.8}_{-1.8}$& $1.38 ^{+1.40}_{-1.07}$\\
		\texttt{SEOBNRv4PHM}& -& -& -& -& -& -& -& -& -& -\\
		\hline
		{GW190521\_074359}&&&&&&&&&&\\
		\texttt{NRSur7dq4}& $75.3 ^{+6.9}_{-4.9}$& $0.73 ^{+0.22}_{-0.17}$& $43.74 ^{+5.94}_{-5.89}$& $31.88 ^{+5.55}_{-5.33}$& $0.34 ^{+0.50}_{-0.31}$& $0.45 ^{+0.45}_{-0.40}$& $0.10 ^{+0.13}_{-0.13}$& $0.41 ^{+0.38}_{-0.29}$& $1.2 ^{+0.4}_{-0.5}$& $2.02 ^{+0.86}_{-1.76}$\\
		\texttt{IMRPhenomXPHM}& $77.3 ^{+6.6}_{-6.3}$& $0.76 ^{+0.19}_{-0.18}$& $43.99 ^{+5.11}_{-5.24}$& $33.63 ^{+5.32}_{-6.44}$& $0.22 ^{+0.49}_{-0.20}$& $0.36 ^{+0.51}_{-0.32}$& $0.07 ^{+0.12}_{-0.11}$& $0.29 ^{+0.39}_{-0.23}$& $0.9 ^{+0.5}_{-0.4}$& $1.48 ^{+1.11}_{-0.96}$\\
		\texttt{SEOBNRv4PHM}& $75.1 ^{+7.0}_{-5.2}$& $0.80 ^{+0.16}_{-0.24}$& $42.30 ^{+6.87}_{-5.08}$& $33.21 ^{+5.07}_{-6.79}$& $0.50 ^{+0.35}_{-0.44}$& $0.51 ^{+0.41}_{-0.45}$& $0.12 ^{+0.12}_{-0.15}$& $0.50 ^{+0.30}_{-0.31}$& $1.4 ^{+0.4}_{-0.6}$& $1.59 ^{+1.26}_{-1.31}$\\
		\hline
		{GW190527\_092055}&&&&&&&&&&\\
		\texttt{NRSur7dq4}& $59.6 ^{+23.9}_{-10.3}$& $0.62 ^{+0.34}_{-0.36}$& $37.11 ^{+20.07}_{-9.52}$& $22.35 ^{+11.00}_{-9.26}$& $0.48 ^{+0.46}_{-0.43}$& $0.47 ^{+0.46}_{-0.42}$& $0.10 ^{+0.28}_{-0.28}$& $0.45 ^{+0.42}_{-0.34}$& $2.5 ^{+2.6}_{-1.3}$& $1.15 ^{+1.64}_{-0.88}$\\
		\texttt{IMRPhenomXPHM}& $59.6 ^{+32.0}_{-10.2}$& $0.59 ^{+0.36}_{-0.38}$& $37.58 ^{+27.65}_{-9.90}$& $21.79 ^{+11.95}_{-10.29}$& $0.46 ^{+0.48}_{-0.41}$& $0.48 ^{+0.45}_{-0.43}$& $0.08 ^{+0.29}_{-0.30}$& $0.44 ^{+0.44}_{-0.34}$& $2.5 ^{+2.6}_{-1.3}$& $1.11 ^{+1.67}_{-0.84}$\\
		\texttt{SEOBNRv4PHM}& $57.2 ^{+9.5}_{-7.9}$& $0.65 ^{+0.28}_{-0.25}$& $34.64 ^{+8.77}_{-6.98}$& $22.32 ^{+6.96}_{-6.38}$& $0.31 ^{+0.50}_{-0.28}$& $0.34 ^{+0.51}_{-0.31}$& $0.10 ^{+0.17}_{-0.14}$& $0.32 ^{+0.43}_{-0.25}$& $2.5 ^{+1.6}_{-1.2}$& $1.15 ^{+1.66}_{-0.88}$\\
		\hline
		{GW191109\_010717}&&&&&&&&&&\\
		\texttt{NRSur7dq4}& $114.8 ^{+9.6}_{-13.3}$& $0.65 ^{+0.20}_{-0.19}$& $69.29 ^{+8.77}_{-8.88}$& $45.28 ^{+9.34}_{-11.06}$& $0.86 ^{+0.12}_{-0.30}$& $0.45 ^{+0.48}_{-0.41}$& $-0.38 ^{+0.21}_{-0.20}$& $0.59 ^{+0.26}_{-0.27}$& $0.9 ^{+0.8}_{-0.3}$& $2.09 ^{+0.49}_{-1.09}$\\
		\texttt{IMRPhenomXPHM}& $111.4 ^{+24.8}_{-16.0}$& $0.73 ^{+0.21}_{-0.24}$& $65.52 ^{+9.88}_{-11.13}$& $46.78 ^{+17.03}_{-13.08}$& $0.81 ^{+0.16}_{-0.62}$& $0.72 ^{+0.25}_{-0.64}$& $-0.31 ^{+0.53}_{-0.32}$& $0.55 ^{+0.33}_{-0.34}$& $1.1 ^{+1.1}_{-0.5}$& $1.87 ^{+0.75}_{-0.83}$\\
		\texttt{SEOBNRv4PHM}& $111.6 ^{+15.0}_{-15.0}$& $0.73 ^{+0.22}_{-0.24}$& $64.65 ^{+12.08}_{-9.97}$& $46.85 ^{+11.40}_{-12.59}$& $0.84 ^{+0.15}_{-0.38}$& $0.55 ^{+0.40}_{-0.50}$& $-0.28 ^{+0.26}_{-0.26}$& $0.70 ^{+0.23}_{-0.37}$& $1.5 ^{+1.1}_{-0.7}$& $2.05 ^{+0.82}_{-1.47}$\\
		\hline
		{GW200129\_065458}&&&&&&&&&&\\
		\texttt{NRSur7dq4}& $62.1 ^{+4.3}_{-2.9}$& $0.62 ^{+0.35}_{-0.21}$& $38.02 ^{+6.86}_{-6.51}$& $23.70 ^{+7.90}_{-5.33}$& $0.77 ^{+0.20}_{-0.63}$& $0.48 ^{+0.45}_{-0.42}$& $0.06 ^{+0.13}_{-0.15}$& $0.76 ^{+0.20}_{-0.47}$& $1.0 ^{+0.2}_{-0.4}$& $0.46 ^{+0.52}_{-0.26}$\\
		\texttt{IMRPhenomXPHM}& $63.3 ^{+4.9}_{-3.5}$& $0.73 ^{+0.23}_{-0.30}$& $37.02 ^{+8.70}_{-5.30}$& $26.51 ^{+5.38}_{-7.68}$& $0.79 ^{+0.18}_{-0.59}$& $0.60 ^{+0.35}_{-0.53}$& $0.11 ^{+0.14}_{-0.19}$& $0.76 ^{+0.20}_{-0.43}$& $1.0 ^{+0.2}_{-0.4}$& $0.54 ^{+0.54}_{-0.31}$\\
		\texttt{SEOBNRv4PHM}& $63.2 ^{+4.3}_{-3.4}$& $0.90 ^{+0.08}_{-0.15}$& $33.58 ^{+3.29}_{-2.45}$& $29.94 ^{+2.69}_{-3.38}$& $0.35 ^{+0.35}_{-0.32}$& $0.41 ^{+0.39}_{-0.35}$& $0.12 ^{+0.09}_{-0.12}$& $0.35 ^{+0.30}_{-0.24}$& $0.8 ^{+0.3}_{-0.3}$& $0.80 ^{+0.82}_{-0.52}$\\
		\hline
	\end{tabular}
	\caption{
	\label{tab:PEresults}
	Inferred source parameter values of the seven highlighted events in Section ~\ref{sec:special_events}.}
\end{table*}

In this paper, we have used \texttt{parallel-bilby}~\cite{Smith:2019ucc} to produce posterior samples.
Here, we describe a convergence issue that we encountered which is due to the way
\texttt{parallel-bilby} draws samples in parallel. This issue does not arise in serial nested sampling,
and our choice of sampler settings has to take into account the number of samples being drawn in
parallel on each iteration.

\begin{figure}
	\includegraphics[width=\columnwidth]{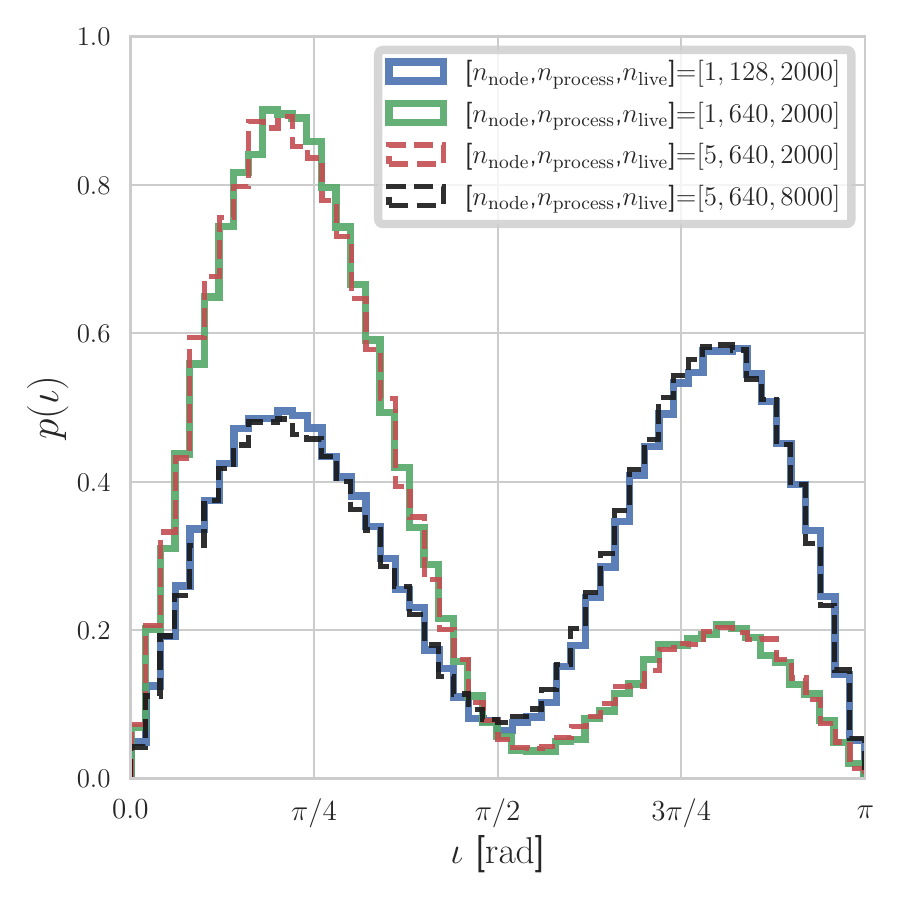}
	\caption{\label{fig:pbilby} We show the marginal posterior distribution for the inclination angle
		and its dependence on the number of processes used to parallelize the nested sampling algorithm
		for one of the representative events: GW190727\_060333. Here $n_{\rm node}$, $n_{\rm process}$ and
		$n_{\rm live}$ denote the number of nodes, processes and live points used in \texttt{dynesty}
		respectively. We generically find that as the number of live-points-per-process increases, the posteriors converge to a unique distribution. The details are in Appendix \ref{sec:pbilby}.
	}
\end{figure}

\texttt{parallel-bilby} uses a parallelized nested sampling algorithm implemented in
\texttt{dynesty}~\cite{speagle2020dynesty}, which parallelizes the step of drawing prior samples
to update the live points at each iteration.
After live points are updated, the prior volume is constrained to lie within a bounding ellipse containing
the new live points, shrinking the prior volume that needs to be sampled at subsequent iterations.
In contrast, at each iteration of serial nested sampling, only one prior sample is drawn at a time,
and a single live point is updated. The differences in how live points are updated affect how bounding
ellipses are drawn at each iteration. In practice, we find the parallel variant can prematurely
exclude regions of the prior with reasonable posterior support and becomes problematic whenever the number
of live-points-per-process becomes too low.

As the number of live-points-per-process increases, the posteriors converge to a
unique distribution. Unfortunately, it is not known ahead of time what this number should be.
We empirically determine the correct number by randomly selecting five events and systematically varying
this value, finding about 16 live-points-per-process is sufficient.

Fig.~\ref{fig:pbilby} shows the posterior's dependence as the number of live-points-per-process is
varied for one of the representative events GW190727\_060333. We consider computations carried out
with 640 processes on 1 node (green solid curve) and five nodes (red dashed curve), which give nearly
identical posteriors as we would expect based on how \texttt{parallel-bilby}'s parallelization is carried out.
For an identical setup using 1 node and 128 processes (blue solid curve), we infer a different posterior distribution, which demonstrates the parallelized sampler's dependence on the number of processes used. We further
show ``convergence'' of the posterior with live-points-per-process. We first use 640 processes distributed
across 5 nodes and set $n_{\rm live}=2000$. This gives about 3 live-points-per-process and the resulting
posteriors are shown as the red dashed curve. We increase the number of live-points-per-process by either
(i) increasing the total number of live points to 8000 (black dashed curve) or (ii) decreasing the number of
processes to 128 (blue solid curve). The blue and black curves visually agree, indicating that the
number of live-points-per-process is sufficiently large for this problem. Increasing the number of
live-points-per-process further still results in posteriors visually identical to the (already
converged) blue and black.

\begin{figure}
	\includegraphics[width=1.0\columnwidth]{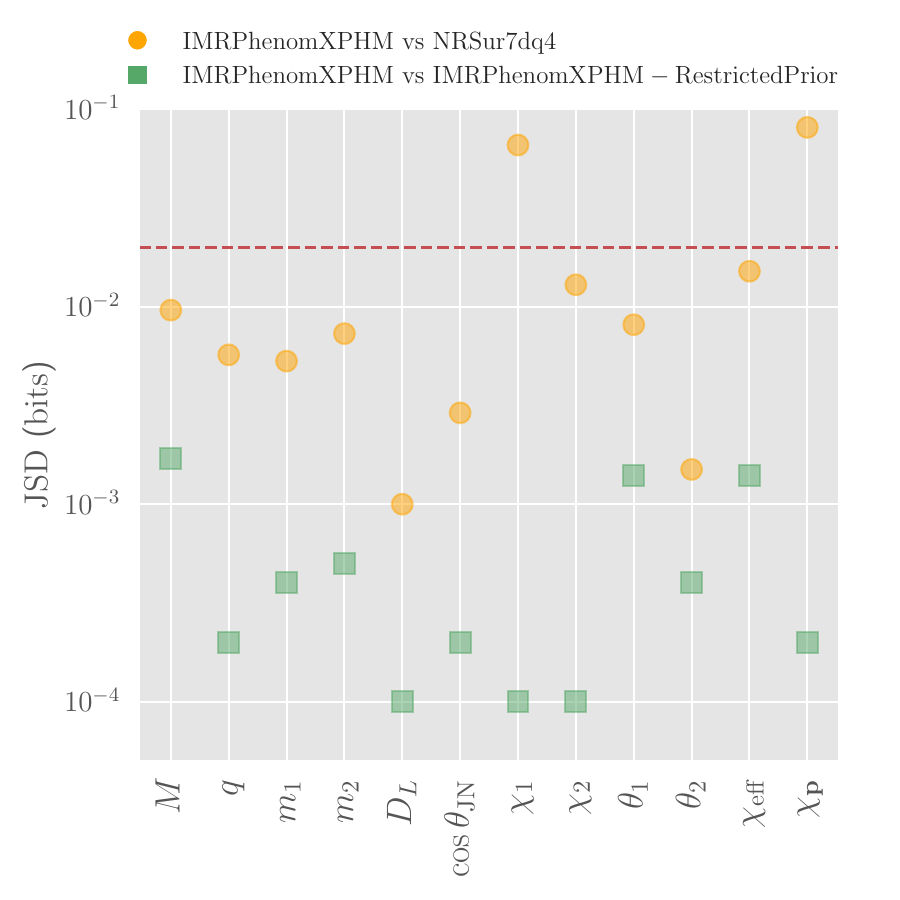}
\caption{\label{fig:jsdivs_in_house}
Jensen-Shannon divergence (JSD) values between the one-dimensional
marginalized posteriors for GW150914\_095045 for a set of parameters obtained using
\texttt{NRSur7dq4}. We compare (shown in green) the public LVK posteriors
(``IMRPhenomXPHM'') and the posteriors computed using our parameter estimation
setup as described in Sections~\ref{sec:settings}, \ref{subsec:prior},
\ref{subsec:strain}, and \ref{subsec:frame_choice}
(``IMRPhenomXPHM--RestrictedPrior''). Dashed red lines correspond to a JS
divergence of 0.02, indicating significant differences between these posteriors.
We find the JDS values (green markers) are well below this threshold, validating
the general correctness of our PE setup. For comparison, we show (shown in
orange) JSD values for LVK's \texttt{IMRPhenomXPHM} results vs our results
obtained with \texttt{NRSur7dq4}. When orange markers are larger than green
ones, we can safely ascribe PE differences to waveform systematics as opposed to
sampler systematics. Further details on GW150914\_095045 are given in
Sec.~\ref{subsubsec:GW15}.
}
\end{figure}

\subsection{Comparison: in-house vs LVK \texttt{IMRPhenomXPHM} posteriors}
\label{sec:in_house}

In Sec.\ref{sec:bayes_factors}, we reanalyzed all 47 events considered in this paper using the \texttt{IMRPhenomXPHM} model, employing the same restricted priors and sampler settings as for the \texttt{NRSur7dq4} runs (see Sec.\ref{sec:settings}). These restricted priors, detailed in Sec.\ref{subsec:prior}, are sufficiently broad to encompass the entire posterior distribution. Consequently, comparing posteriors obtained with our setup to the publicly available LVK posteriors\cite{LIGOScientific:2021usb, LIGOScientific:2021djp, GWTC2.1_PE, GWTC3_PE} using the same \texttt{IMRPhenomXPHM} model serves as a robust test of our parameter estimation (PE) workflow, including sampler settings, PSD computation, data handling, and software versions.

Our primary comparison method involves calculating the Jensen-Shannon divergence (JSD) between various one-dimensional marginalized posteriors. Consistent with the convention used throughout this paper, we interpret JSD values exceeding 0.02 bits as evidence of significant differences between the posteriors from our setup and those published by the LVK. Since the same models and matched parameter estimation (PE) settings are used, we expect most JSD values to remain below this threshold. Fig.~\ref{fig:jsdivs_in_house} presents this comparison for GW150914\_095045. The very small JSD values between the LVK posteriors and our own when using the \texttt{IMRPhenomXPHM} model (green) confirm the correctness of our PE setup. In contrast, the JSD values between the \texttt{NRSur7dq4} run and the LVK posteriors are significantly larger for a subset of parameters, indicating that the differences originate from waveform systematics rather than the PE setup. Similar results hold for the reanalyzed events; for example, over all 47 events, the 95\% JSD value for total mass is 0.005 (LVK vs our PE setup when using \texttt{IMRPhenomXPHM}) and 0.04 (\texttt{NRSur7dq4} posteriors and the LVK posteriors).

An important question arises: which PE results, ours or the LVK's, are ``correct''? Unfortunately, this is difficult to resolve. Even within the LVK's analyses, posteriors from different models (\texttt{IMRPhenomXPHM}, \texttt{SEOBNRv4PHM}, and combined samples) frequently exhibit JSD values exceeding the 0.02 threshold. At least for when comparing \texttt{IMRPhenomXPHM} and \texttt{NRSur7dq4}, we have controlled for sampler systematics by using the same PE setup as the LVK and verifying the correctness of this setup. Yet any events analyzed in this paper reveal tensions between \texttt{IMRPhenomXPHM} and \texttt{NRSur7dq4} models when interpreting gravitational wave signals.

Ultimately, model selection using Bayes factors is needed to assess which set of results is more likely to be correct~\footnote{When compared to precessing NR simulations the \texttt{NRSur7dq4} is more accurate than \texttt{IMRPhenomXPHM}. Hence one could argue that prior model odds (and hence the Bayes factors) should reflect this. We have taken a conservative approach and assume all models are equally likely when setting prior model odds.}. As reported in Sec.~\ref{sec:bayes_factors}, the Bayes factors slightly favor the \texttt{NRSur7dq4} model. However, this preference is not conclusive.

\begin{figure*}
	\includegraphics[width=1.0\textwidth]{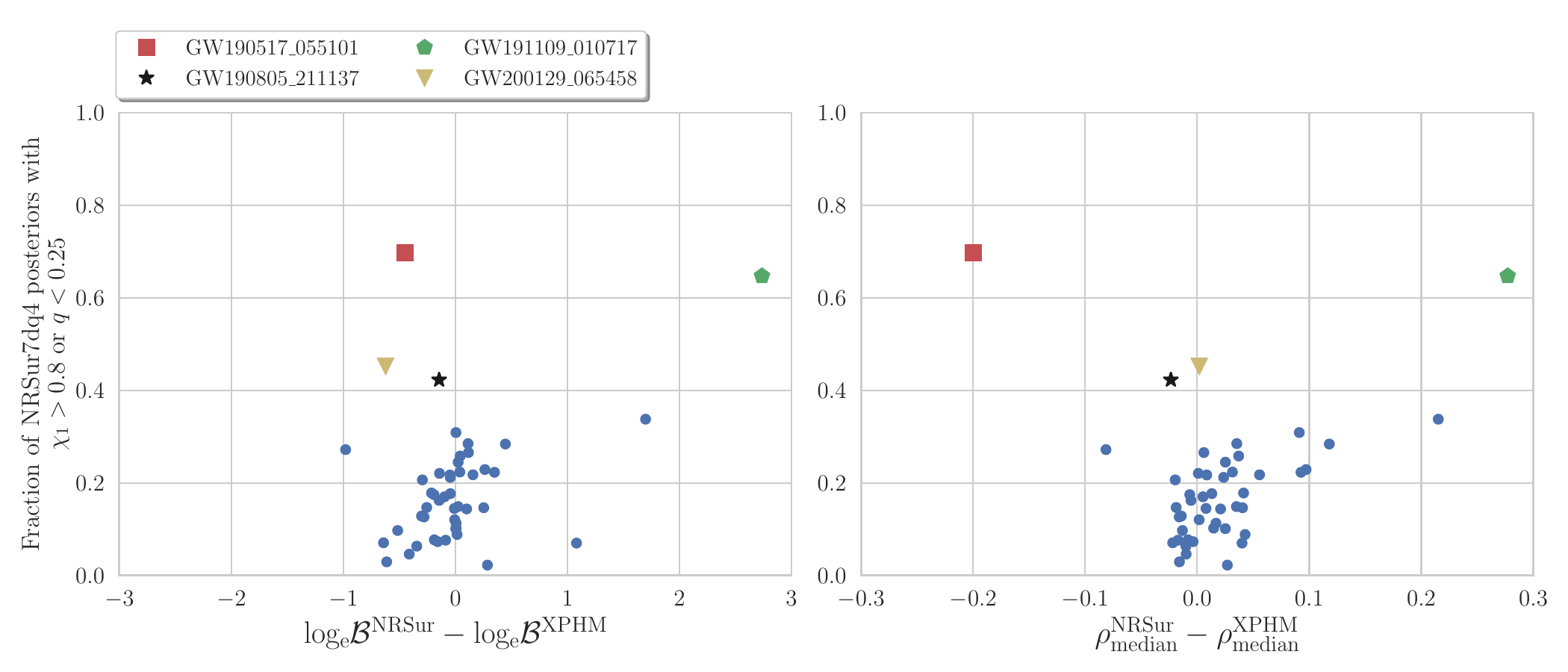}
	\caption{\label{fig:extrapolation_fraction}
		Fraction of \texttt{NRSur7dq4} posteriors with either $\chi_1 > 0.8$ or
		$q<1/4$ as a function of the differences in the recovered log Bayes factors
		(\textit{left panel}) and the median SNR (\textit{right panel}) between the
		\texttt{NRSur7dq4} (referred to as `NRSur') and
		\texttt{IMRPhenomXPHM} models (referred to as `XPHM') for
		all 47 events considered in this work. The events with a fraction $>0.4$ are
		highlighted. Further details are given in Section \ref{sec:bayes_factors} and
		Appendix ~\ref{sec:outliers_extrapolation}.
	}
\end{figure*}

\section{Understanding the outliers and correlations in model-selection diagnostics}
\label{sec:outliers}

In Sec.~\ref{sec:bayes_factors} we noted that while
there is no outright strong preference for either \texttt{NRSur7dq4} or \texttt{IMRPhenomXPHM}
when considering log Bayes factors or the SNR, the data seems to
show a mild preference
for \texttt{NRSur7dq4} over \texttt{IMRPhenomXPHM}.
In this appendix, we explore possible explanations for trends and outliers found while
performing the model selection.

\subsection{Impact due to model extrapolation}
\label{sec:outliers_extrapolation}
The \texttt{NRSur7dq4} model has been trained on a parameter domain defined by
$q\geq1/4$ and $\chi_{1},\chi_2\leq 0.8$, yet throughout this paper, we have used
it over the expanded region $q\geq1/6$ and $\chi_1, \chi_2 \leq 0.99$ (see Sec.\ref{subsec:prior}).
We note, however, that \texttt{IMRPhenomXPHM}
is also not well calibrated the $q<1/4$ or $\chi_1>0.8$ region, and wherever comparisons to NR are possible,
\texttt{NRSur7dq4} is more accurate than existing waveform models~\cite{varma2019surrogate, Varma:2022pld, Walker:2022zob}.

To investigate the possibility that the extrapolated model may produce inaccurate waveforms,
we compute the fraction of \texttt{NRSur7dq4}
posteriors that require extrapolation in either the mass ratio or the primary
spin magnitude - i.e. $q<1/4$ or $\chi_1>0.8$,\footnote{We do not include
extrapolation in $\chi_2$ in this fraction as $\chi_2$ is poorly measured and
is therefore prior dominated (see Fig.\ref{fig:violin_spin}).} and plot this
fraction in Fig.\ref{fig:extrapolation_fraction} as a function of the differences
in the log Bayes factor (left panel) and the median SNR (right panel) between
\texttt{NRSur7dq4} and \texttt{IMRPhenomXPHM} for all 47 events considered.
For the events where this fraction is relatively small ($<0.4$), we find a mild
correlation where \texttt{NRSur7dq4} is doing better than \texttt{IMRPhenomXPHM}
with increasing fractions.
For more extreme events (where the fraction is above 0.4 -- these are highlighted
in Fig.\ref{fig:extrapolation_fraction}), there is no clear correlation, and with only
4 events its challenging to say anything meaningful.
Synthetic NR injection studies could be used to systematically explore model fidelity in these
extreme regions of parameter space.

\begin{figure*}
	\includegraphics[width=1.0\textwidth]{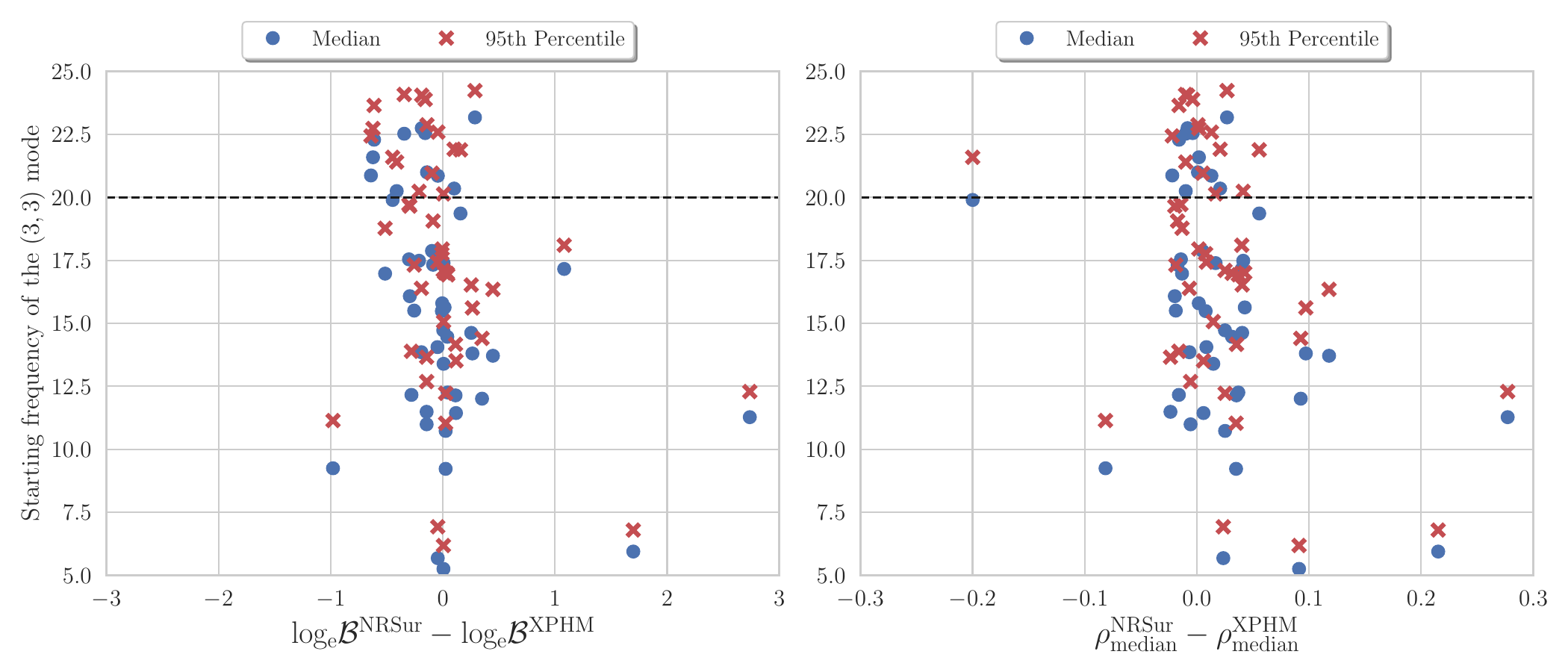}
	\caption{\label{fig:33_omega0}
		Median (blue circles) and 95th percentile (red crosses) values of the starting
		frequency of the $(3,3)$ mode of \texttt{NRSur7dq4} as a function of
		the differences in the recovered log Bayes factors (\textit{left panel}) and
		the median SNR (\textit{right panel}) between \texttt{NRSur7dq4} (referred to
		as `NRSur') and \texttt{IMRPhenomXPHM} models (referred to as
		`XPHM') for all 47 events considered in this work.  Black dashed
		lines indicate the low frequency cutoff ($f_{\rm low}=20$ Hz) used for the
		overlap integral. Further details are given in Section \ref{sec:bayes_factors} and
		Appendix ~\ref{sec:outliers_duration}.}
\end{figure*}

\subsection{Impact due to model duration}
\label{sec:outliers_duration}
Another possibility is that the \texttt{NRSur7dq4} model's limited length causes it to miss some of the
higher mode content near $f_{\rm low}=20$ Hz, resulting in a smaller Bayes Factor or SNR for some events.

As explained in Sec.\ref{sec:settings},
we use $f_{\rm low}=20$ Hz for the overlap integral, and our
\texttt{NRSur7dq4} evaluations return the full length of the surrogate
(about 20 orbits). Given this length restriction, the $(2,2)$ mode
for a $M_{{\rm det}} \sim 60 M_{\odot}$ BBH system will start at 20 Hz~\cite{varma2019surrogate}.
This means the higher harmonics will start at multiples of this frequency -- for example,
the next most important harmonic, the $(3,3)$ mode, will start at 30 Hz.
And in general, if the $(2,2)$ mode's starting frequency is $f_{22}$
the initial frequency of the $(\ell, m)$ waveform is  $m/2 \times f_{22}$ Hz.

By using the full length of the surrogate, the (2,2) mode is guaranteed to start
below $20$ Hz (for $M_{{\rm det}} \gtrsim 60 M_{\odot}$), while the higher
harmonics with $m>2$ may start above $20$ Hz. Here, we consider if the missing lower
frequency content of certain subdominant modes
could impact the Bayes Factor and/or the SNR recovered by the surrogate.

To investigate this possibility, for each event, we randomly choose 5000 samples
from the \texttt{NRSur7dq4} posteriors and compute the starting frequency of
the $(3,3)$ mode for the \texttt{NRSur7dq4} model in the co-precessing frame.
Figure~\ref{fig:33_omega0} shows the median and 95 percentile values of the $(3,3)$
mode starting frequency as a function of the differences in the log Bayes factor
(left panel) and the median SNR (right panel) between \texttt{NRSur7dq4} and
\texttt{IMRPhenomXPHM} for all 47 events considered. We find no clear
correlations suggesting that any missing low-frequency content in the higher
harmonics of \texttt{NRSur7dq4} may not be the cause for the
\texttt{IMRPhenomXPHM} model recovering a higher SNR for GW190517\_055101
and a higher Bayes factor for GW190706\_222641.

While there could be numerous other causes, another possibility worth noting is
that noise fluctuations can also cause the data to prefer one model over
another. Determining this will require more observations and/or careful NR
synthetic injection studies in realistic noise. As detector sensitivity
improves and the number of events increases, comparisons like
Figs.~\ref{fig:SNR}, \ref{fig:extrapolation_fraction} and \ref{fig:33_omega0}
can help determine which waveform models best describe the observed data.

\section*{References}
\bibliography{References.bib}

\end{document}